\documentclass{aa}  

\newcommand{\magphys}{\textsc{Magphys}}
\newcommand{\Lradio}{\ensuremath{L_\mathrm{150\,MHz}}}

\usepackage{graphicx}
\usepackage{subfigure}
\usepackage{txfonts}
\begin{document}

   \title{The LOFAR Two-metre Sky Survey Deep fields: }

   \subtitle{The star formation rate -- radio luminosity relation at low frequencies}

   \author{D.\,J.\,B. Smith
          \inst{1}
          \and
          P. Haskell\inst{1}
          \and
          G. G\"urkan\inst{2}
          \and
          P.\,N. Best\inst{3}
          \and
          M.\,J. Hardcastle\inst{1}
          \and
          R. Kondapally\inst{3}
          \and
          W. Williams\inst{4}
          \and
          K.\,J. Duncan\inst{4,3}
          \and
          R.\,K. Cochrane\inst{5}
          \and
          I. McCheyne\inst{6}
          \and
          H.\,J.\,A\, R\"ottgering\inst{4}
	 \and
	 J. Sabater\inst{3}
	 \and
	 T.\,W.\,Shimwell\inst{7,4}
	 \and
	 C. Tasse\inst{8,9}
          \and
          M. Bonato\inst{10,11,12}
          \and\\
          M. Bondi\inst{10}
          \and
          M.\,J. Jarvis\inst{13,14}
          \and 
          S.\,K. Leslie\inst{4}
          \and
          I. Prandoni\inst{10}
          \and
          L. Wang\inst{15,16}
        }

   \institute{Centre for Astrophysics Research, University of Hertfordshire, Hatfield, AL10 9AB, UK\\
              \email{d.j.b.smith@herts.ac.uk},
         \and
         CSIRO Astronomy and Space Science, PO Box 1130, Bentley WA 6102, Australia
         \and
        SUPA, Institute for Astronomy, Royal Observatory, Blackford Hill, Edinburgh, EH9 3HJ, UK
        \and 
        Leiden Observatory, Leiden University, PO Box 9513, BL-2300 RA Leiden, The Netherlands
        \and
        Harvard-Smithsonian Center for Astrophysics, 60 Garden St, Cambridge, MA 02138, USA
        \and
        Astronomy Centre, Department of Physics \& Astronomy, University of Sussex, Brighton, BN1 9QH, UK
        \and
        ASTRON, Netherlands Institute for Radio Astronomy, Oude Hoogeveensedijk 4, 7991 PD, Dwingeloo, The Netherlands
        \and
        GEPI \&\ USN, Observatoire de Paris, Universit\'e PSL, CNRS, 5 Place Jules Jannsen, 92190 Meudon, France 
        \and
        Department of Physics \&\ Electronics, Rhodes University, PO Box 94, Grahamstown, 6140, South Africa
        \and
        INAF - Istituto di Radioastronomia, Via P. Gobetti 101, 40129, Bologna, Italy
        \and
        Italian ALMA Regional Centre, Via Gobetti 101, I-40129, Bologna, Italy
        \and
        INAF-Osservatorio Astronomico di Padova, Vicolo dell'Osservatorio 5, I-35122, Padova, Italy
        \and 
        Astrophysics, Department of Physics, Keble Road, Oxford, OX1 3RH, UK
        \and
	Department of Physics \& Astronomy, University of the Western Cape, Private Bag X17, Bellville, Cape Town, 7535, South Africa
	\and
	SRON Netherlands Institute for Space Research, Landleven 12, 9747 AD, Groningen, The Netherlands
	\and
	Kapteyn Astronomical Institute, University of Groningen, Postbus 800, 9700 AV Groningen, the Netherlands
        }

   \date{Draft version \today}
 
  \abstract
    {In this paper, we investigate the relationship between 150\,MHz luminosity and star formation rate -- the SFR-\Lradio\ relation -- using 150\,MHz measurements for a near-infrared selected sample of 118,517 $z<1$ galaxies. 
    New radio survey data offer compelling advantages for studying star formation in galaxies, with huge increases in sensitivity, survey speed and resolution over previous generation surveys, and remaining impervious to extinction. The LOFAR Surveys Key Science Project is transforming our understanding of the low-frequency radio sky, with the 150\,MHz data over the ELAIS-N1 field reaching an RMS sensitivity of 20\,$\mu$Jy\slash beam over $10\,$deg$^2$ at 6 arcsec resolution. 
    All of the galaxies studied have SFR and stellar mass estimates derived from energy balance spectral energy distribution fitting, using  redshifts and aperture-matched forced photometry from the LOFAR Two-metre Sky Survey (LoTSS) deep fields data release. The impact of active galactic nuclei (AGN) is minimised by leveraging the deep ancillary data in the LoTSS deep fields data release, alongside median-likelihood methods that we demonstrate are resistant to AGN contamination.
    We find a linear and non-evolving SFR-\Lradio\ relation, apparently consistent with expectations based on calorimetric arguments, down to the lowest SFRs $< 0.01 M_\odot\,$yr$^{-1}$. However, we also recover compelling evidence for stellar mass dependence in line with previous work on this topic, in the sense that higher mass galaxies have a larger 150\,MHz luminosity at a given SFR, suggesting that the overall agreement with calorimetric arguments may be a coincidence.
  We conclude that, in the absence of AGN, 150\,MHz observations can be used to measure accurate galaxy SFRs out to $z = 1$ at least, but it is necessary to account for stellar mass in the estimation in order to obtain 150\,MHz-derived SFRs accurate to better than 0.5\,dex. Our best-fit relation is $\log_{10} (\Lradio\ / W\,Hz^{-1}) = (0.90\pm 0.01) \log_{10}(\psi/M_\odot\,\mathrm{yr}^{-1}) + (0.33 \pm 0.04) \log_{10} (M/10^{10}M_\odot) + 22.22 \pm 0.02$.}
    

   \maketitle
%
%-------------------------------------------------------------------

\section{Introduction}\label{sec:intro}

Observations at radio wavelengths have great advantages for studying star formation across cosmic history, in particular being impervious to the effects of the dust obscuration which blights star formation rate (SFR) measures at optical wavelengths \citep[e.g.][]{kennicutt1998}. In addition, as we embark on the construction of the Square Kilometre Array \citep[the SKA; e.g.][]{carilli2004,dewdney2009}, radio observations are undergoing an explosion of capabilities, including huge increases in survey speed, spatial resolution and sensitivity. The Low Frequency Array  \citep[LOFAR;][]{vanhaarlem2013} is already revolutionising our understanding of the low-frequency radio sky, and the LOFAR Surveys Key Science Project \citep[LSKSP;][]{rottgering2011} aims to survey the entire northern sky with an unprecedented combination of sensitivity and angular resolution. Huge progress has already been made; the first data release of the LOFAR Two-metre Sky Survey \citep[LoTSS:][]{shimwell2017,shimwell2019,duncan2019,williams2019} covered an area of 424\,deg$^2$ with a median sensitivity of 71\,$\mu$Jy at 150\,MHz and with 6 arcsec resolution, while the forthcoming second data release will cover 5,200\,deg$^2$ of the northern sky with similar sensitivity. Within the LSKSP, the Low Band Antenna (LBA) is also being used to carry out a sister survey to LoTSS at 60\,MHz -- the LOFAR LBA Sky Survey (LoLSS; de Gasperin et al. {\it in preparation}). 

As well as the 150\,MHz survey covering the entire northern sky, the LSKSP is also producing deeper observations of multiple 10\,deg$^2$-scale regions with the best multi-wavelength data, known as the LoTSS Deep Fields. The first three of these fields, Bo\"otes, Lockman Hole and ELAIS-N1 (which are the subject of the first deep field data release -- fully described in four papers: \citealt{tasse_lotss}, \citealt{sabater_lotss}, \citealt{kondapally_lotss} and \citealt{duncan_lotss}) reach sensitivities between 20-35\,$\mu$Jy RMS, a factor of 2-3$\times$ deeper than the standard LoTSS observations. 
At these flux densities the source counts are increasingly dominated by star-forming galaxies \citep[e.g.][]{retana2018,williams2019} in which the synchrotron radiation is attributed to electrons accelerated by the remnants of supernovae that are the end-points of the evolution of short-lived, massive stars. The association between particle acceleration and supernova rate has been used to directly calibrate the synchrotron luminosity as a star formation rate indicator \citep[e.g.][]{condon1992,cram1998}.

Further support for the use of radio frequency observations to study star formation in galaxies comes in the form of the far-infrared radio correlation (the FIRC), which many works have shown to be a tight and constant relationship, which persists over several orders of magnitude in luminosity \citep[e.g.][]{vanderkruit1971,dejong1985,helou1985,yun2001,appleton2004,jarvis2010,ivison2010a,ivison2010b,sargent2010,bourne2011,delhaize2017}. However, relying on the FIRC to underpin the SFR calibration of radio luminosity is sub-optimal, in the sense that although a constant FIRC can be explained by calorimetry arguments, conspiracies \citep[i.e. the precise balance between disparate phenomena such as the cosmic ray electron escape fraction and the optical depth to UV photons; e.g.][]{lisenfeld1996,bell2003,lacki2010} are required to explain the correlation itself. Furthermore, there is good empirical evidence that the FIRC is not constant, varying in different galaxy types \citep[e.g.][]{molnar2018,read2018}, as a function of dust temperature \citep{smith2014}, and redshift \citep[e.g.][]{magnelli2015}.  

Clearly then, it is essential that we use the best observations available to test the efficacy of radio continuum data as a star formation rate indicator, and to determine the best functional form to use. 
Many observational works have looked directly at the SFR -- radio luminosity relation \citep[e.g.][]{condon1992,cram1998,garn2009,kennicutt2009,murphy2011,tabatabaei2017}, finding broad support for proportionality. On the other hand, \citet{bell2003} determined a downturn in the relation at low SFR (in the sense of lower luminosity per unit SFR) on the pragmatic basis of a decreasing non-thermal fraction at 1.4\,GHz in less luminous sources,  couched within a wider discussion of possible physical reasons for the variation, including the possibility of increasing cosmic ray escape in this regime \citep[e.g.][]{chi1990,murphy2008}. Similarly, \citet{lacki2010a} and \citet{murphy2009} predict a deviation in the relation at high-$z$ due to inverse Compton losses as a result of a significantly increased cosmic microwave background photon density at earlier points in cosmic history. \citet{davies2017} used data from the GAMA \citep{driver2011} and FIRST \citep{becker1995} surveys to study the SFR radio luminosity relation, while \citet{hodge2008} used the SDSS reprocessed spectroscopy \citep{york2000} in the MPA-JHU catalogue \citep{brinchmann2004} alongside FIRST data to do likewise, with both studies uncovering clear evidence for a super-linear slope\footnote{Note that the slope of $0.75\pm0.03$ quoted by \citet{davies2017} uses the inverse definition of the SFR radio luminosity relation, which is why the value quoted is lower than unity.}. Each of these presents strong evidence for deviation from the linear (i.e. gradient of unity) form expected on the basis of calorimetric models. 

As well as the benefits of survey speed, sensitivity and resolution that LOFAR provides, LoTSS observations have far superior sensitivity to extended emission than FIRST due to the short baselines sampled \citep[e.g.][]{hardcastle2019a,mahatma2019,tasse_lotss}. In addition, the low frequencies may also be preferable for studying the relationship between star formation rate and radio luminosity, since free-free emission  \citep[which can dominate at GHz frequencies, e.g.][]{condon1992,murphy2011} makes a negligible contribution to the 150\,MHz luminosity. Low-frequency observations are therefore potentially ``cleaner'' than the much
better studied GHz frequency range \citep[provided that we do not observe to such low frequencies that the spectra become self-absorbed, e.g.][]{kellermann1988}. With the advent of LOFAR, several works have looked at this issue at low frequencies \citep[e.g.][]{brown2017,calistro2017,gurkan2018,wang2019}. \citet{brown2017} found a slightly super-linear SFR-relationship between SFR and 150\,MHz luminosity (hereafter "SFR-\Lradio" relation) with a gradient of $1.14 \pm 0.05$, using 150\,MHz data from the TGSS \citep{intema2017}. \citet{wang2019} found an even steeper relationship (gradient of $1.3-1.4$) using a 60\,$\mu$m-selected sample from the revised IRAS Faint Source Survey Redshift Catalogue \citep{wang2014} matched to LOFAR data from LoTSS DR1. \citet{calistro2017} used a 150\,MHz-selected sample of 750 $z < 2.5$ star-forming galaxies identified in LoTSS data over 7\,deg$^2$ of the Bo\"otes field, alongside SFRs from SED fitting to find evidence for significant evolution in the FIRC, curved radio spectra, and an increase in the observed radio luminosity for a given SFR at higher redshift. 

Of particular relevance, \citet[][hereafter G18]{gurkan2018} studied SFR-\Lradio\ using a spectroscopically-classified sample from the MPA-JHU catalogue containing $\sim15$k galaxies with the first LOFAR observations of the HATLAS NGP field \citep[from][reaching an RMS of 100\,$\mu$Jy near the centre of the field but with non-uniform coverage]{hardcastle2016}. G18 used SFRs and stellar masses based on 14-band spectral energy distribution fitting using the \magphys\ code \citep{dacunha2008}, and found compelling evidence for (a) significant scatter about the maximum-likelihood SFR-\Lradio\ relation, (b) a strong preference for a mass-dependence in SFR-\Lradio, and perhaps most intriguingly (c) an upturn towards larger radio luminosity at SFRs, $\log_{10} (\psi/M_\odot$\,yr$^{-1}) < 0$, in the opposite sense to that expected by calorimetry arguments. This low-SFR \Lradio\ excess was also found by Read et al. (\textit{in preparation}) using 150\,MHz data from LoTSS DR1 over the HETDEX field alongside SFRs derived from the MPA-JHU catalogue. The cause of this extra radio luminosity in the galaxies with the lowest SFRs is of potentially great interest, since the rapid increase in radio survey capabilities mean that future radio surveys will access this regime ever more readily. Possible mechanisms for providing additional cosmic rays at low SFRs include pulsars, type Ia supernovae, residual contamination by active galactic nuclei (AGN), or varying magnetic field properties \citep[see also][]{sudoh2020}. 

Putting these issues to one side, perhaps the principal reason \textit{not} to use low-frequency radio observations for studying star formation is that other physical phenomena can also be bright at radio frequencies. The main issue for this type of study is activity due to AGN, where the synchrotron radio luminosity is linked not to star formation but to accretion, with the necessary relativistic electrons instead coming from particle acceleration in jets. While \citet{sabater2019} showed that the most massive radio-selected galaxies are always radio loud, AGN are far from ubiquitous in optically-selected samples. However, AGN contamination is virtually impossible to eradicate completely from samples of star-forming galaxies at \textit{any} wavelength, since AGN can be highly dust obscured \citep[e.g.][]{antonucci1993,lacy2004,martinez2005}, since accretion processes are variable \citep[e.g.][]{read2020} and since AGN are highly multimodal \citep[e.g.][]{hardcastle2007,best2012}. In addition, radio AGN have a very large range of power \citep[e.g.][]{hardcastle2019} extending down to values that would be typical of star forming galaxies \citep[e.g.][]{sadler2002,hardcastle2016,lofthouse2018}, making simple cuts in luminosity suboptimal. 

Despite the issues that AGN contamination present, the potential gains from using radio survey data for studying star formation are very large, and this is reflected by the wide range of authors who have done so \citep[e.g.][]{haarsma2000,hopkins2003,pannella2009,karim2011,zwart2014,pannella2015,bonato2017,novak2017,upjohn2019,leslie2020}. In this paper we revisit the low-frequency radio-luminosity star formation rate relation, using data from the LOFAR deep fields. The structure of this work is as follows: in section \ref{sec:data} we introduce the different data sets that we use for our analysis, while in section \ref{sec:results} we discuss our analysis and present the results. In section \ref{sec:conclusions} we present some concluding remarks. We assume a standard cosmology with $H_0 = 70\,$km\,s$^{-1}$\,Mpc$^{-1}$, $\Omega_m=0.3$ and $\Omega_\Lambda = 0.7$.
    
\section{Data}\label{sec:data}

In this work we focus on the 6.7\,deg$^2$ of the ELAIS-N1 field where the best multi-wavelength data exist \citep{kondapally_lotss}. We describe the key data sets that we use in sections \ref{sec:mwdata} -- \ref{sec:agn}.

\subsection{Multi-wavelength data}
\label{sec:mwdata}

\begin{figure}
   \centering
	\includegraphics[width=1.01\columnwidth]{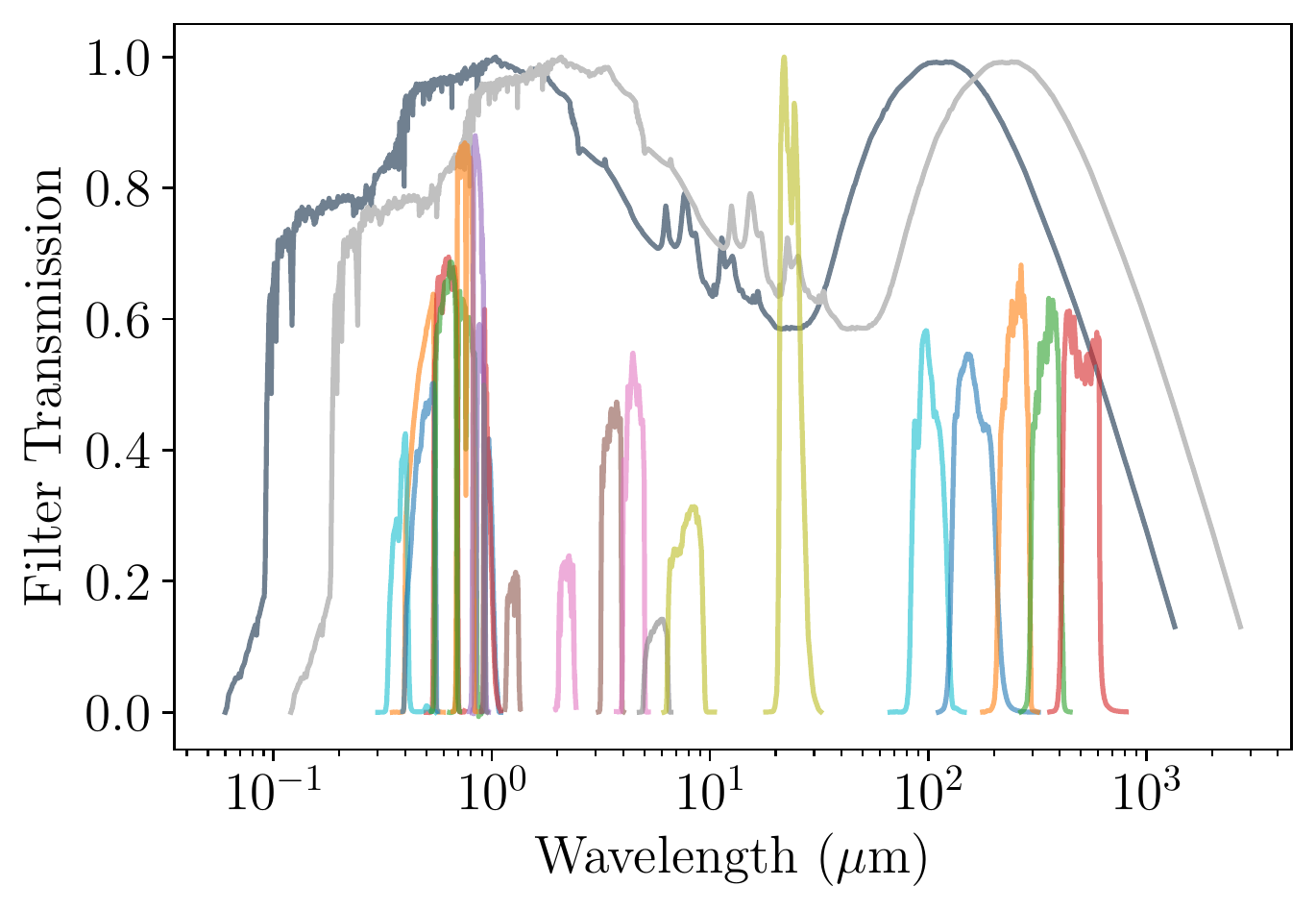}
	\caption{Filter curves of the assembled multi-wavelength data set (coloured lines), overlaid with a galaxy template from \citet{smith2012}. The SED has been normalised to have a peak flux of 1, and is shown both at $z = 0$ (in grey), and with the wavelength axis shifted to $z = 1$ (in silver) to give the reader an impression of how the excellent multi-wavelength coverage in ELAIS-N1 samples a galaxy's SED.}
	\label{fig:filters_sed}
\end{figure}

We use the aperture-matched photometry from \citet{kondapally_lotss}. In the ELAIS-N1 field, the photometric bands include SpARCS u band  \citep{wilson2009,muzzin2009}, PanSTARRS grizy \citep{chambers2016}, grizy and NB921 narrow-band data from HSC-SSP public data release 1 \citep{aihara2018}, J \&\ K  band data from the DXS \citep{lawrence2007}, IRAC channels 1-4 data from SWIRE \citep{lonsdale2003} and channels 1 \&\ 2 data from SERVS \citep{mauduit2012}. These data have been reprocessed onto a common pixel scale, and used to produce consistent aperture forced photometry, as well as state-of-the-art photometric redshift information from \citet{duncan_lotss} for every source in the catalogue. In this paper we use the compiled spectroscopic redshifts where available and reliable, and the Duncan et al. photometric redshift estimates otherwise (1.5 percent of our sample have spectroscopic redshifts). We have also used the SWIRE 24\,$\mu$m data as provided by the HELP project \citep{shirley2019}, which also includes far-infrared data in the 100, 160, 250, 350 and 500\,$\mu$m bands from HerMES \citep{oliver2012}, probabilistically deblended using the XID+ tool \citep{hurley2017} as described by McCheyne et al. \textit{in preparation}. To demonstrate the full wavelength coverage of the extensive multi-wavelength dataset that has been assembled, in figure \ref{fig:filters_sed} we show the filter curves overlaid on an indicative galaxy spectrum template from \citet{smith2012}. The template is shown both at $z = 0$ and with the wavelength axis shifted to $z = 1$ to illustrate the range spanned by our sample; it is clear that there is excellent coverage all the way from the near-ultraviolet to the far-infrared wavelengths. 

\begin{table}[]
\caption{Percentage of the area covered in this study by each set of ancillary data used for the SED fitting, along with the mean percentage of sources detected at $\ge 3\sigma$ in the most sensitive band of each data set. For the depth of the data in each band, please refer to \citet{kondapally_lotss}. }
\centering
\begin{tabular}{lll}
%\hline 
Band & \%\ Coverage & $\% \ge 3\sigma$ \\
\hline
SpARCS u & 100 & 68 \\
PanSTARRS grizy & 100 & 100 \\
HSC grizy, nb921 & 94 & 94 \\
DXS JK & 100 & 100 \\
SWIRE IRAC ch1 & 100 & 100 \\
SWIRE IRAC ch 2-4 & 100 & 96 \\
SERVS IRAC 1\,\&\,2 & 34 & 34\\
MIPS 24\,$\mu$m & 85 & 54 \\
\textit{Herschel} PACS & 100 & 1 \\
\textit{Herschel} SPIRE & 100 & 15  
%\hline
\end{tabular}
\label{tab:coverage}
\end{table}

\begin{table}[]
\caption{Percentage of sources in our parent sample with measured photometry (middle column) and a $\ge 3\sigma$ detection in \textit{at least} N bands. For example, 98 percent of our sample has measured photometry in 17 or more bands, and more than half have $\ge 3\sigma$ detections in at least 16 bands. }
\centering
\begin{tabular}{lll}
Number of bands & \%\ with photometry & \% $\ge 3\sigma$ \\
\hline
7  &  100  &  100  \\
8  &  100  &  99  \\
9  &  100  &  98  \\
10  &  100  &  96  \\
11  &  100  &  94  \\
12  &  99  &  92  \\
13  &  98  &  89  \\
14  &  98  &  86  \\
15  &  98  &  77  \\
16  &  98  &  62  \\
17  &  98  &  44  \\
18  &  84  &  28  \\
19  &  72  &  16  \\
20  &  67  &  8  \\
21  &  61  &  4  \\
22  &  61  &  2  \\
23  &  61  &  1  \\
24  &  25  &  0  \\
25  &  21  &  0  
\end{tabular}
\label{tab:nbands}
\end{table}

\subsubsection{Sample definition}

To avoid the possible influence of radio selection on our results (see Appendix \ref{sec:detections} for further details), we begin with a sample identified in the 3.6\,$\mu$m data from SWIRE. These data are not only very sensitive, but they are also less susceptible to the influence of the dust obscuration that could introduce bias into samples identified at shorter wavelengths. We select those sources with measured 3.6\,$\mu$m flux density $> 10\,\mu$Jy, approximately equal to the $5\sigma$ detection threshold in the SWIRE 3.6\,$\mu$m data, and as recommended on the SWIRE webpages\footnote{\url{https://irsa.ipac.caltech.edu/data/SPITZER/SWIRE/}}. We include only those sources with the necessary flags in the band-merged catalogue, as recommended in \citet{kondapally_lotss} to ensure we use only those sources with the highest-quality photometry, best wavelength coverage and most reliable photometric redshifts\footnote{As noted in \citet{kondapally_lotss}, in the ELAIS-N1 field, the necessary flags to apply are {\texttt{flag\_clean = 1 \&\ flag\_overlap = 7}}}. This leaves us with a parent sample of 142,037 3.6\,$\mu$m-selected sources at $z < 1$, where Duncan et al. show that the photometric redshifts have an average scatter $\sigma_\mathrm{NMAD} < 0.04(1+z)$ and an outlier fraction around 5\%\footnote{The scatter is measured in terms of the normalised median absolute deviation, which is defined as $\sigma_\mathrm{NMAD} = 1.48 \times \mathrm{median}(|\Delta z|/(1+z_\mathrm{spec}))$, while outliers are defined as those sources where $|\Delta z|/(1+z_\mathrm{spec}) > 0.15$}.

%flag_clean == 1 & i_e_raw > 0 & ch2_swire_e_raw > 0 & z_best_1 < 1.5 & (ch1_swire_f > 1e-5 || ch2_swire_f > 1.2e-5) & flag_overlap == 7

\subsection{LOFAR data}

We use the new deep LOFAR observations of the ELAIS-N1 field \citep{sabater_lotss} which cover an area of 10\,deg$^2$ with an RMS below 30\,$\mu$Jy, and reaching 20\,$\mu$Jy in the deepest regions, making them the most sensitive data in the LoTSS Deep Fields first data release.  Although the area covered is significantly narrower than that studied in G18, the 150\,MHz data are around five times deeper. Importantly, due to the very high quality of the multi-wavelength data in this field, more than 97 per cent of the sources in the ELAIS-N1 LoTSS 150\,MHz catalogue have counterparts in at least one band, which have been reliably identified using a new colour-dependent implementation \citet{kondapally_lotss} of the Likelihood Ratio method \citep[][]{sutherland1992,smith2011,mcalpine2012,nisbetthesis,williams2019}. 
We use the \citet{sabater_lotss} catalogue total 150\,MHz flux densities and uncertainties where they exist, however for the $\sim 91$\,percent of IRAC sources which are not identified as the counterparts of sources in the 150\,MHz catalogue we use pixel flux densities, specifically the flux density measured in the 150\,MHz map at the pixel corresponding to the coordinates of each IRAC source. These values indicate the maximum likelihood estimate of the integrated flux density for sources which are unresolved on the scale of the 6\,arcsec LoTSS beam, and are strongly preferred over aperture photometry in these deep data since they are less susceptible to influence from neighbouring sources. Including these sources is especially important, since although they are not individually detected (in the sense that they have a signal-to-noise ratio $< 3$) they are numerically dominant, and together can provide significant diagnostic power, as we shall demonstrate. To limit the potential influence of IRAC sources with 150\,MHz emission more extended than 6\,arcsec, we consider only those 130,689 sources at $z > 0.05$. The Kondapally et al. catalogue includes size information for sources in the $\chi^2$ image used for source detection, which we have used to estimate the extent of the sources in our sample. The mean FWHM of galaxies along their major axes decreases from 2.7\,arcsec at the lowest redshifts to around 1.2\,arcsec by $z = 1$. In practice therefore, this cut does not have significant impact on our results since $>99.5$\,per cent of $z < 1$ sources have FWHM smaller than 6\,arcsec in the multiwavelength data, and $>98$\,percent even at $z < 0.2$.  We also include 150\,MHz flux density uncertainties measured from the corresponding pixel of the RMS map. 

\subsection{SFR and Mass estimation}
We use the panchromatic energy balance SED fitting code \magphys\ to fit the multi-wavelength matched-aperture photometry and determine the physical properties of each source. \magphys\ is fully described in \citet{dacunha2008}, but to summarize, it uses an energy balance criterion to link the stellar emission that dominates at optical\slash near-infrared wavelengths with the dust emission that dominates in the far-infrared. The stellar emission is modelled using the templates from \citet{bruzual2003} assuming an initial mass function (IMF) from \citet{chabrier2003}, attenuated using a two-component dust model from \citet{charlot2000}, and the energy absorbed is then re-radiated using a multi-component dust model (including dust grains of various sizes and temperatures, as well as polycyclic aromatic hydrocarbons). The energy balance criterion is used to combine an optical\slash near-infrared library based on 50,000 SEDs representing exponentially-declining star formation histories with stochastic bursts superposed, with a library of 50,000 dust SEDs with a realistic range of physical properties. The energy balance criterion is designed to ensure that only physical combinations of the stellar and dust SEDs are used to model the input photometry. \magphys\ does not account for possible additional dust heating due to AGN \citep[though see e.g.][for an attempt to include it]{berta2013}. As well as producing best-fit SEDs for every source, we also obtained best-fit physical parameters, with marginalised probability distribution functions (PDFs) for each parameter. These PDFs are used to derive median-likelihood parameter estimates (which are the values corresponding to the 50th percentile of the PDF) as well as the 16th and 84th percentiles which are equivalent to the $\pm 1\sigma$ values for each parameter in the limit of Gaussian statistics.

\begin{figure}
   \centering
	\includegraphics[width=1.01\columnwidth]{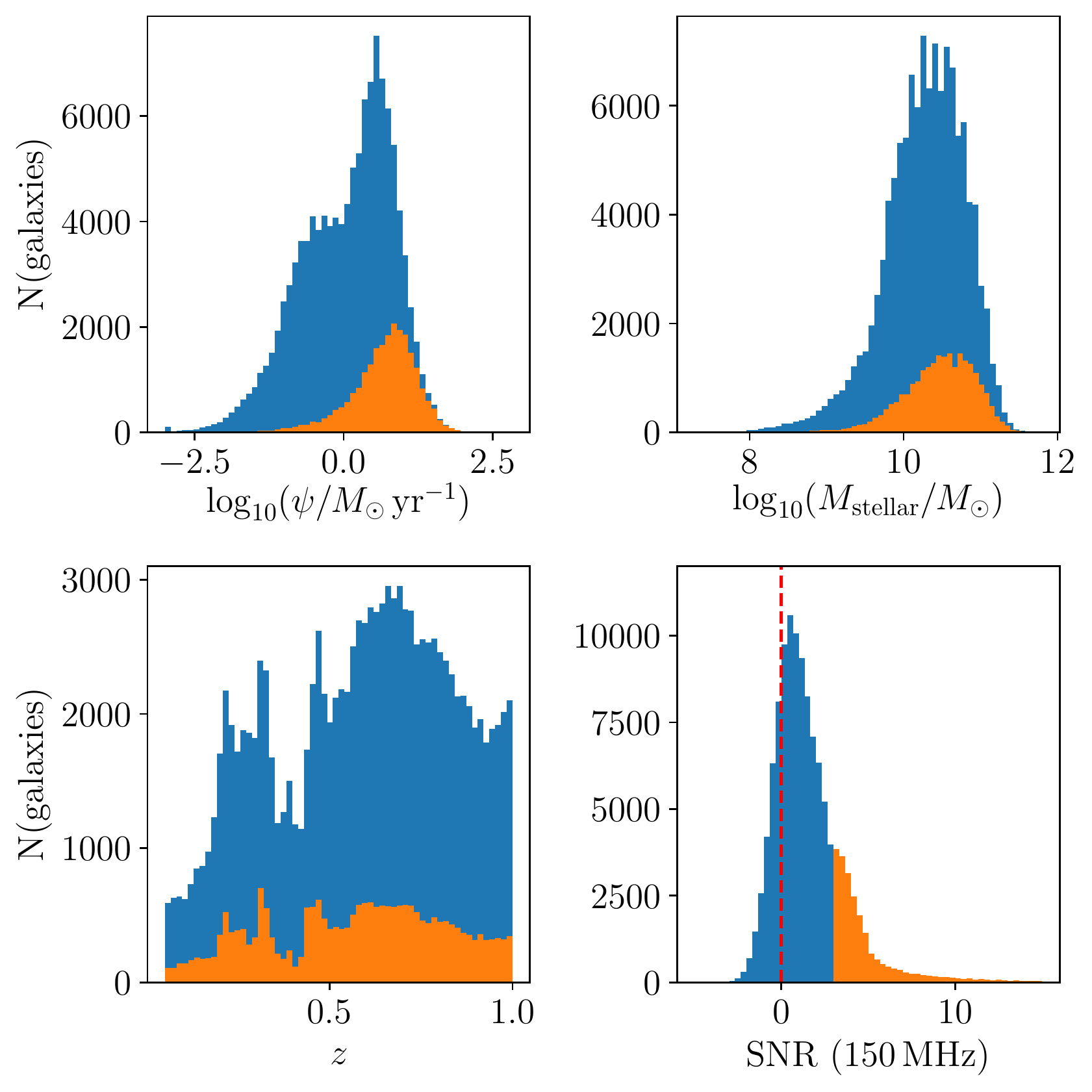}
	\caption{Histograms (clockwise from the top-left) showing the SFR ($\psi$), stellar mass, 150\,MHz signal-to-noise ratio and redshift distributions for the 118,517 galaxies in our 3.6\,$\mu$m-selected $z < 1$ sample after applying the cuts discussed in section \ref{sec:data}. The overlaid orange histograms show the corresponding distributions for the subset of the sample that is detected to $\ge 3\sigma$ at 150\,MHz. }
	\label{fig:sample}
\end{figure}

Since the multi-wavelength data over the EN1 field have been compiled from a wide range of sources with different coverage (see section \ref{sec:mwdata}), not every source has measured photometry in every band. Table \ref{tab:coverage} shows the fraction of sources in our sample that has coverage and\slash or a $\ge 3\sigma$ detection in each band, while Table \ref{tab:nbands} shows the fraction of sources that have measured photometry and $\ge 3\sigma$ detections in at least $N$ photometric bands. While Table \ref{tab:coverage} highlights that 100 per cent of our sample has \textit{Herschel} coverage, only $\sim 15$\%\ are detected at $\ge 3\sigma$ in the most-sensitive 250\,$\mu$m band, with 22\% $\ge 2\sigma$, and 43\% $\ge 1\sigma$. Indeed, only $\sim 65$\%\ of the sample has been assigned \textit{Herschel} photometry by XID+; that the remaining 3.6\,$\mu$m sources were not assigned any measurable \textit{Herschel} flux density, despite the absence of an SED prior in the version of XID+ we have used \citep[cf.][]{pearson2017}, is a strong indication that they are fainter than the confusion noise in each of the PACS and SPIRE bands \citep{hurley2017}. To include this ``upper limit'' information in the \magphys\ parameter estimation, we assign these sources uncertainties in the PACS and SPIRE bands equal to the median uncertainty for the sources that do have measured flux densities (these values are 10.3, 14.1, 2.6, 2.8 \&\ 3.7\,mJy in the 100\slash160\slash250\slash350\slash500\,$\mu$m bands, respectively), alongside a small flux density (equal to 0.1\%\ of the median uncertainty). \footnote{We have repeated all of the following analysis both with and without including these sources in the analysis, and the results do not show significant variation.}

The principal quantities of interest for this work are the stellar mass and the star formation rate averaged over the last 100\,Myr, for which \magphys\ estimates have been shown to be reliable in a range of different situations \citep[e.g.][]{hayward2015,smith2018,dud2020}. To ensure that the \magphys\ parameters are as reliable as possible, we further refine our sample to include only those 123,425 galaxies for which \magphys\ has been able to produce an acceptable fit. We define ``acceptable" in the sense that the $\chi^2$ parameter comparing the goodness of fit between the model and the observed photometry is below the 99 percent confidence threshold derived for the number of bands of photometry available for each object, as discussed in \citet{smith2012}. Though we are unable to repeat their emission line classification, our sample is roughly a factor of eight times larger than the one used by G18. In addition, the multi-wavelength dataset is far deeper; the HSC $i$-band data in ELAIS-N1 are four magnitudes deeper (and the comparison is similar in the other HSC bands), while the IRAC 3.6$\mu$m data are $>20 \times$ as sensitive as the closest comparable band in that work. Not only are the individual observations more sensitive, but there are also many more bands of photometry available for us to use. We demonstrate this in table \ref{tab:nbands}, which shows that more than half of our sample has photometry in 23 or more bands, as compared with a maximum of 14 that were available in the HATLAS-NGP area used by G18. 

\begin{figure}
   \centering
\subfigure{\includegraphics[width=1.01\columnwidth, trim=0cm 0cm 0cm 0cm,clip=true]{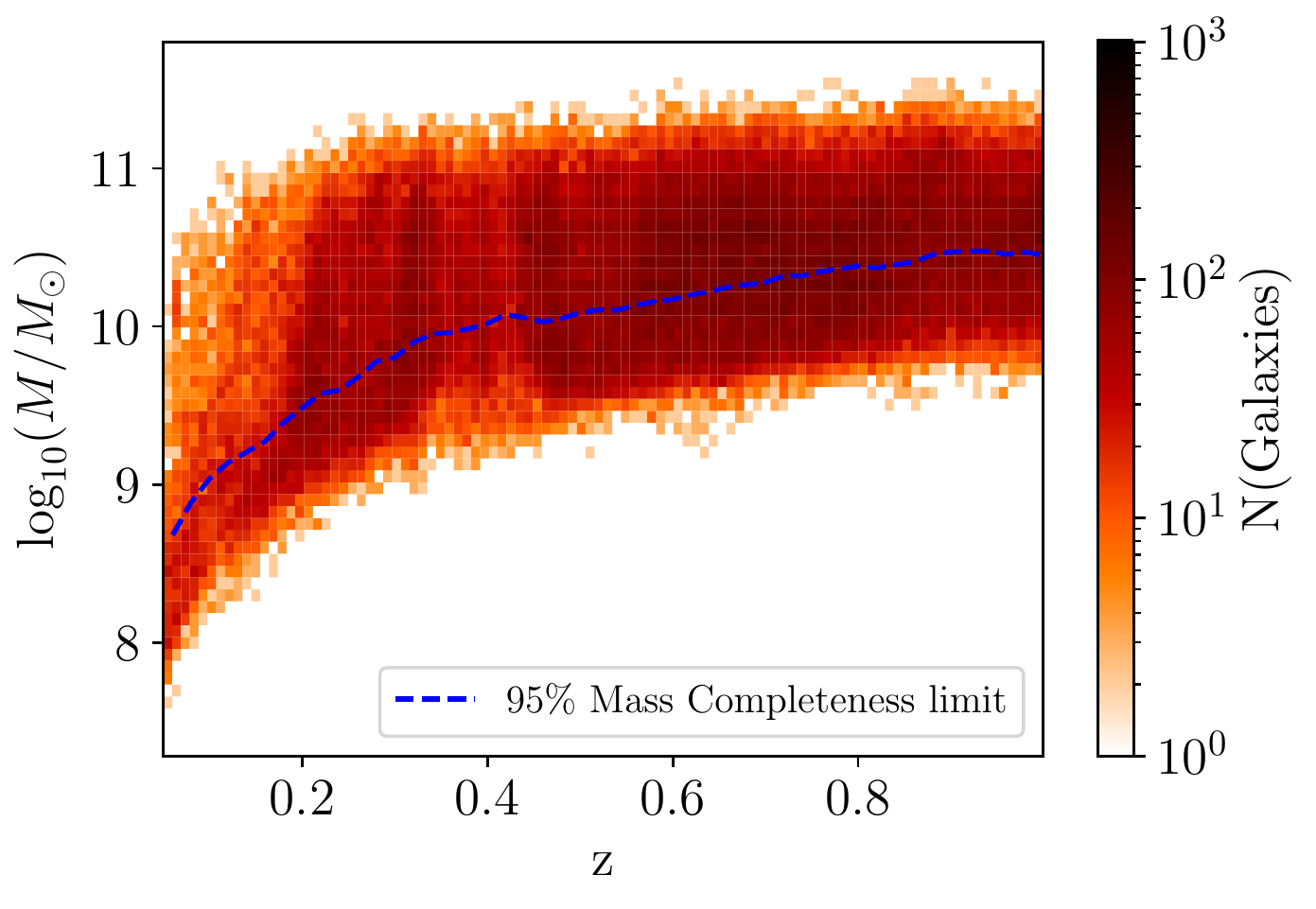}}
\subfigure{\includegraphics[width=1.01\columnwidth, clip=true]{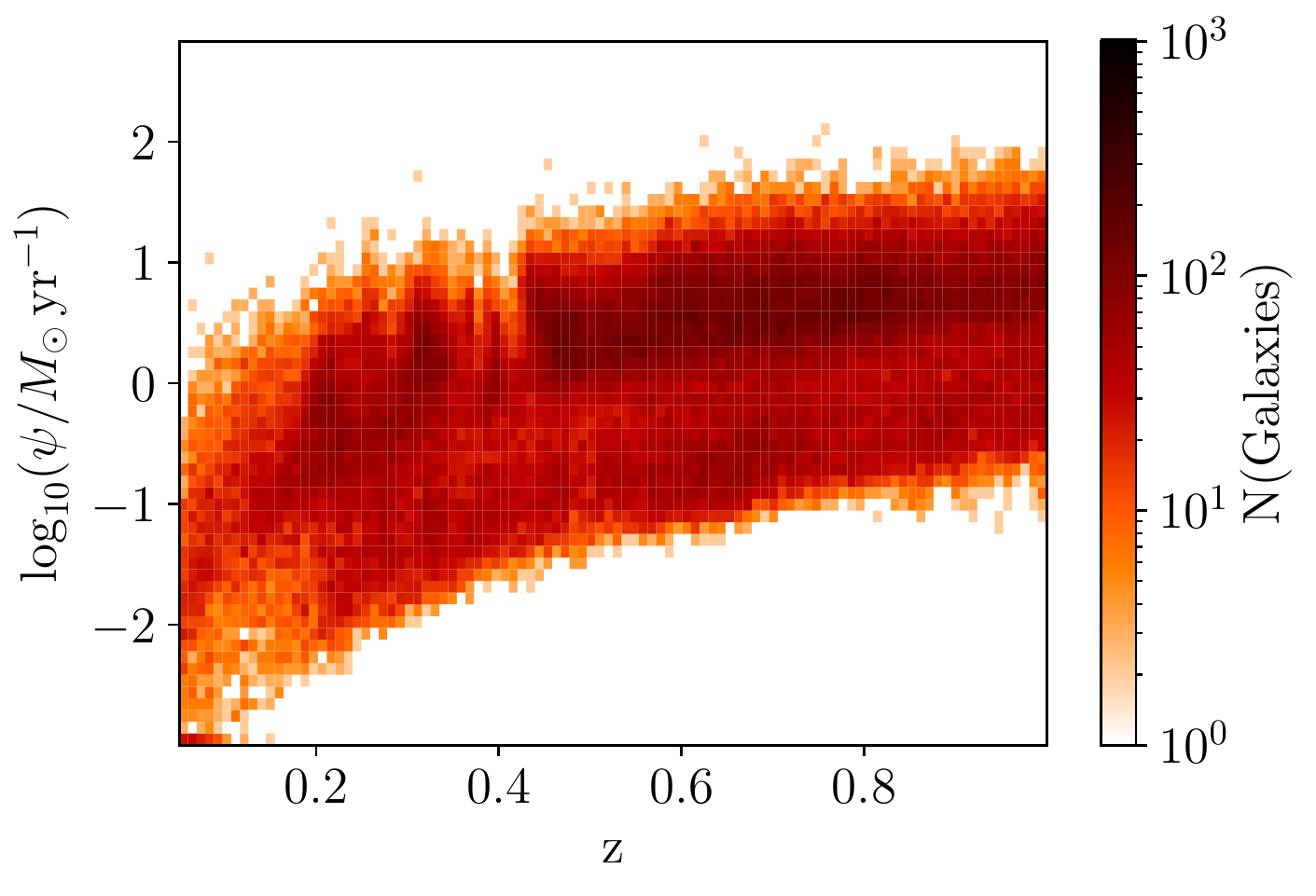}}
      \caption{Heatmaps showing the variation in median-likelihood stellar mass in $M_\odot$ (top) and SFR $(\psi)$ in units of $M_\odot$\,yr$^{-1}$ (bottom) as a function of redshift in our IRAC-selected sample. The redshift axis is common to both plots, and the number of galaxies in each bin is indicated by the colour-scales to the right. In the upper panel, the blue dashed line indicates the 95\%\ mass completeness limit as a function of redshift, derived using the method of \citet{pozzetti2010}.}
         \label{fig:msfrz}
\end{figure}

\subsection{Sample properties and AGN contamination}
\label{sec:agn}

As recommended by \citet{duncan_lotss}, we remove the most obvious AGN in our mass-selected sample using the flags supplied in the input catalogue, which include sources in the Million Quasar Catalog \citep{flesch2019} or spectroscopically classified as an AGN in the literature, those sources that fall in the \citet{donley2012} infrared colour space dominated by AGN, and those with bright X-ray counterparts (see Duncan et al. for details), leaving us with 122,646 sources. However, for those galaxies with 150\,MHz detections, we also remove additional AGN identified using the method of \citet{best_lotss}, which relies on the results of a comprehensive multi-wavelength SED fitting analysis using our \magphys\ results alongside those derived using \textsc{Bagpipes} \citep{carnall2018}, \textsc{AGNfitter} \citep{calistro2016} and \textsc{Cigale} \citep{burgarella2005}. The details of the Best et al. method are complex, however the general idea is to compare the results of those SED fitting codes that account for AGN (\textsc{AGNfitter} and \textsc{Cigale}) with those that are focussed on normal star-forming\slash passive galaxies (\magphys\ and \textsc{Bagpipes}) considering all of the available information, include comparing the goodness-of-fit that each code produces, accounting for the best-fit AGN fractions, and identifying sources with a clear radio excess. A comparison between the Best et al. AGN flagging procedure and the results of removing the bad \magphys\ fits using the aforementioned $\chi^2$ threshold method from \citet{smith2012} suggests that applying the \magphys\ threshold alone removes $\sim 96$\,per cent of the flagged AGN. This is especially useful for this work since the Best et al. AGN flags have only been derived for the sources detected in the 150\,MHz catalogue\footnote{This is not unreasonable, given the immense SED-fitting effort required to replicate them for the full IRAC-selected sample.}. After applying these cuts we are left with 120,232 $z < 1$ galaxies, of which 9,298 (25,777) are detected at $\ge 5 \sigma\ (\ge 3\sigma)$ in the deep LoTSS data. We also identified 1,715 further sources not flagged as AGN in the Best et al. radio excess sample, but with a clear radio excess in our pixel flux density measurements ($> 5\sigma$ 150\,MHz detections and \Lradio\ more than 1\,dex larger than the SFR-\Lradio\ relation from G18; this is 1.4\,percent of our sample) as likely undiagnosed AGN, and removed them from our sample, leaving us with 118,517 galaxies on which to base our analyses. We note that repeating the following analyses the results are not significantly changed whether we include these sources or not, whether we use a 1\,dex excess or a 0.7\,dex radio excess to identify residual AGN, or whether we instead use the mass-dependent SFR-\Lradio\ relation from G18. 

Histograms showing the median-likelihood SFR, stellar mass, redshift and 150\,MHz signal-to-noise (SNR) ratio distribution for the galaxies in our sample are shown in Figure \ref{fig:sample}. In Figure \ref{fig:msfrz} we display the stellar mass (top) and SFR (bottom) distribution as a function of redshift, colour-coded by the number of galaxies in each bin as indicated by the colour-bar to the right.

\begin{figure}
   \centering
\includegraphics[width=1.01\columnwidth]{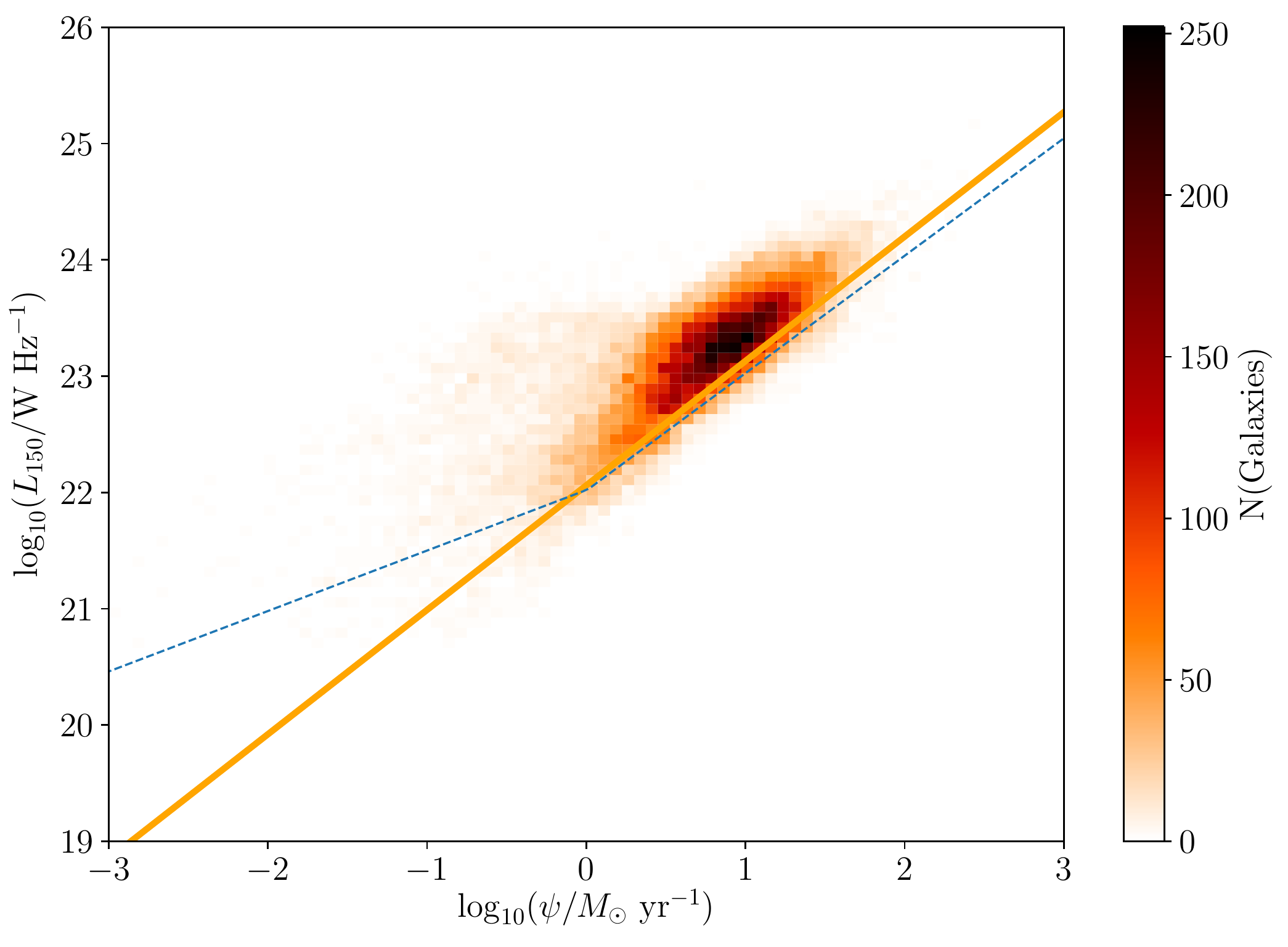}
      \caption{SFR-\Lradio\ plane populated by the 25,777 sources with $\ge 3\sigma$ detections in the 150\,MHz data, with the colour bar to the right indicating the bin occupancy. The best-fit relations from G18 are overlaid for comparison; the mass-independent relation is shown in orange, while the broken power law relation evaluated at $10^{10}\,M_\odot$ is shown as the blue dashed line. 
              }
         \label{fig:visualisation}
\end{figure}

In Figure \ref{fig:visualisation} we show the location of the SFR and 150\,MHz luminosity for those 25,777 sources which are detected at $> 3\sigma$ in the deep LoTSS data, overlaid with the mass-independent SFR-\Lradio\ relation (orange line) and broken power-law parameterisation (blue dashed line) found by G18 (which was based on a sample identified based on the MPA-JHU catalogue at optical wavelengths). The apparent offset to larger \Lradio\ in this sample relative to the G18 relations is an artefact of the data having been censored by the $3\sigma$ threshold in \Lradio, and does not account for the large majority of sources in our sample that fall below it. The formally non-detected (i.e. $< 3\sigma$) radio sources have huge potential diagnostic power due to their numerical dominance \citep[e.g.][]{zwart2015,malefahlo2020}, but they are challenging to visualise in the SFR-\Lradio\ plane. As the lower-right panel of Figure \ref{fig:sample} shows, the vast majority of sources in our IRAC-selected sample have 150\,MHz flux densities with $< 3\sigma$ significance, and a substantial minority have negative measured flux densities. It is critical that we also account for these sources, since failure to do so clearly truncates the luminosity distribution at a given SFR, and left unchecked this would have the potential to introduce significant bias into our results. 

Finally, to test the possible influence of sample incompleteness on our results, we used the method of \citet{pozzetti2010} to identify those galaxies with mass in excess of the 95\%\ mass completeness limit as a function of redshift (shown as the blue dashed line in the upper panel of Figure \ref{fig:msfrz}). We repeated all of the following analyses considering only those galaxies with mass in excess of the 95\%\ completeness limit, finding that our results were unchanged once the uncertainties were taken into account. We therefore conclude that incompleteness does not exert significant influence on our results. 

\section{Results}\label{sec:results}

\subsection{The SFR-\Lradio\ relation}
\label{sec:sfrl150}

\begin{figure}
   \centering
   \includegraphics[width=1.01\columnwidth]{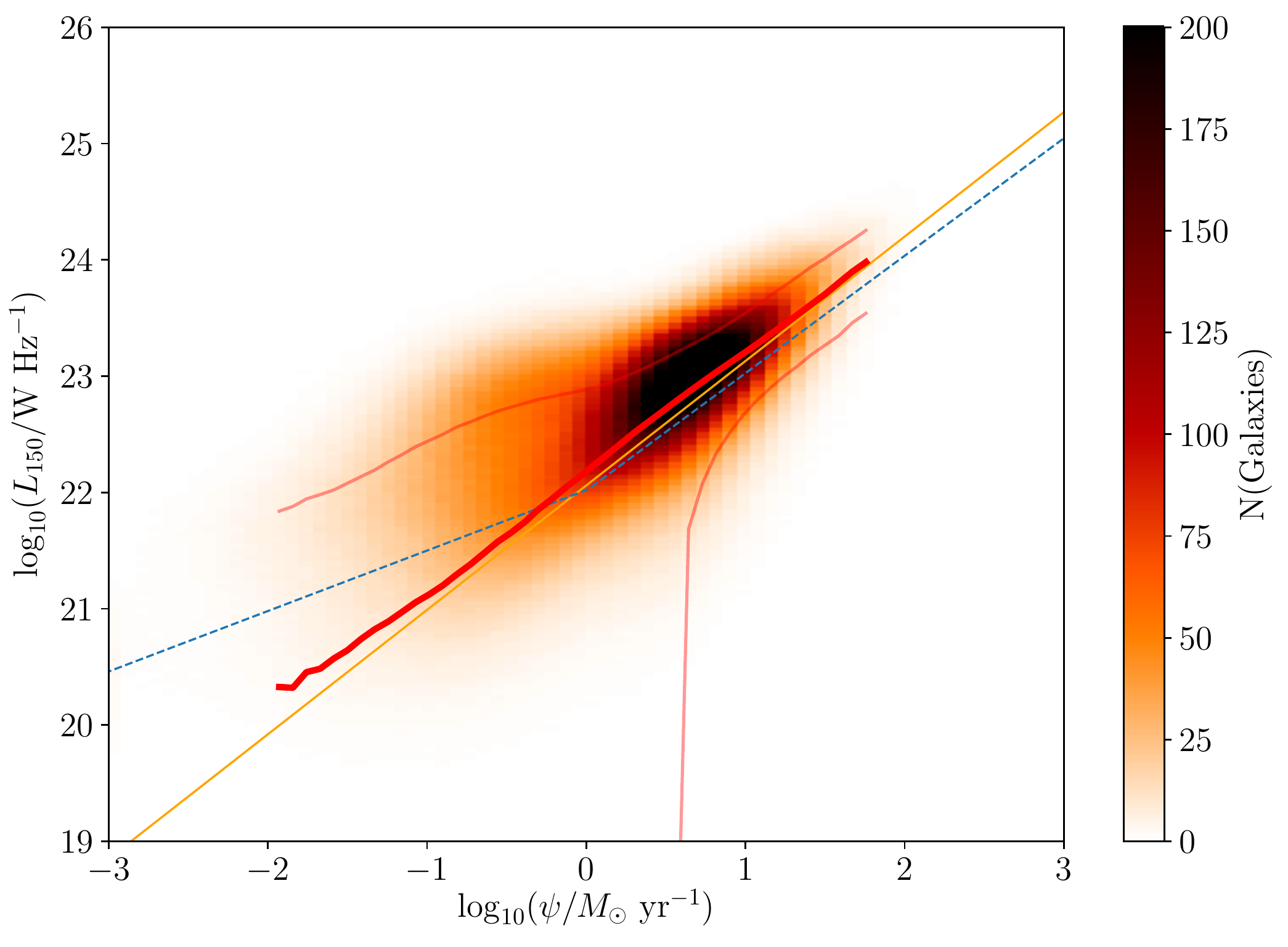}
      \caption{Stacked heatmap showing the two-dimensional probability distribution function for SFR and \Lradio\ for all 118,517 galaxies in our sample, including the uncertainties on the SFR (derived based on the percentiles of the \magphys\ SFR PDF) and on \Lradio\ (using the pixel flux densities and RMS values at the redshift in the \citealt{duncan_lotss} catalogue). The \textit{effective} number of galaxies in each bin -- recall that each galaxy is sampled one hundred times (with each sample representing 0.01 galaxies) and can therefore contribute to multiple pixels in the stack -- is indicated by the colour-bar to the right. The best-fit SFR-\Lradio\ relationship from G18 is overlaid as the orange solid line, along with the broken power law G18 relation (which we have evaluated at a canonical stellar mass of $10^{10}\,M_\odot$, and shown as the dashed blue line) along with (in red) the 16th, 50th and 84th percentiles of the \Lradio\ distribution at each SFR. 
              }
         \label{fig:sfrl150}
\end{figure}

We determine the SFR-\Lradio\ relation in ELAIS-N1, whilst accounting for uncertainties on both SFR and \Lradio, as follows. First, for each source in our sample, we create a two-dimensional PDF in the SFR-\Lradio\ parameter space with logarithmic axes in both directions, with 70 equally spaced bins of SFR between $-3 < \log_{10} (\psi/M_\odot yr^{-1}) < 3$, and 180 equally log-spaced bins of \Lradio\ between $17 < \log_{10}\,(L_\mathrm{150\,MHz}/W\,Hz^{-1}) < 26$. We generate each source's PDF by creating a histogram using the aforementioned bins, populated by 100 samples each in the SFR and \Lradio\ directions, assuming that the uncertainties in SFR and \Lradio\ are uncorrelated. For the \Lradio\ values we adopt a normally distributed error distribution (in linear space), with median and standard deviation equal to those derived using the measured flux densities and RMS from the LoTSS data at each source's redshift in our adopted cosmology. To ensure that the low-signal to noise and negative values of \Lradio\ are included in the PDF -- essential to avoid biasing them by censure -- we arbitrarily assign samples with $\log_{10} \Lradio < 17$ (including the negative values) to the lowest bin of the \Lradio\ PDF. For the SFRs we adopt an asymmetric error distribution with median equal to the median likelihood value of the \magphys\ SFR PDF, and a different standard deviation either side of the median, equal to the difference between the 84th (16th) and 50th percentiles of the SFR PDF for the positive (negative) wings of the error distribution. We then sum these PDFs over the whole sample to arrive at a stacked PDF showing the distribution of SFR and \Lradio\ including the uncertainties for the whole sample in both directions. This is shown as the heatmap in the background of Figure \ref{fig:sfrl150}, with the colour bar to the right indicating the number of galaxies in each bin.\footnote{Appendix \ref{sec:sampling_pdfs} describes an illustrative example of this method, along with more of the intermediate steps for the interested reader.}

One of the most interesting results from G18 was the discovery of an upturn in the SFR-\Lradio\ relation at low SFRs, below $\log_{10}\,(\psi / M_\odot\,\mathrm{yr}^{-1}) \approx 0$, and the G18 best-fit broken power-law relation is shown as the dashed blue line in figure \ref{fig:sfrl150} (for a stellar mass of $10^{10}\,M_\odot$, typical of galaxies in our sample). 
In order to identify the SFR-\Lradio\ relation in our data, we use a non-parametric approach. Non-parametric methods have the implicit advantage of being agnostic about the precise form of any relation that they may recover, and are therefore ideal for determining whether the data support an upturn at low SFR of the type seen in G18, or otherwise. At each SFR, we calculate the 16th, 50th and 84th percentiles of the \Lradio\ distribution in the stacked 2D PDF. The 50th percentile of the distribution then corresponds to our median-likelihood estimate of the SFR-\Lradio\ relation at that particular SFR, while the 16th and 84th percentiles also depend on the combination of the uncertainties on the luminosity estimates, plus any intrinsic scatter in the relation itself. We estimate the uncertainties on the median-likelihood value in each SFR bin by using the median statistics method of \citet{gott2001}. A further appealing feature of a median-likelihood estimate is that it has a degree of built-in resistance to outliers, such as might be expected from e.g. some residual minority of sources hosting unidentified radio excess due to AGN activity. We also note that deriving the best-fit relation in this way does not require us to account for the intrinsic dispersion of the relation itself, $\sigma_L$, to which we will return in section \ref{sec:scatter}. 

The results are shown as the red lines overlaid on Figure \ref{fig:sfrl150}, with the thick line corresponding to the median likelihood estimate of SFR-\Lradio\ over the range $-2 < \log_{10} (\psi/M_\odot\,\mathrm{yr}^{-1}) < 2$, with the 16th and 84th percentiles shown as the thin red lines. It is immediately clear that the median-likelihood relation is similar to the mass-independent relation found by G18, albeit slightly flatter, and that the recovered values appear consistent with a power-law relationship across the full range of SFRs spanned by our sample, with no evidence of an upturn in \Lradio\ at low SFRs. To determine the best-fit parameters, we adopt the form of the SFR-\Lradio\ relation from G18, specifically: 

\begin{equation}
\Lradio = L_1\ \psi^\beta,
\label{eq:sfrl150}
\end{equation}

\noindent and find best-fit values of $\log_{10} L_1 = 22.181\pm 0.005$ and $\beta = 1.041\pm0.007$, where the uncertainties have been determined using the \texttt{emcee} \citep{foreman2013} Monte Carlo Markov Chain algorithm with sixteen walkers and a chain length of 10,000 samples.

In Appendix \ref{sec:methodtests}, we present a suite of simulations conducted to test how well we can recover a known SFR-\Lradio\ relation using this method, and find that the best-fit estimates are likely to be systematically offset by $\Delta \log_{10} L_1 \approx 0.040$ and $\Delta \beta \approx 0.016$. Correcting for these effects gives our best estimate of the overall SFR-\Lradio\ relation, which is $\log_{10} \Lradio = (22.221 \pm 0.008) + (1.058 \pm 0.007) \log_{10} (\psi / M_\odot \mathrm{yr}^{-1})$.  Although these values are formally inconsistent with the results of G18, who obtained $\log_{10} L_1 = 22.06 \pm 0.01$, $\beta = 1.07\pm0.01$, the fact that it is this close is encouraging, given the large differences in methodology and sample definition.

Figure \ref{fig:sfrl150} shows a clear offset between the apparent peak of the stacked PDF (background colour scale) and the median-likelihood values (thick red lines) corresponding to our estimate of the SFR-\Lradio\ relation. This effect results from sampling linear \Lradio\ values in a logarithmic PDF, which means that the lower half of the PDF is effectively spread out over a larger number of bins than the higher half.\footnote{Consider for example that the $\pm 2\sigma$ confidence interval in log values for a $3\sigma$ source with a true luminosity of $10^{24}$\,W\,Hz$^{-1}$ are $23.52 < \log_{10} (\Lradio / W\,Hz^{-1}) < 24.22$. The range is more than twice as large on the negative side than the positive side.} This effect is also clearly apparent in the simulations of our method discussed in Appendix \ref{sec:methodtests}, which underlines that the non-Gaussianity in the PDF does not stop the median-likelihood values (of the individual sources and of the population) from being able to recover the true SFR-\Lradio\ relation.

The third important thing to notice about Figure \ref{fig:sfrl150} is that the 16th percentile of the \Lradio\ distribution (shown as the lower red line) rapidly decreases below the bottom of the plotting window at $\log_{10} \psi < 0.7$. This highlights the importance of including the formally undetected (e.g. those with 150\,MHz flux density $< 3\sigma$) sources in this study, and this importance of course increases as we move to lower SFR, where an increasing fraction of objects have 150\,MHz flux densities below $1\sigma$. 
The simulations discussed in Appendix \ref{sec:methodtests} further highlight the importance of accounting for the negative \Lradio\ samples, since not doing so would introduce significant positive bias in the median flux density.

It is tempting to also determine the SFR-\Lradio\ relation by finding the `ridge-line' in the stacked 2D PDF, corresponding to the modal value of \Lradio\ at a given SFR. If we do this, we obtain good agreement with the above median-likelihood estimates at $\log_{10} (\psi\,/\,M_\odot\,\mathrm{yr}^{-1}) > 1$; however, as we move to lower SFRs we begin to see positive bias introduced, similar to the upturn seen in G18. This is another facet of the aforementioned issue with increasingly sampling the noise distribution as we move to lower SFRs - if all sources were at the same redshift then we are effectively attempting to plot a histogram of the logarithm of a normal distribution in luminosity which is centred very close to zero, and the ridge-line becomes increasingly biased.

\subsection{Scatter on SFR-\Lradio}
\label{sec:scatter}

\begin{figure}
   \centering
   \subfigure{\includegraphics[width=1.0\columnwidth]{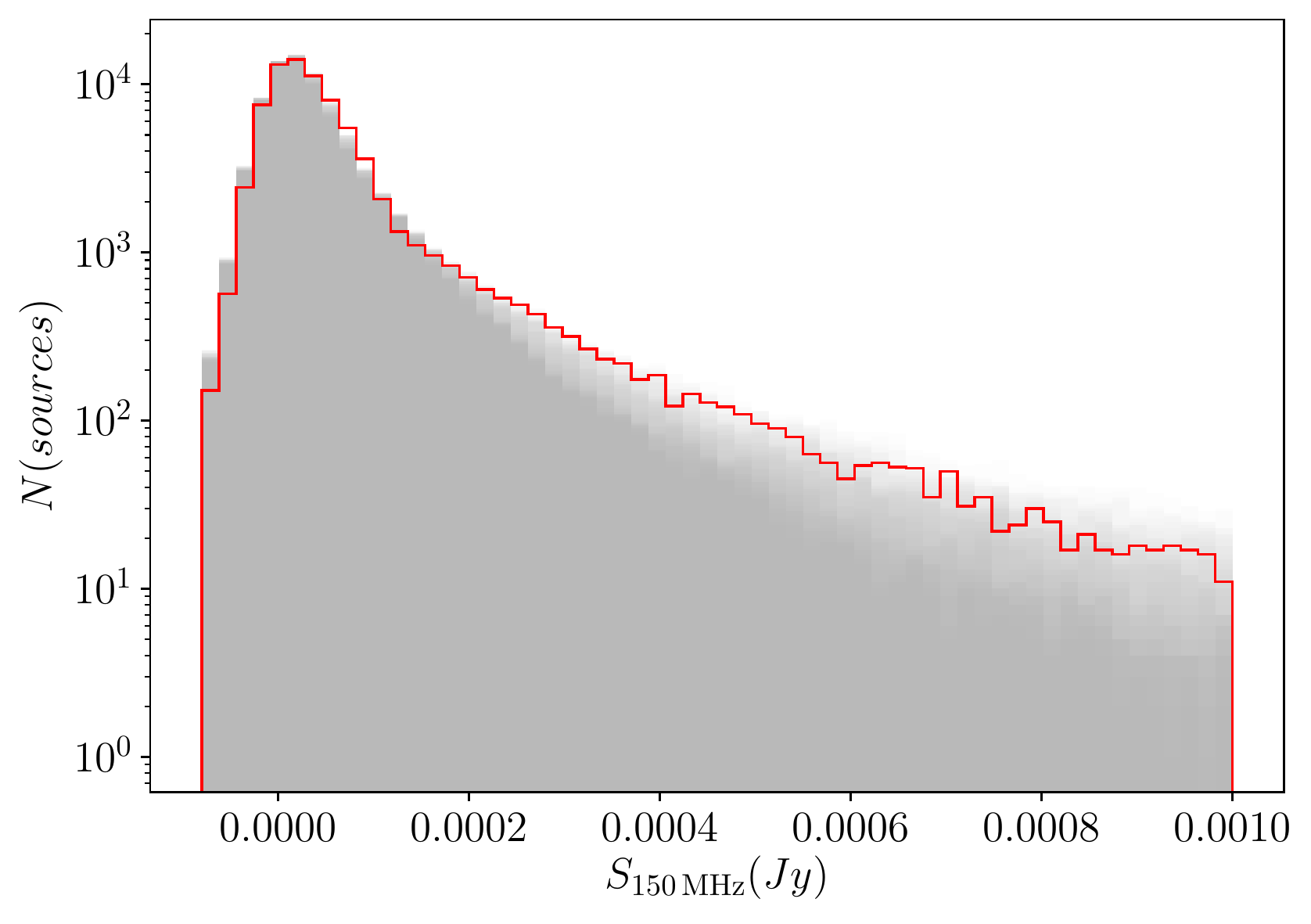}}   
        \caption{Histogram with logarithmic ordinates, showing the true observed flux density distribution (in red) overlaid on individual Monte Carlo realisations (shown in grey with transparency, such that lighter shading indicates the range of outputs) based on assuming the best-fit SFR-\Lradio\ relation from section \ref{sec:sfrl150} with the full range of intrinsic scatter $0.0 < \sigma_L < 0.5$. 
              }
         \label{fig:scatter_overall}
\end{figure} 

As well as the form of the SFR-\Lradio\ relation, it is also of interest to determine the scatter on the relation itself, $\sigma_L$, usually quoted in logarithmic terms (dex). Whatever the cause, $\sigma_L$ is important since it forms a key part of the process of identifying AGN in 150\,MHz samples on the basis of a SFR-dependent radio excess \citep{best_lotss}. It is also of interest when studying the so-called ``main sequence" of star formation -- the relationship between the stellar mass and SFR of star-forming galaxies \citep[e.g.][]{noeske2007,johnston2015,schreiber2015} using radio observations \citep[e.g.][]{leslie2020}. The width of the main sequence has been interpreted as a manifestation of variation in a star forming galaxy's gas supply \citep{tacchella2016}, therefore an additional source of scatter in the SFR indicator itself has the potential to bias the results if it is not measured and accounted for.

To measure the scatter on SFR-\Lradio\ in our data we use Monte Carlo simulations. We do this by creating multiple realisations of our sample, using the best-fit redshifts alongside random draws from the \magphys\ SFRs (assuming the same asymmetric error distribution as in Section \ref{sec:sfrl150}) and from the best-fit relation (Equation \ref{eq:sfrl150}), with a scatter in the range $0.0 < \sigma_L < 0.5$\,dex, to calculate a model \Lradio\ for each source. We use these values to derive ``true'' flux densities, and simulate measurement errors based on realisations of the values in the RMS map at the position of that source in the real data. For each simulation we conduct two-sample Kolmogorov-Smirnov tests to compare the model flux densities with the real flux density distribution for any choice of SFR and redshift, to determine the degree of support for the hypothesis that the simulated flux densities are consistent with being drawn from the same distribution as the real values, as a function of $\sigma_L$. Figure \ref{fig:scatter_overall} shows the true pixel flux density distribution in red, overlaid on one set of Monte Carlo simulations covering the full range in scatter (shaded grey). The individual grey histograms are transparent, such that lighter grey regions reveal the variation in the flux density distribution for the range of scatter considered. Following Macfarlane et al. \textit{submitted}, we truncate the distribution above a flux density of 1\,mJy to avoid giving undue influence to residual undiagnosed AGN in our sample (this doesn't make a significant difference to the results). We determine the mean $P$ returned by the KS test (averaging over all 10,000 Monte Carlo simulations) as a function of $\sigma_L$, and marginalise the resulting distribution. We are then able to calculate median likelihood estimates of $\sigma_L$, alongside uncertainties by estimating the 16th, 50th and 84th percentiles of the PDF, as well as Bayesian estimates calculated according to $\sum \sigma_L P(\sigma_L)$.

\begin{figure}
   \centering
   \includegraphics[width=\columnwidth, trim=16cm 0cm 0cm 0cm, clip=true]{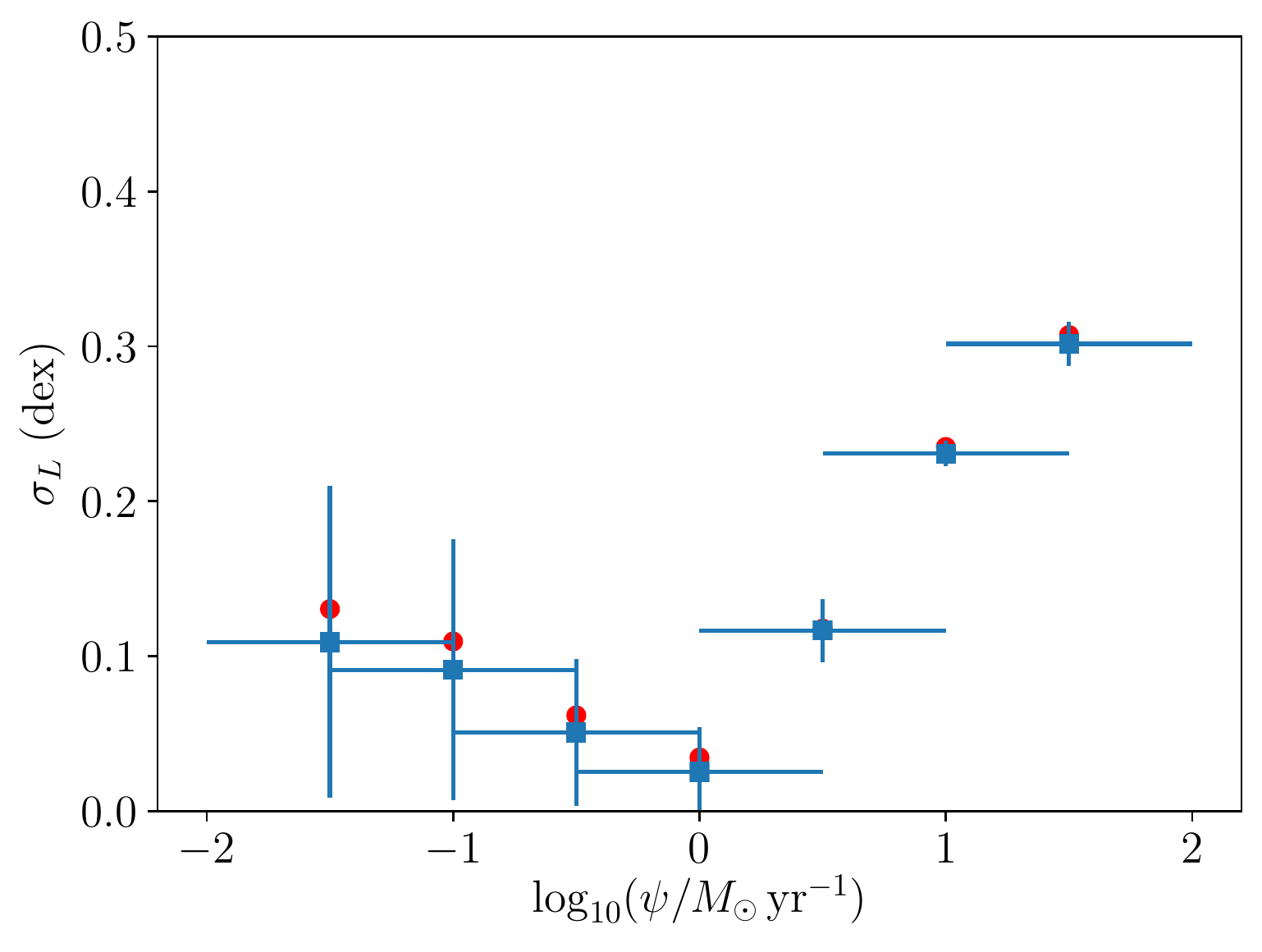}
      \caption{Variation in median likelihood estimates of $\sigma_L$ as a function of the central value of each SFR bin. The error bars in the vertical direction are derived from the 16th and 84th percentiles of the PDF, while in the horizontal direction they indicate the bin width. Also overlaid are red circles, indicating the Bayesian estimates of scatter from the PDF, calculated according to the product of $\sum \sigma_L P(\sigma_L)$.
              }
         \label{fig:scatter_sfrdep}
\end{figure} 

Figure \ref{fig:scatter_sfrdep} shows our results, using the KS tests to determine the level of support in the data for different values of $\sigma_L$ as a function of SFR in bins with a constant width of 1 dex in SFR on a sliding scale from $-2 < \log_{10} (\psi / M_\odot\,\mathrm{yr}^{-1}) < 2$. We detect significant scatter about SFR-\Lradio\ only at $\log_{10} (\psi / M_\odot\,\mathrm{yr}^{-1}) > 0$,  with $\sigma_L$ apparently increasing with SFR and reaching $0.31\pm0.01$\,dex at  $1 < \log_{10} (\psi / M_\odot\,\mathrm{yr}^{-1}) < 2$. This may be because as we approach lower SFRs the scatter in the flux density distribution is increasingly dominated by the sensitivity of the LOFAR observations, rather than the physical effect that is most visible at larger SFR. It will be of great interest to see whether this effect persists with larger samples of $\log_{10} (\psi / M_\odot\,\mathrm{yr}^{-1}) < 0$ galaxies from the wider-area LoTSS second data release once it is available, and in due course with the huge increase in sensitivity that the SKA will provide.

\subsection{Evolution of SFR-\Lradio\ and $\sigma_L$}
\label{sec:evolution}

\begin{figure*}
   \centering
   \includegraphics[width=0.8\textwidth]{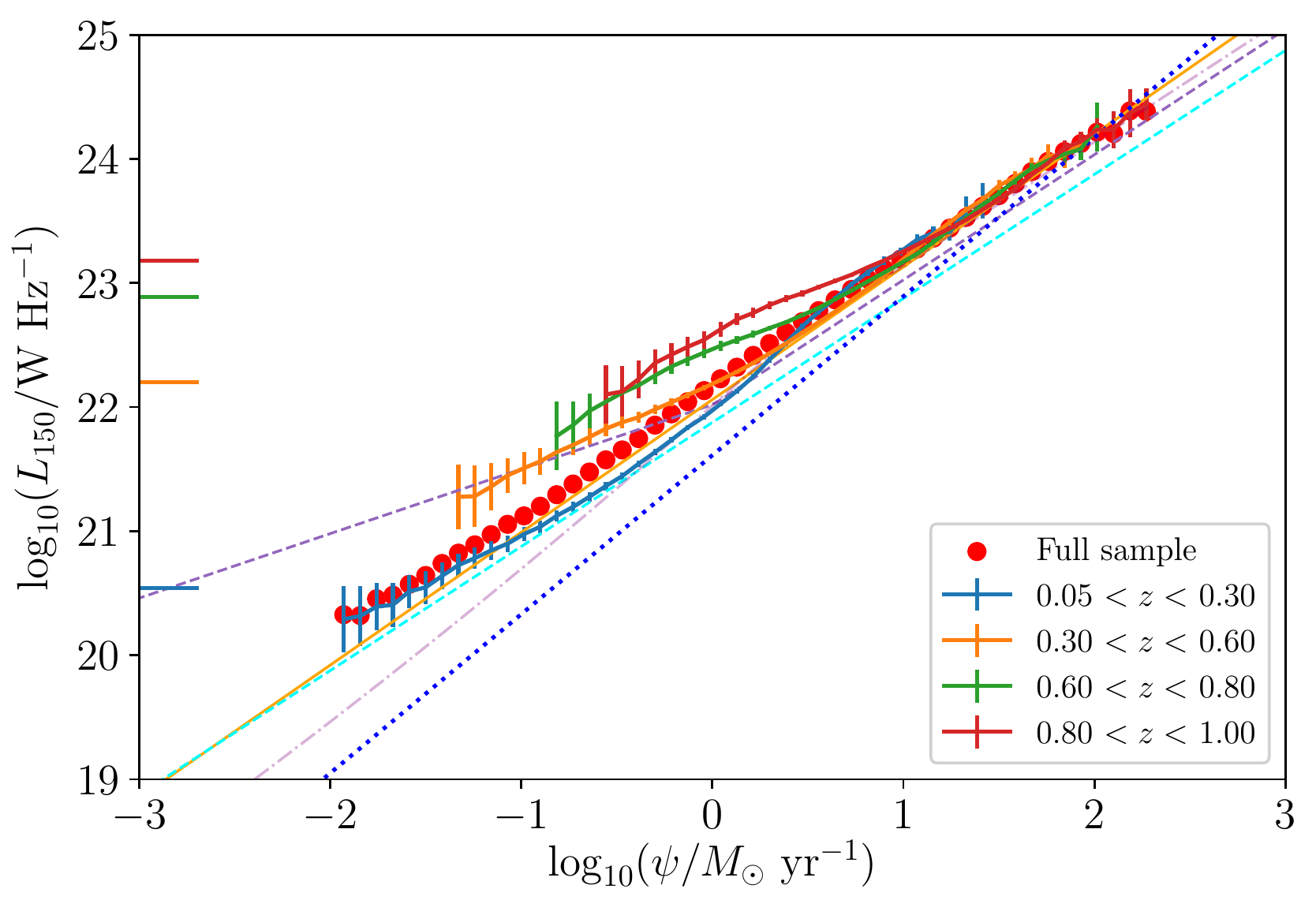}
      \caption{Median likelihood estimates of the SFR-\Lradio\ relation for the whole sample (red circles), and in four redshift bins ($0.05 < z < 0.30$, $0.30 < z < 0.60$, $0.60 < z < 0.80$ and $0.80 < z < 1.00$ are shown as the blue, orange, green and red crosses, respectively). Error bars on the median \Lradio\ are calculated using the median statistics method of \citet{gott2001}. Also overlaid are the SFR-\Lradio\ relationships from \citet[dot-dashed purple line]{bell2003} and \citet[dashed cyan line]{murphy2011} converted to 150\,MHz assuming a canonical spectral index $\alpha = 0.7$, and the 150\,MHz relations from G18 (mass independent as the orange solid line, broken power law evaluated at $10^{10}\,M_\odot$ as the dashed purple line) and from \citet{wang2019}, which is shown as the dotted blue line. The literature calibrations have been converted to our adopted IMF from \citet{chabrier2003} using the factors recommended in \citet{madau2014}. The horizontal coloured lines immediately to the right of the left-hand vertical axis indicate the luminosity corresponding to the $60\,\mu$Jy at the lower redshift bound of each bin indicated by the colour (see text for details).
              }
         \label{fig:slope_evolution}
\end{figure*} 

In order to investigate the possibility of evolution in the derived properties of SFR-\Lradio, we split the sample into four redshift bins, and repeat the analysis of the previous two sections. Figure \ref{fig:slope_evolution} shows the median likelihood SFR-\Lradio\ derived over the full redshift range (in red), and overlaid with the values derived for each redshift bin (coloured as in the legend). For the purposes of comparison we have also again overlaid the G18 relation (in orange) as well as the empirical relations from \citet[]{bell2003} and \citet{murphy2011}, both converted from the 1.4\,GHz expectations assuming a canonical spectral index, $\alpha = 0.7$ \citep[similar to values in the literature e.g.][]{mauch2013,prescott2016}, and the relation from \citet{wang2019}. All of these literature relations have been converted to the \citet{chabrier2003} IMF used in this work where necessary, using the corrections provided in \citet{madau2014}. To indicate the flux density scale in each redshift bin, we have also overlaid coloured horizontal lines adjacent to the left-hand vertical axis, which indicate values of \Lradio\ corresponding to 60\,$\mu$Jy (i.e. the approximate $3\sigma$ limit in the deepest regions of the 150\,MHz data) at the lowest redshift bound of each bin (indicated by the colour of the line).

At $\log_{10}\,(\psi/M_\odot\,\mathrm{yr}^{-1}) > 1$ there is negligible difference apparent between the SFR-\Lradio\ relation derived over the full redshift range, and the values in each redshift bin. However at $\log_{10} (\psi/M_\odot\,\mathrm{yr}^{-1}) < 1$, there is the first evidence of variation, and perhaps for an excess of radio luminosity which increases towards the lowest SFR end of the data sampled in each redshift range, where the individual galaxies' 150\,MHz flux densities are formally undetected. To determine whether this effect is real or instrumental, we have conducted further simulations, repeating the analysis using a known input SFR-\Lradio\ relation, sampling the observed $z$ and SFR distributions and uncertainties, alongside a realistic model for mass-dependent AGN contamination following \citet{sabater2019}, and modelling the effects of noise in the 150\,MHz flux densities by sampling from the real data set. The simulations are discussed in detail in Appendix  \ref{sec:methodtests}, and using a mass-independent SFR-\Lradio\ relation of the form given by Equation \ref{eq:sfrl150} we are unable to reproduce variation in SFR-\Lradio\ like the possible upturn seen at at $\log_{10} (\psi/M_\odot\,\mathrm{yr}^{-1}) < 1$. If this effect is real, it may point to possible mass-dependence in SFR-\Lradio\ as found by G18. We will return to this topic in Section \ref{sec:massdep}.

To make a better comparison with G18, who used a $z < 0.3$ sample defined based on SDSS spectroscopy, we have also derived best-fit parameters for the corresponding redshift range in our sample (our lowest redshift bin). We find best fit parameters of $\log_{10} L_1 = 22.14 \pm 0.01$ and $\beta = 1.22\pm0.01$, corrected for residual bias in the same way as in Section \ref{sec:sfrl150}. Although our $z < 0.3$ $L_1$ estimate is comparable, the value for $\beta$ that we obtain in the lowest redshift bin is significantly steeper than G18, who found $L_1 = 22.06 \pm 0.01$ and $\beta = 1.07\pm 0.01$. As mentioned in Section \ref{sec:sfrl150}, there are significant differences in the methodology and especially in the selection function used by the two works, meaning that we are not necessarily comparing like with like. 

\begin{figure}
   \centering
	\includegraphics[width=1.01\columnwidth, trim=16cm 0cm 0cm 0cm, clip=true]{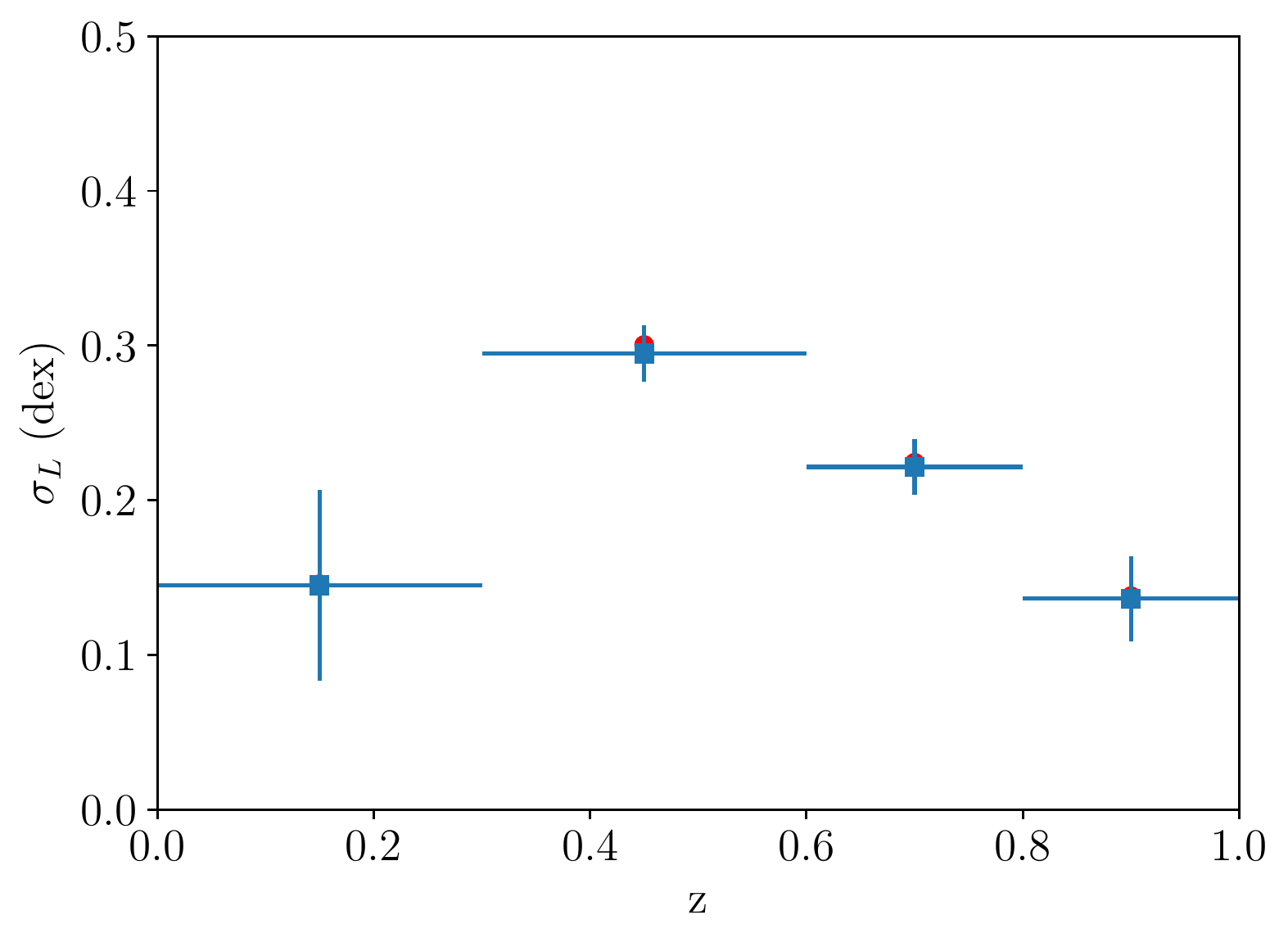}
      \caption{Variation in $\sigma_L$ for galaxies with $0.5 <  \log_{10}  \psi < 1.5$, $10 < \log_{10} (M\slash M_\odot) < 11$ as a function of redshift, with error bars derived from the 16th and 84th percentiles of the PDF, and centred on the median-likelihood values. The error bars in the redshift (horizontal) direction indicate the bounds of each redshift bin. }
         \label{fig:scatter_evolution}
\end{figure} 

We also search for evidence of evolution in the scatter using the same method as in section \ref{sec:scatter}, but limiting the SFR range to $0.5 < \log_{10} (\psi / M_\odot\,\mathrm{yr}^{-1}) < 1.5$ (to mitigate possible variation in scatter over SFR), and limiting the stellar mass range to $10 < \log_{10} (M\slash M_\odot) < 11$ (to mitigate any possible influence of the mass dependence in SFR-\Lradio\ found by G18), using the same redshift bins as above. In Figure \ref{fig:scatter_evolution}, we plot the derived values as a function of redshift, along with their error bars. 
While some variation in $\sigma_L$ is possible within the uncertainties, the data do not show any evidence for linear evolution in $\sigma_L$, at least out to $z = 1$. 

\subsection{Mass dependence of SFR-\Lradio}
\label{sec:massdep}

\begin{figure*}
   \centering
   \includegraphics[width=\textwidth, trim=0cm 12.5cm 0cm 0cm, clip=true]{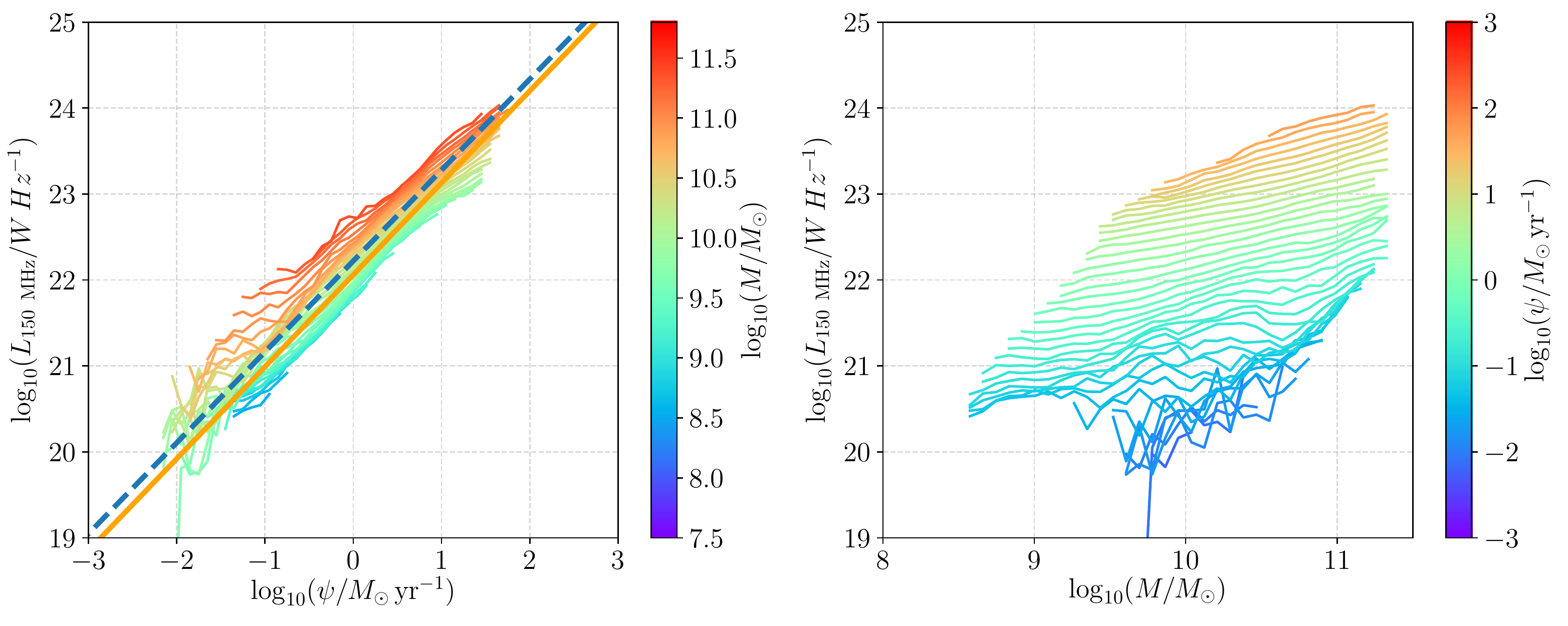}
      \caption{Left: SFR-\Lradio\ relation as a function of stellar mass, indicated by the colour of the line and relative to the colour-bar, overlaid on the mass-independent relation from G18 (solid orange line), and our best-fit estimate from section \ref{sec:sfrl150} (dashed blue line). Each coloured line shows the median-likelihood \Lradio\ at a given SFR and stellar mass. Right: relationship between stellar mass and median-likelihood \Lradio, coloured as a function of SFR with the scale indicated by the bar to the right.}
         \label{fig:massdep}
\end{figure*}

Finally, we also consider the possibility of stellar mass dependence in SFR-\Lradio\ as discussed by G18, \citet[][and \textit{in preparation}]{readthesis}. To do this, we introduce a three-dimensional version of the method we used in section \ref{sec:sfrl150}, now producing 100 samples in each of the SFR, stellar mass and \Lradio\ directions, to make a three-dimensional PDF for each source. Once again, we account for possible asymmetry in the error bars of SFR and stellar mass by using the 16th, 50th and 84th percentiles of the \magphys\ estimates, and sampling from a linear space in the uncertainty on \Lradio. As before, we then sum together across all 118,517 sources in our sample. We use fifty equally-spaced logarithmic bins of stellar mass between $7.5 < \log_{10} (M \slash M_\odot) < 11.8$, sixty equally-spaced logarithmic bins of SFR between $-3 < \log_{10} (\psi / M_\odot\,\mathrm{yr}^{-1}) < 3$, and 180 equally-spaced logarithmic bins of \Lradio\ between $17 < \log_{10} (\Lradio\ / W\,Hz^{-1}) < 26$ to calculate the median likelihood \Lradio\ in each bin. The left panel of Figure \ref{fig:massdep} shows values of median-likelihood \Lradio\ as a function of SFR at a constant stellar mass indicated by the colour bar, effectively slicing through our three-dimensional stellar mass\slash SFR\slash \Lradio\ distribution. Similarly, the right panel shows corresponding values of \Lradio\ as a function of stellar mass at fixed SFR (with the SFR indicated for each line by the color bar). In both panels, we show only those bins populated by at least 15 galaxies (after accounting for the fact that each galaxy is sampled 100 times), though there are still some effects of small number statistics visible in the lowest SFR bins, particularly in the right panel.
Nevertheless, the left panel reveals variation of at least 0.5\,dex in \Lradio\ for a given SFR, depending on the stellar mass, while the right panel shows more than 2\,dex of SFR-dependence in \Lradio\ at a fixed stellar mass. These effects are clearly large enough to potentially account for the difference in SFR-\Lradio\ that we observe in our lowest redshift bin relative to G18 and which we discussed in section \ref{sec:evolution}. 

To quantify our results, we use the mass-dependent parameterisation from G18:

\begin{equation}
L_\mathrm{150} = L_C\ \psi^\beta\ \left(\frac{M}{10^{10}M_\odot}\right)^\gamma,
\label{eq:sfmass}
\end{equation}

\noindent and obtain best-fit values of $\log_{10} L_C = 22.111\pm 0.004$, $\beta = 0.850 \pm 0.005$ and $\gamma = 0.402\pm0.005$, where G18 obtained $\log_{10} L_C = 22.13 \pm 0.01$, $\beta = 0.77\pm 0.01$ and $\gamma = 0.43 \pm 0.01$. We conducted an additional set of simulations which we discuss in Appendix \ref{sec:methodtests} to determine how well we are able to recover a known relation of the form given by Equation \ref{eq:sfmass}. We find that using this method the best-fit estimates are likely to be offset by a residual bias of $\Delta \beta = 0.053$, $\Delta \log_{10} L_C = 0.107$ and $\Delta \gamma = -0.072$ giving best estimates of $\beta = 0.903 \pm 0.012$, $\log_{10} L_C = 22.218 \pm 0.016$ and $\gamma = 0.332 \pm 0.037$, where we have propagated the uncertainties based on the systematic corrections by adding in quadrature with those values derived from our MCMC fitting. 

\begin{figure*}
   \centering
   \includegraphics[width=\textwidth, trim=0cm 0cm 0cm 12.5cm, clip=true]{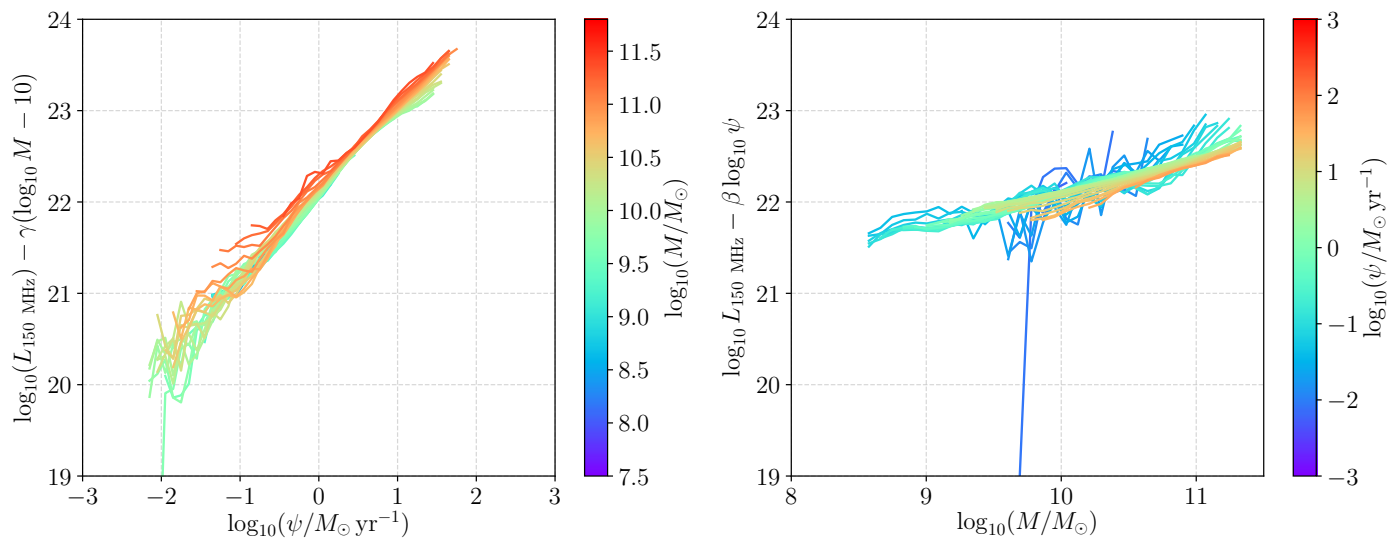}
      \caption{Left: SFR-\Lradio\ relationship with the mass dependence taken out using equation \ref{eq:sfmass} and the best-fit parameters. Stellar mass is indicated by the colour bar to the right. Right: as left, but relationship between stellar mass and radio luminosity normalised by the SFR using equation \ref{eq:sfmass}.}
         \label{fig:massdep_removed}
\end{figure*} 

To visualise the improvement in accuracy resulting from equation \ref{eq:sfmass}, in Figure \ref{fig:massdep_removed} we show the same data as in Figure \ref{fig:massdep}, but with the mass and SFR dependency taken out (in the left and right panels respectively), using the best-fit parameters above in equation \ref{eq:sfmass}. Together, these plots reveal compelling evidence for mass dependence in SFR-\Lradio, in the sense that more massive galaxies have a larger radio luminosity at a fixed star formation rate. That the mass dependence appears constant, and that it is clearly evident even in galaxies with stellar mass below $10^{10}\,M_\odot$, suggests that it is unlikely to be caused by some undiagnosed AGN contamination, given the mass-dependence in AGN fraction shown by \citet{sabater2019} for example. 

Interestingly, we repeated the Monte Carlo simulations discussed in Section \ref{sec:scatter} to see whether assuming a mass-dependent SFR-\Lradio\ relation, and including the \magphys\ stellar mass information in the simulations, made any difference to our measurements of $\sigma_L$. The results on both the SFR and redshift dependence in $\sigma_L$ are consistent within the uncertainties whether we use a mass-dependent or independent SFR-\Lradio, suggesting that mass dependence -- at least as parameterised in Equation \ref{eq:sfmass} -- cannot explain the scatter on SFR-\Lradio.

We have also used a mass-dependent SFR-\Lradio\ relation as an input to the simulations discussed in Appendix \ref{sec:upturn_test}, and find that it allows us to recover an excess \Lradio\ at low SFRs, very similar to the possible excess observed in the redshift bins of figure \ref{fig:slope_evolution}. Mass dependence on SFR-\Lradio\ may therefore be able to explain the possible variation revealed in Figure \ref{fig:slope_evolution}, and may also provide an explanation for the radio excess apparent at low-SFRs first noticed by G18. We intend to revisit this issue in a future work.

\section{Discussion \&\ Conclusions}\label{sec:conclusions}

We have studied the relationship between SFR and 150\,MHz luminosity using new, sensitive deep field observations from the LoTSS Deep Fields first data release. Starting from a near-infrared selected sample, we leverage the multi-wavelength aperture-matched forced photometry and state-of-the-art photometric redshifts alongside the new LoTSS maps of the EN1 field, to produce stellar mass and SFR estimates using energy balance SED fits using the \magphys\ package. We use 150\,MHz flux densities from the ELAIS-N1 catalogue, plus pixel flux densities for the remaining 109,206 IRAC sources that are not identified as counterparts to catalogued 150\,MHz sources, to estimate the median likelihood \Lradio\ as a function of star formation rate and stellar mass. 

The 150\,MHz data used in this study are $5\times$ more sensitive than those used by G18, the sample size is $8\times$ larger, and the multi-wavelength coverage in this field is far superior. This is true both in terms of depth (e.g. the HSC $i$ band data over ELAIS-N1 reach a $5\sigma$ magnitude fainter than 26\,mag, around four magnitudes deeper than the SDSS imaging used in G18), and in terms of the number of photometric bands which are available (Table \ref{tab:nbands} reveals that more than half of our sample has $\ge 3\sigma$ flux densities in at least 16 bands, as compared to a maximum of 14 in G18). The LoTSS deep field data in ELAIS-N1 are around $35\times$ deeper than the FIRST data that have been used for previous studies of this topic \citep[e.g.][]{hodge2008,garn2009,davies2017}, assuming a standard canonical spectral index value of $\alpha = 0.7$, and are comparable to the deepest degree-scale interferometric radio data in existence \citep[e.g.][]{smolcic2017} but cover an area of sky $5\times$ larger.

Using a non-parametric approach, we find an apparently linear relationship between SFR and \Lradio\ over the range in SFR $-2 < \log_{10} (\psi / M_\odot \mathrm{yr}^{-1}) < 2$ of the form $\Lradio\ = L_1 \psi^\beta$, with best-fit parameters equal to $\log_{10} L_1 = 22.221 \pm 0.008$ and $\beta = 1.058 \pm 0.007$. We find an SFR-dependent scatter about the SFR-\Lradio\ relation, reaching $\sigma_L \approx 0.31\pm0.01$\,dex at $1 < \log_{10} (\psi / M_\odot\,\mathrm{yr}^{-1})< 2$. Our inability to detect significant scatter at lower SFRs $\log_{10} (\psi/M_\odot\,\mathrm{yr}^{-1}) < 0$ may be a limitation of sensitivity in the 150\,MHz data, even though they are the deepest in existence. Neither the scatter nor the high-SFR end ($\log_{10} (\psi/M_\odot\,\mathrm{yr}^{-1}) >1$) of the best-fit relation show significant evidence for redshift evolution, with the latter finding in agreement with e.g. \citet{garn2009} and \citet{duncan2020}, out to larger SFRs and redshifts. Our results also agree with \citet[who used an 150\,MHz and $i$-band selected sample]{calistro2017} out to our $z < 1$ limit, though they do see evidence for evolution in SFR-\Lradio\ in more distant sources.

The (close to) unitary slope that we determine for the best-fit SFR-\Lradio\ relation is apparently consistent with expectations based on calorimetry models \citep[e.g.][]{chi1990,yun2001,lacki2010}, and it is similar to previous low frequency results (e.g. \citealt{brown2017} who found $\beta = 1.14 \pm 0.05$, and to G18, who found $\beta = 1.07 \pm 0.01$). However, like G18 and \citet{readthesis}, we find a clear mass-dependence in the SFR-\Lradio\ relation, in the sense that higher mass galaxies have a larger \Lradio\ at a fixed SFR. Our best-fit mass-dependent relation is $\log_{10} \Lradio\ = (0.90\pm 0.01) \log_{10}(\psi/M_\odot\,\mathrm{yr}^{-1}) + (0.33 \pm 0.04) \log_{10} (M/10^{10}M_\odot) + 22.22 \pm 0.02$. Using a suite of realistic simulations, we have shown that the mass-dependence can explain the possible observed deviation from linearity in the redshift-binned SFR-\Lradio\ relation, as well as potentially the radio excess in low-SFR galaxies found by G18. This implies that the unitary slope we recovered in the overall SFR-\Lradio\ may be a coincidence, and -- assuming direct proportionality in the relationship between the far-infrared luminosity and SFR as described e.g. in \citet{kennicutt2012} -- we expect to observe similar mass-dependence in the far-infrared radio correlation.

One possibility for explaining these results could be a mass-dependent cosmic ray escape fraction, allowing particles to remove energy from the galaxy before it can be radiated away at radio frequencies, especially in lower mass (smaller) galaxies. However it is also important to consider whether undiagnosed AGN contamination can play a role, since a radio excess that increases with growing stellar mass would be in keeping with previous results on the mass-dependence of the AGN fraction \citep{sabater2019}. However, the mass-dependence apparent in our data is clear and consistent across the whole sample, including in galaxies with stellar mass below $10^{10}\,M_\odot$, implying that it is unlikely to be due to undiagnosed AGN contamination. At the same time, we are unable to rule out the presence of undiagnosed AGN in our sample altogether, especially low-excitation systems \citep[which manifest as a large accretion-related radio excess but with little or no evidence for AGN in the multi-wavelength SED, e.g.][]{hardcastle2007,best2012}, but this type of undiagnosed AGN is unlikely to explain our results. These issues relating to mass dependence are also discussed by \citet{molnar2018} in the context of a variation in the FIRC in bulge-dominated galaxies (which also tend to be more massive than pure disk-dominated systems). We defer a more detailed discussion of this aspect of our results to a future work. 

Irrespective of the cause, the size of this trend is such that failing to account for the stellar mass dependence can introduce systematic offsets on 150\,MHz-derived SFRs with a magnitude around 0.5\,dex in either direction \citep[consistent with the results shown in][]{readthesis}, which are therefore potentially larger than the scatter inherent in SFR-\Lradio. This value is also large enough that mass effects may explain the redshift evolution in the SFR-\Lradio\ relation found by \citet{calistro2017}, although it is also possible the evolution they report is partly due to their selection function (as underlined by our tests in Appendix \ref{sec:detections}), and\slash or only detectable at the higher redshifts probed by that work. 

We are continuing to obtain new 150\,MHz data over the ELAIS-N1 field, with the ultimate goal of reaching a further factor of two greater sensitivity over the coming years. Complete optical spectroscopy will be obtained for every ELAIS-N1 150\,MHz source brighter than 100\,$\mu$Jy as part of the WEAVE-LOFAR survey \citep{smith2016}, which is scheduled to begin in the second Quarter of 2021. Over five initial years of survey operations, WEAVE-LOFAR will obtain around a million spectra of LOFAR-selected sources in the best-studied extragalactic fields in the Northern hemisphere of every scale (ranging from the deep fields such as ELAIS-N1, Lockman Hole and Bo\"otes, covering the whole of the H-ATLAS NGP field used by G18, and thousands of square degrees at high galactic latitudes, for example) providing precise redshifts for virtually every source placed in a WEAVE fibre at $z < 1$. These new data will enable the use of extensive emission line classifications and Balmer-decrement derived SFRs to study the SFR-\Lradio\ relation that will enable us to significantly improve on this work, including the full coverage of the luminosity-redshift plane sampled by the wide and deep fields simultaneously, with highly uniform spectroscopy.

\begin{acknowledgements}
The authors would like to thank the anonymous reviewer for a positive and constructive report which has improved the quality of the paper. MJH acknowledges support from the UK Science and Technology Facilities Council (ST/R000905/1).
RK acknowledges support from the Science and Technology Facilities Council (STFC) through an STFC studentship via grant ST/R504737/1.
KJD and HR acknowledge support from the ERC Advanced Investigator programme NewClusters 321271.
IM acknowledges support from STFC via grant ST/R505146/1.
PNB and JS are grateful for support from the UK STFC via grant ST/R000972/1.
MJJ acknowledges support from the UK Science and Technology Facilities Council (ST/N000919/1) and the Oxford Hintze Centre for Astrophysical Surveys which is funded through generous support from the Hintze Family Charitable Foundation.
MBo acknowledges support from INAF under PRIN SKA/CTA FORECaST and from the Ministero degli Affari Esteri della Cooperazione Internazionale - Direzione Generale per la Promozione del Sistema Paese Progetto di Grande Rilevanza ZA18GR02.
IP acknowledges support from INAF under the SKA/CTA PRIN ``FORECaST'' and the PRIN MAIN STREAM ``SAuROS'' projects. This research has made use of NASA's Astrophysics Data System Bibliographic Services.

LOFAR \citep{vanhaarlem2013} is the Low Frequency Array designed and constructed by ASTRON. It has observing, data processing, and data storage facilities in several countries, which are owned by various parties (each with their own funding sources), and that are collectively operated by the ILT foundation under a joint scientific policy. The ILT resources have benefited from the following recent major funding sources: CNRS-INSU, Observatoire de Paris and Universit\'e d`Orl\'eans, France; BMBF, MIWF-NRW, MPG, Germany; Science Foundation Ireland (SFI), Department of Business, Enterprise and Innovation (DBEI), Ireland; NWO, The Netherlands; The Science and Technology Facilities Council, UK; Ministry of Science and Higher Education, Poland; The Istituto Nazionale di Astrofisica (INAF), Italy.

This research made use of the Dutch national e-infrastructure with support of the SURF Cooperative (e-infra 180169) and the LOFAR e-infra group. The J\"ulich LOFAR Long Term Archive and the German LOFAR network are both coordinated and operated by the J\"ulich Supercomputing Centre (JSC), and computing resources on the supercomputer JUWELS at JSC were provided by the Gauss Centre for Supercomputing e.V. (grant CHTB00) through the John von Neumann Institute for Computing (NIC).

This research made use of the University of Hertfordshire high-performance computing facility and the LOFAR-UK computing facility located at the University of Hertfordshire and supported by STFC [ST/P000096/1], and of the Italian LOFAR IT computing infrastructure supported and operated by INAF, and by the Physics Department of Turin University (under an agreement with Consorzio Interuniversitario per la Fisica Spaziale) at the C3S Supercomputing Centre, Italy.

\end{acknowledgements}


\bibliographystyle{aa}
\bibliography{sfr_l150}

\begin{thebibliography}{126}
\expandafter\ifx\csname natexlab\endcsname\relax\def\natexlab#1{#1}\fi

\bibitem[{{Aihara} {et~al.}(2018){Aihara}, {Arimoto}, {Armstrong}, {Arnouts},
  {Bahcall}, {Bickerton}, {Bosch}, {Bundy}, {Capak}, {Chan}, {Chiba}, {Coupon},
  {Egami}, {Enoki}, {Finet}, {Fujimori}, {Fujimoto}, {Furusawa}, {Furusawa},
  {Goto}, {Goulding}, {Greco}, {Greene}, {Gunn}, {Hamana}, {Harikane},
  {Hashimoto}, {Hattori}, {Hayashi}, {Hayashi}, {He{\l}miniak}, {Higuchi},
  {Hikage}, {Ho}, {Hsieh}, {Huang}, {Huang}, {Ikeda}, {Imanishi}, {Inoue},
  {Iwasawa}, {Iwata}, {Jaelani}, {Jian}, {Kamata}, {Karoji}, {Kashikawa},
  {Katayama}, {Kawanomoto}, {Kayo}, {Koda}, {Koike}, {Kojima}, {Komiyama},
  {Konno}, {Koshida}, {Koyama}, {Kusakabe}, {Leauthaud}, {Lee}, {Lin}, {Lin},
  {Lupton}, {Mand elbaum}, {Matsuoka}, {Medezinski}, {Mineo}, {Miyama},
  {Miyatake}, {Miyazaki}, {Momose}, {More}, {More}, {Moritani}, {Moriya},
  {Morokuma}, {Mukae}, {Murata}, {Murayama}, {Nagao}, {Nakata}, {Niida},
  {Niikura}, {Nishizawa}, {Obuchi}, {Oguri}, {Oishi}, {Okabe}, {Okamoto},
  {Okura}, {Ono}, {Onodera}, {Onoue}, {Osato}, {Ouchi}, {Price}, {Pyo}, {Sako},
  {Sawicki}, {Shibuya}, {Shimasaku}, {Shimono}, {Shirasaki}, {Silverman},
  {Simet}, {Speagle}, {Spergel}, {Strauss}, {Sugahara}, {Sugiyama}, {Suto},
  {Suyu}, {Suzuki}, {Tait}, {Takada}, {Takata}, {Tamura}, {Tanaka}, {Tanaka},
  {Tanaka}, {Tanaka}, {Terai}, {Terashima}, {Toba}, {Tominaga}, {Toshikawa},
  {Turner}, {Uchida}, {Uchiyama}, {Umetsu}, {Uraguchi}, {Urata}, {Usuda},
  {Utsumi}, {Wang}, {Wang}, {Wong}, {Yabe}, {Yamada}, {Yamanoi}, {Yasuda},
  {Yeh}, {Yonehara}, \& {Yuma}}]{aihara2018}
{Aihara}, H., {Arimoto}, N., {Armstrong}, R., {et~al.} 2018, \pasj, 70, S4

\bibitem[{{Antonucci}(1993)}]{antonucci1993}
{Antonucci}, R. 1993, \araa, 31, 473

\bibitem[{{Appleton} {et~al.}(2004){Appleton}, {Fadda}, {Marleau}, {Frayer},
  {Helou}, {Condon}, {Choi}, {Yan}, {Lacy}, {Wilson}, {Armus}, {Chapman},
  {Fang}, {Heinrichson}, {Im}, {Jannuzi}, {Storrie-Lombardi}, {Shupe},
  {Soifer}, {Squires}, \& {Teplitz}}]{appleton2004}
{Appleton}, P.~N., {Fadda}, D.~T., {Marleau}, F.~R., {et~al.} 2004, \apjs, 154,
  147

\bibitem[{{Becker} {et~al.}(1995){Becker}, {White}, \& {Helfand}}]{becker1995}
{Becker}, R.~H., {White}, R.~L., \& {Helfand}, D.~J. 1995, \apj, 450, 559

\bibitem[{{Bell}(2003)}]{bell2003}
{Bell}, E.~F. 2003, \apj, 586, 794

\bibitem[{{Berta} {et~al.}(2013){Berta}, {Lutz}, {Santini}, {Wuyts}, {Rosario},
  {Brisbin}, {Cooray}, {Franceschini}, {Gruppioni}, {Hatziminaoglou}, {Hwang},
  {Le Floc'h}, {Magnelli}, {Nordon}, {Oliver}, {Page}, {Popesso}, {Pozzetti},
  {Pozzi}, {Riguccini}, {Rodighiero}, {Roseboom}, {Scott}, {Symeonidis},
  {Valtchanov}, {Viero}, \& {Wang}}]{berta2013}
{Berta}, S., {Lutz}, D., {Santini}, P., {et~al.} 2013, \aap, 551, A100

\bibitem[{{Best} \& {Heckman}(2012)}]{best2012}
{Best}, P.~N. \& {Heckman}, T.~M. 2012, \mnras, 421, 1569

\bibitem[{{Best} {et~al.}(2020){Best}, {Kondapally}, {Duncan}, {Authors},
  {Authors}, \& {Authors}}]{best_lotss}
{Best}, P.~N., {Kondapally}, R., {Duncan}, K., {et~al.} 2020, \aap, Submitted
  (LoTSS SI)

\bibitem[{{Bonato} {et~al.}(2017){Bonato}, {Negrello}, {Mancuso}, {De Zotti},
  {Ciliegi}, {Cai}, {Lapi}, {Massardi}, {Bonaldi}, {Sajina},
  {Smol{\v{c}}i{\'c}}, \& {Schinnerer}}]{bonato2017}
{Bonato}, M., {Negrello}, M., {Mancuso}, C., {et~al.} 2017, \mnras, 469, 1912

\bibitem[{{Bourne} {et~al.}(2011){Bourne}, {Dunne}, {Ivison}, {Maddox},
  {Dickinson}, \& {Frayer}}]{bourne2011}
{Bourne}, N., {Dunne}, L., {Ivison}, R.~J., {et~al.} 2011, \mnras, 410, 1155

\bibitem[{{Brinchmann} {et~al.}(2004){Brinchmann}, {Charlot}, {White},
  {Tremonti}, {Kauffmann}, {Heckman}, \& {Brinkmann}}]{brinchmann2004}
{Brinchmann}, J., {Charlot}, S., {White}, S.~D.~M., {et~al.} 2004, \mnras, 351,
  1151

\bibitem[{{Brown} {et~al.}(2017){Brown}, {Moustakas}, {Kennicutt}, {Bonne},
  {Intema}, {de Gasperin}, {Boquien}, {Jarrett}, {Cluver}, {Smith}, {da Cunha},
  {Imanishi}, {Armus}, {Brandl}, \& {Peek}}]{brown2017}
{Brown}, M. J.~I., {Moustakas}, J., {Kennicutt}, R.~C., {et~al.} 2017, \apj,
  847, 136

\bibitem[{{Bruzual} \& {Charlot}(2003)}]{bruzual2003}
{Bruzual}, G. \& {Charlot}, S. 2003, \mnras, 344, 1000

\bibitem[{{Burgarella} {et~al.}(2005){Burgarella}, {Buat}, \&
  {Iglesias-P{\'a}ramo}}]{burgarella2005}
{Burgarella}, D., {Buat}, V., \& {Iglesias-P{\'a}ramo}, J. 2005, \mnras, 360,
  1413

\bibitem[{{Calistro Rivera} {et~al.}(2016){Calistro Rivera}, {Lusso},
  {Hennawi}, \& {Hogg}}]{calistro2016}
{Calistro Rivera}, G., {Lusso}, E., {Hennawi}, J.~F., \& {Hogg}, D.~W. 2016,
  \apj, 833, 98

\bibitem[{{Calistro Rivera} {et~al.}(2017){Calistro Rivera}, {Williams},
  {Hardcastle}, {Duncan}, {R{\"o}ttgering}, {Best}, {Br{\"u}ggen}, {Chy{\.z}y},
  {Conselice}, {de Gasperin}, {Engels}, {G{\"u}rkan}, {Intema}, {Jarvis},
  {Mahony}, {Miley}, {Morabito}, {Prandoni}, {Sabater}, {Smith}, {Tasse}, {van
  der Werf}, \& {White}}]{calistro2017}
{Calistro Rivera}, G., {Williams}, W.~L., {Hardcastle}, M.~J., {et~al.} 2017,
  \mnras, 469, 3468

\bibitem[{{Carilli} \& {Rawlings}(2004)}]{carilli2004}
{Carilli}, C.~L. \& {Rawlings}, S. 2004, \nar, 48, 979

\bibitem[{{Carnall} {et~al.}(2018){Carnall}, {McLure}, {Dunlop}, \&
  {Dav{\'e}}}]{carnall2018}
{Carnall}, A.~C., {McLure}, R.~J., {Dunlop}, J.~S., \& {Dav{\'e}}, R. 2018,
  \mnras, 480, 4379

\bibitem[{{Chabrier}(2003)}]{chabrier2003}
{Chabrier}, G. 2003, \pasp, 115, 763

\bibitem[{{Chambers} {et~al.}(2016){Chambers}, {Magnier}, {Metcalfe},
  {Flewelling}, {Huber}, {Waters}, {Denneau}, {Draper}, {Farrow}, {Finkbeiner},
  {Holmberg}, {Koppenhoefer}, {Price}, {Rest}, {Saglia}, {Schlafly}, {Smartt},
  {Sweeney}, {Wainscoat}, {Burgett}, {Chastel}, {Grav}, {Heasley}, {Hodapp},
  {Jedicke}, {Kaiser}, {Kudritzki}, {Luppino}, {Lupton}, {Monet}, {Morgan},
  {Onaka}, {Shiao}, {Stubbs}, {Tonry}, {White}, {Ba{\~n}ados}, {Bell},
  {Bender}, {Bernard}, {Boegner}, {Boffi}, {Botticella}, {Calamida},
  {Casertano}, {Chen}, {Chen}, {Cole}, {Deacon}, {Frenk}, {Fitzsimmons},
  {Gezari}, {Gibbs}, {Goessl}, {Goggia}, {Gourgue}, {Goldman}, {Grant},
  {Grebel}, {Hambly}, {Hasinger}, {Heavens}, {Heckman}, {Henderson}, {Henning},
  {Holman}, {Hopp}, {Ip}, {Isani}, {Jackson}, {Keyes}, {Koekemoer}, {Kotak},
  {Le}, {Liska}, {Long}, {Lucey}, {Liu}, {Martin}, {Masci}, {McLean}, {Mindel},
  {Misra}, {Morganson}, {Murphy}, {Obaika}, {Narayan}, {Nieto-Santisteban},
  {Norberg}, {Peacock}, {Pier}, {Postman}, {Primak}, {Rae}, {Rai}, {Riess},
  {Riffeser}, {Rix}, {R{\"o}ser}, {Russel}, {Rutz}, {Schilbach}, {Schultz},
  {Scolnic}, {Strolger}, {Szalay}, {Seitz}, {Small}, {Smith}, {Soderblom},
  {Taylor}, {Thomson}, {Taylor}, {Thakar}, {Thiel}, {Thilker}, {Unger},
  {Urata}, {Valenti}, {Wagner}, {Walder}, {Walter}, {Watters}, {Werner},
  {Wood-Vasey}, \& {Wyse}}]{chambers2016}
{Chambers}, K.~C., {Magnier}, E.~A., {Metcalfe}, N., {et~al.} 2016, arXiv
  e-prints, arXiv:1612.05560

\bibitem[{{Charlot} \& {Fall}(2000)}]{charlot2000}
{Charlot}, S. \& {Fall}, S.~M. 2000, \apj, 539, 718

\bibitem[{{Chi} \& {Wolfendale}(1990)}]{chi1990}
{Chi}, X. \& {Wolfendale}, A.~W. 1990, \mnras, 245, 101

\bibitem[{{Condon}(1992)}]{condon1992}
{Condon}, J.~J. 1992, \araa, 30, 575

\bibitem[{{Cram} {et~al.}(1998){Cram}, {Hopkins}, {Mobasher}, \&
  {Rowan-Robinson}}]{cram1998}
{Cram}, L., {Hopkins}, A., {Mobasher}, B., \& {Rowan-Robinson}, M. 1998, \apj,
  507, 155

\bibitem[{{da Cunha} {et~al.}(2008){da Cunha}, {Charlot}, \&
  {Elbaz}}]{dacunha2008}
{da Cunha}, E., {Charlot}, S., \& {Elbaz}, D. 2008, \mnras, 388, 1595

\bibitem[{{Davies} {et~al.}(2017){Davies}, {Huynh}, {Hopkins}, {Seymour},
  {Driver}, {Robotham}, {Baldry}, {Bland-Hawthorn}, {Bourne}, {Bremer},
  {Brown}, {Brough}, {Cluver}, {Grootes}, {Jarvis}, {Loveday}, {Moffet},
  {Owers}, {Phillipps}, {Sadler}, {Wang}, {Wilkins}, \& {Wright}}]{davies2017}
{Davies}, L.~J.~M., {Huynh}, M.~T., {Hopkins}, A.~M., {et~al.} 2017, \mnras,
  466, 2312

\bibitem[{{de Jong} {et~al.}(1985){de Jong}, {Klein}, {Wielebinski}, \&
  {Wunderlich}}]{dejong1985}
{de Jong}, T., {Klein}, U., {Wielebinski}, R., \& {Wunderlich}, E. 1985, \aap,
  147, L6

\bibitem[{{Delhaize} {et~al.}(2017){Delhaize}, {Smol{\v{c}}i{\'c}},
  {Delvecchio}, {Novak}, {Sargent}, {Baran}, {Magnelli}, {Zamorani},
  {Schinnerer}, {Murphy}, {Aravena}, {Berta}, {Bondi}, {Capak}, {Carilli},
  {Ciliegi}, {Civano}, {Ilbert}, {Karim}, {Laigle}, {Le F{\`e}vre}, {Marchesi},
  {McCracken}, {Salvato}, {Seymour}, \& {Tasca}}]{delhaize2017}
{Delhaize}, J., {Smol{\v{c}}i{\'c}}, V., {Delvecchio}, I., {et~al.} 2017, \aap,
  602, A4

\bibitem[{{Dewdney} {et~al.}(2009){Dewdney}, {Hall}, {Schilizzi}, \&
  {Lazio}}]{dewdney2009}
{Dewdney}, P.~E., {Hall}, P.~J., {Schilizzi}, R.~T., \& {Lazio}, T.~J.~L.~W.
  2009, IEEE Proceedings, 97, 1482

\bibitem[{{Donley} {et~al.}(2012){Donley}, {Koekemoer}, {Brusa}, {Capak},
  {Cardamone}, {Civano}, {Ilbert}, {Impey}, {Kartaltepe}, {Miyaji}, {Salvato},
  {Sanders}, {Trump}, \& {Zamorani}}]{donley2012}
{Donley}, J.~L., {Koekemoer}, A.~M., {Brusa}, M., {et~al.} 2012, \apj, 748, 142

\bibitem[{{Driver} {et~al.}(2011){Driver}, {Hill}, {Kelvin}, {Robotham},
  {Liske}, {Norberg}, {Baldry}, {Bamford}, {Hopkins}, {Loveday}, {Peacock},
  {Andrae}, {Bland -Hawthorn}, {Brough}, {Brown}, {Cameron}, {Ching},
  {Colless}, {Conselice}, {Croom}, {Cross}, {de Propris}, {Dye}, {Drinkwater},
  {Ellis}, {Graham}, {Grootes}, {Gunawardhana}, {Jones}, {van Kampen},
  {Maraston}, {Nichol}, {Parkinson}, {Phillipps}, {Pimbblet}, {Popescu},
  {Prescott}, {Roseboom}, {Sadler}, {Sansom}, {Sharp}, {Smith}, {Taylor},
  {Thomas}, {Tuffs}, {Wijesinghe}, {Dunne}, {Frenk}, {Jarvis}, {Madore},
  {Meyer}, {Seibert}, {Staveley-Smith}, {Sutherland}, \& {Warren}}]{driver2011}
{Driver}, S.~P., {Hill}, D.~T., {Kelvin}, L.~S., {et~al.} 2011, \mnras, 413,
  971

\bibitem[{{Dudzevi{\v{c}}i{\={u}}t{\.{e}}}
  {et~al.}(2020){Dudzevi{\v{c}}i{\={u}}t{\.{e}}}, {Smail}, {Swinbank}, {Stach},
  {Almaini}, {da Cunha}, {An}, {Arumugam}, {Birkin}, {Blain}, {Chapman},
  {Chen}, {Conselice}, {Coppin}, {Dunlop}, {Farrah}, {Geach}, {Gullberg},
  {Hartley}, {Hodge}, {Ivison}, {Maltby}, {Scott}, {Simpson}, {Simpson},
  {Thomson}, {Walter}, {Wardlow}, {Weiss}, \& {van der Werf}}]{dud2020}
{Dudzevi{\v{c}}i{\={u}}t{\.{e}}}, U., {Smail}, I., {Swinbank}, A.~M., {et~al.}
  2020, \mnras, 494, 3828

\bibitem[{{Duncan} {et~al.}(2020{\natexlab{a}}){Duncan}, {Kondapally}, {Brown},
  {Authors}, {Authors}, \& {Authors}}]{duncan_lotss}
{Duncan}, K., {Kondapally}, R., {Brown}, M., {et~al.} 2020{\natexlab{a}}, \aap,
  Submitted (LoTSS SI)

\bibitem[{{Duncan} {et~al.}(2019){Duncan}, {Sabater}, {R{\"o}ttgering},
  {Jarvis}, {Smith}, {Best}, {Callingham}, {Cochrane}, {Croston}, {Hardcastle},
  {Mingo}, {Morabito}, {Nisbet}, {Prandoni}, {Shimwell}, {Tasse}, {White},
  {Williams}, {Alegre}, {Chy{\.z}y}, {G{\"u}rkan}, {Hoeft}, {Kondapally},
  {Mechev}, {Miley}, {Schwarz}, \& {van Weeren}}]{duncan2019}
{Duncan}, K.~J., {Sabater}, J., {R{\"o}ttgering}, H.~J.~A., {et~al.} 2019,
  \aap, 622, A3

\bibitem[{{Duncan} {et~al.}(2020{\natexlab{b}}){Duncan}, {Shivaei}, {Shapley},
  {Reddy}, {Mobasher}, {Coil}, {Kriek}, \& {Siana}}]{duncan2020}
{Duncan}, K.~J., {Shivaei}, I., {Shapley}, A.~E., {et~al.} 2020{\natexlab{b}},
  arXiv e-prints, arXiv:2008.04329

\bibitem[{{Flesch}(2019)}]{flesch2019}
{Flesch}, E.~W. 2019, arXiv e-prints, arXiv:1912.05614

\bibitem[{{Foreman-Mackey} {et~al.}(2013){Foreman-Mackey}, {Hogg}, {Lang}, \&
  {Goodman}}]{foreman2013}
{Foreman-Mackey}, D., {Hogg}, D.~W., {Lang}, D., \& {Goodman}, J. 2013, \pasp,
  125, 306

\bibitem[{{Garn} {et~al.}(2009){Garn}, {Green}, {Riley}, \& {Alexand
  er}}]{garn2009}
{Garn}, T., {Green}, D.~A., {Riley}, J.~M., \& {Alexand er}, P. 2009, \mnras,
  397, 1101

\bibitem[{{Gott} {et~al.}(2001){Gott}, {Vogeley}, {Podariu}, \&
  {Ratra}}]{gott2001}
{Gott}, J.~Richard, I., {Vogeley}, M.~S., {Podariu}, S., \& {Ratra}, B. 2001,
  \apj, 549, 1

\bibitem[{{G{\"u}rkan} {et~al.}(2018){G{\"u}rkan}, {Hardcastle}, {Smith},
  {Best}, {Bourne}, {Calistro-Rivera}, {Heald}, {Jarvis}, {Prandoni},
  {R{\"o}ttgering}, {Sabater}, {Shimwell}, {Tasse}, \& {Williams}}]{gurkan2018}
{G{\"u}rkan}, G., {Hardcastle}, M.~J., {Smith}, D.~J.~B., {et~al.} 2018,
  \mnras, 475, 3010

\bibitem[{{Haarsma} {et~al.}(2000){Haarsma}, {Partridge}, {Windhorst}, \&
  {Richards}}]{haarsma2000}
{Haarsma}, D.~B., {Partridge}, R.~B., {Windhorst}, R.~A., \& {Richards}, E.~A.
  2000, \apj, 544, 641

\bibitem[{{Hardcastle} {et~al.}(2019{\natexlab{a}}){Hardcastle}, {Croston},
  {Shimwell}, {Tasse}, {G{\"u}rkan}, {Morganti}, {Murgia}, {R{\"o}ttgering},
  {van Weeren}, \& {Williams}}]{hardcastle2019a}
{Hardcastle}, M.~J., {Croston}, J.~H., {Shimwell}, T.~W., {et~al.}
  2019{\natexlab{a}}, \mnras, 488, 3416

\bibitem[{{Hardcastle} {et~al.}(2007){Hardcastle}, {Evans}, \&
  {Croston}}]{hardcastle2007}
{Hardcastle}, M.~J., {Evans}, D.~A., \& {Croston}, J.~H. 2007, \mnras, 376,
  1849

\bibitem[{{Hardcastle} {et~al.}(2016){Hardcastle}, {G{\"u}rkan}, {van Weeren},
  {Williams}, {Best}, {de Gasperin}, {Rafferty}, {Read}, {Sabater}, {Shimwell},
  {Smith}, {Tasse}, {Bourne}, {Brienza}, {Br{\"u}ggen}, {Brunetti},
  {Chy{\.z}y}, {Conway}, {Dunne}, {Eales}, {Maddox}, {Jarvis}, {Mahony},
  {Morganti}, {Prandoni}, {R{\"o}ttgering}, {Valiante}, \&
  {White}}]{hardcastle2016}
{Hardcastle}, M.~J., {G{\"u}rkan}, G., {van Weeren}, R.~J., {et~al.} 2016,
  \mnras, 462, 1910

\bibitem[{{Hardcastle} {et~al.}(2019{\natexlab{b}}){Hardcastle}, {Williams},
  {Best}, {Croston}, {Duncan}, {R{\"o}ttgering}, {Sabater}, {Shimwell},
  {Tasse}, {Callingham}, {Cochrane}, {de Gasperin}, {G{\"u}rkan}, {Jarvis},
  {Mahatma}, {Miley}, {Mingo}, {Mooney}, {Morabito}, {O'Sullivan}, {Prandoni},
  {Shulevski}, \& {Smith}}]{hardcastle2019}
{Hardcastle}, M.~J., {Williams}, W.~L., {Best}, P.~N., {et~al.}
  2019{\natexlab{b}}, \aap, 622, A12

\bibitem[{{Hayward} \& {Smith}(2015)}]{hayward2015}
{Hayward}, C.~C. \& {Smith}, D. J.~B. 2015, \mnras, 446, 1512

\bibitem[{{Helou} {et~al.}(1985){Helou}, {Soifer}, \&
  {Rowan-Robinson}}]{helou1985}
{Helou}, G., {Soifer}, B.~T., \& {Rowan-Robinson}, M. 1985, \apjl, 298, L7

\bibitem[{{Hodge} {et~al.}(2008){Hodge}, {Becker}, {White}, \& {de
  Vries}}]{hodge2008}
{Hodge}, J.~A., {Becker}, R.~H., {White}, R.~L., \& {de Vries}, W.~H. 2008,
  \aj, 136, 1097

\bibitem[{{Hopkins} {et~al.}(2003){Hopkins}, {Miller}, {Nichol}, {Connolly},
  {Bernardi}, {G{\'o}mez}, {Goto}, {Tremonti}, {Brinkmann}, {Ivezi{\'c}}, \&
  {Lamb}}]{hopkins2003}
{Hopkins}, A.~M., {Miller}, C.~J., {Nichol}, R.~C., {et~al.} 2003, \apj, 599,
  971

\bibitem[{{Hurley} {et~al.}(2017){Hurley}, {Oliver}, {Betancourt}, {Clarke},
  {Cowley}, {Duivenvoorden}, {Farrah}, {Griffin}, {Lacey}, {Le Floc'h},
  {Papadopoulos}, {Sargent}, {Scudder}, {Vaccari}, {Valtchanov}, \&
  {Wang}}]{hurley2017}
{Hurley}, P.~D., {Oliver}, S., {Betancourt}, M., {et~al.} 2017, \mnras, 464,
  885

\bibitem[{{Intema} {et~al.}(2017){Intema}, {Jagannathan}, {Mooley}, \&
  {Frail}}]{intema2017}
{Intema}, H.~T., {Jagannathan}, P., {Mooley}, K.~P., \& {Frail}, D.~A. 2017,
  \aap, 598, A78

\bibitem[{{Ivison} {et~al.}(2010{\natexlab{a}}){Ivison}, {Alexander}, {Biggs},
  {Brand t}, {Chapin}, {Coppin}, {Devlin}, {Dickinson}, {Dunlop}, {Dye},
  {Eales}, {Frayer}, {Halpern}, {Hughes}, {Ibar}, {Kov{\'a}cs}, {Marsden},
  {Moncelsi}, {Netterfield}, {Pascale}, {Patanchon}, {Rafferty}, {Rex},
  {Schinnerer}, {Scott}, {Semisch}, {Smail}, {Swinbank}, {Truch}, {Tucker},
  {Viero}, {Walter}, {Wei{\ss}}, {Wiebe}, \& {Xue}}]{ivison2010a}
{Ivison}, R.~J., {Alexander}, D.~M., {Biggs}, A.~D., {et~al.}
  2010{\natexlab{a}}, \mnras, 402, 245

\bibitem[{{Ivison} {et~al.}(2010{\natexlab{b}}){Ivison}, {Magnelli}, {Ibar},
  {Andreani}, {Elbaz}, {Altieri}, {Amblard}, {Arumugam}, {Auld}, {Aussel},
  {Babbedge}, {Berta}, {Blain}, {Bock}, {Bongiovanni}, {Boselli}, {Buat},
  {Burgarella}, {Castro-Rodr{\'\i}guez}, {Cava}, {Cepa}, {Chanial}, {Cimatti},
  {Cirasuolo}, {Clements}, {Conley}, {Conversi}, {Cooray}, {Daddi},
  {Dominguez}, {Dowell}, {Dwek}, {Eales}, {Farrah}, {F{\"o}rster Schreiber},
  {Fox}, {Franceschini}, {Gear}, {Genzel}, {Glenn}, {Griffin}, {Gruppioni},
  {Halpern}, {Hatziminaoglou}, {Isaak}, {Lagache}, {Levenson}, {Lu}, {Lutz},
  {Madden}, {Maffei}, {Magdis}, {Mainetti}, {Maiolino}, {Marchetti},
  {Morrison}, {Mortier}, {Nguyen}, {Nordon}, {O'Halloran}, {Oliver}, {Omont},
  {Owen}, {Page}, {Panuzzo}, {Papageorgiou}, {Pearson}, {P{\'e}rez-Fournon},
  {P{\'e}rez Garc{\'\i}a}, {Poglitsch}, {Pohlen}, {Popesso}, {Pozzi},
  {Rawlings}, {Raymond}, {Rigopoulou}, {Riguccini}, {Rizzo}, {Rodighiero},
  {Roseboom}, {Rowan-Robinson}, {Saintonge}, {Sanchez Portal}, {Santini},
  {Schulz}, {Scott}, {Seymour}, {Shao}, {Shupe}, {Smith}, {Stevens}, {Sturm},
  {Symeonidis}, {Tacconi}, {Trichas}, {Tugwell}, {Vaccari}, {Valtchanov},
  {Vieira}, {Vigroux}, {Wang}, {Ward}, {Wright}, {Xu}, \&
  {Zemcov}}]{ivison2010b}
{Ivison}, R.~J., {Magnelli}, B., {Ibar}, E., {et~al.} 2010{\natexlab{b}}, \aap,
  518, L31

\bibitem[{{Jarvis} {et~al.}(2010){Jarvis}, {Smith}, {Bonfield}, {Hardcastle},
  {Falder}, {Stevens}, {Ivison}, {Auld}, {Baes}, {Baldry}, {Bamford}, {Bourne},
  {Buttiglione}, {Cava}, {Cooray}, {Dariush}, {de Zotti}, {Dunlop}, {Dunne},
  {Dye}, {Eales}, {Fritz}, {Hill}, {Hopwood}, {Hughes}, {Ibar}, {Jones},
  {Kelvin}, {Lawrence}, {Leeuw}, {Loveday}, {Maddox}, {Micha{\l}owski},
  {Negrello}, {Norberg}, {Pohlen}, {Prescott}, {Rigby}, {Robotham},
  {Rodighiero}, {Scott}, {Sharp}, {Temi}, {Thompson}, {van der Werf}, {van
  Kampen}, {Vlahakis}, \& {White}}]{jarvis2010}
{Jarvis}, M.~J., {Smith}, D.~J.~B., {Bonfield}, D.~G., {et~al.} 2010, \mnras,
  409, 92

\bibitem[{{Johnston} {et~al.}(2015){Johnston}, {Vaccari}, {Jarvis}, {Smith},
  {Giovannoli}, {H{\"a}u{\ss}ler}, \& {Prescott}}]{johnston2015}
{Johnston}, R., {Vaccari}, M., {Jarvis}, M., {et~al.} 2015, \mnras, 453, 2540

\bibitem[{{Karim} {et~al.}(2011){Karim}, {Schinnerer},
  {Mart{\'\i}nez-Sansigre}, {Sargent}, {van der Wel}, {Rix}, {Ilbert},
  {Smol{\v{c}}i{\'c}}, {Carilli}, {Pannella}, {Koekemoer}, {Bell}, \&
  {Salvato}}]{karim2011}
{Karim}, A., {Schinnerer}, E., {Mart{\'\i}nez-Sansigre}, A., {et~al.} 2011,
  \apj, 730, 61

\bibitem[{{Kellermann} \& {Owen}(1988)}]{kellermann1988}
{Kellermann}, K.~I. \& {Owen}, F.~N. 1988, {Radio galaxies and quasars.},
  563--602

\bibitem[{{Kennicutt}(1998)}]{kennicutt1998}
{Kennicutt}, Robert~C., J. 1998, \araa, 36, 189

\bibitem[{{Kennicutt} {et~al.}(2009){Kennicutt}, {Hao}, {Calzetti},
  {Moustakas}, {Dale}, {Bendo}, {Engelbracht}, {Johnson}, \&
  {Lee}}]{kennicutt2009}
{Kennicutt}, Robert~C., J., {Hao}, C.-N., {Calzetti}, D., {et~al.} 2009, \apj,
  703, 1672

\bibitem[{{Kennicutt} \& {Evans}(2012)}]{kennicutt2012}
{Kennicutt}, R.~C. \& {Evans}, N.~J. 2012, \araa, 50, 531

\bibitem[{{Kondapally} {et~al.}(2020){Kondapally}, {Best}, {Hardcastle},
  {Authors}, {Authors}, \& {Authors}}]{kondapally_lotss}
{Kondapally}, R., {Best}, P., {Hardcastle}, M., {et~al.} 2020, \aap, Submitted
  (LoTSS SI)

\bibitem[{{Lacki} \& {Thompson}(2010)}]{lacki2010a}
{Lacki}, B.~C. \& {Thompson}, T.~A. 2010, \apj, 717, 196

\bibitem[{{Lacki} {et~al.}(2010){Lacki}, {Thompson}, \& {Quataert}}]{lacki2010}
{Lacki}, B.~C., {Thompson}, T.~A., \& {Quataert}, E. 2010, \apj, 717, 1

\bibitem[{{Lacy} {et~al.}(2004){Lacy}, {Storrie-Lombardi}, {Sajina},
  {Appleton}, {Armus}, {Chapman}, {Choi}, {Fadda}, {Fang}, {Frayer},
  {Heinrichsen}, {Helou}, {Im}, {Marleau}, {Masci}, {Shupe}, {Soifer},
  {Surace}, {Teplitz}, {Wilson}, \& {Yan}}]{lacy2004}
{Lacy}, M., {Storrie-Lombardi}, L.~J., {Sajina}, A., {et~al.} 2004, \apjs, 154,
  166

\bibitem[{{Lawrence} {et~al.}(2007){Lawrence}, {Warren}, {Almaini}, {Edge},
  {Hambly}, {Jameson}, {Lucas}, {Casali}, {Adamson}, {Dye}, {Emerson},
  {Foucaud}, {Hewett}, {Hirst}, {Hodgkin}, {Irwin}, {Lodieu}, {McMahon},
  {Simpson}, {Smail}, {Mortlock}, \& {Folger}}]{lawrence2007}
{Lawrence}, A., {Warren}, S.~J., {Almaini}, O., {et~al.} 2007, \mnras, 379,
  1599

\bibitem[{{Leslie} {et~al.}(2020){Leslie}, {Schinnerer}, {Liu}, {Magnelli},
  {Algera}, {Karim}, {Davidzon}, {Gozaliasl}, {Jim{\'e}nez-Andrade}, {Lang},
  {Sargent}, {Novak}, {Groves}, {Smol{\v{c}}i{\'c}}, {Zamorani}, {Vaccari},
  {Battisti}, {Vardoulaki}, {Peng}, \& {Kartaltepe}}]{leslie2020}
{Leslie}, S., {Schinnerer}, E., {Liu}, D., {et~al.} 2020, arXiv e-prints,
  arXiv:2006.13937

\bibitem[{{Lisenfeld} {et~al.}(1996){Lisenfeld}, {Voelk}, \&
  {Xu}}]{lisenfeld1996}
{Lisenfeld}, U., {Voelk}, H.~J., \& {Xu}, C. 1996, \aap, 306, 677

\bibitem[{{Lofthouse} {et~al.}(2018){Lofthouse}, {Kaviraj}, {Smith}, \&
  {Hardcastle}}]{lofthouse2018}
{Lofthouse}, E.~K., {Kaviraj}, S., {Smith}, D.~J.~B., \& {Hardcastle}, M.~J.
  2018, \mnras, 479, 807

\bibitem[{{Lonsdale} {et~al.}(2003){Lonsdale}, {Smith}, {Rowan-Robinson},
  {Surace}, {Shupe}, {Xu}, {Oliver}, {Padgett}, {Fang}, {Conrow},
  {Franceschini}, {Gautier}, {Griffin}, {Hacking}, {Masci}, {Morrison},
  {O'Linger}, {Owen}, {P{\'e}rez-Fournon}, {Pierre}, {Puetter}, {Stacey},
  {Castro}, {Polletta}, {Farrah}, {Jarrett}, {Frayer}, {Siana}, {Babbedge},
  {Dye}, {Fox}, {Gonzalez-Solares}, {Salaman}, {Berta}, {Condon}, {Dole}, \&
  {Serjeant}}]{lonsdale2003}
{Lonsdale}, C.~J., {Smith}, H.~E., {Rowan-Robinson}, M., {et~al.} 2003, \pasp,
  115, 897

\bibitem[{{Madau} \& {Dickinson}(2014)}]{madau2014}
{Madau}, P. \& {Dickinson}, M. 2014, \araa, 52, 415

\bibitem[{{Magnelli} {et~al.}(2015){Magnelli}, {Ivison}, {Lutz}, {Valtchanov},
  {Farrah}, {Berta}, {Bertoldi}, {Bock}, {Cooray}, {Ibar}, {Karim}, {Le
  Floc'h}, {Nordon}, {Oliver}, {Page}, {Popesso}, {Pozzi}, {Rigopoulou},
  {Riguccini}, {Rodighiero}, {Rosario}, {Roseboom}, {Wang}, \&
  {Wuyts}}]{magnelli2015}
{Magnelli}, B., {Ivison}, R.~J., {Lutz}, D., {et~al.} 2015, \aap, 573, A45

\bibitem[{{Mahatma} {et~al.}(2019){Mahatma}, {Hardcastle}, {Williams}, {Best},
  {Croston}, {Duncan}, {Mingo}, {Morganti}, {Brienza}, {Cochrane},
  {G{\"u}rkan}, {Harwood}, {Jarvis}, {Jamrozy}, {Jurlin}, {Morabito},
  {R{\"o}ttgering}, {Sabater}, {Shimwell}, {Smith}, {Shulevski}, \&
  {Tasse}}]{mahatma2019}
{Mahatma}, V.~H., {Hardcastle}, M.~J., {Williams}, W.~L., {et~al.} 2019, \aap,
  622, A13

\bibitem[{{Malefahlo} {et~al.}(2020){Malefahlo}, {Santos}, {Jarvis}, {White},
  \& {Zwart}}]{malefahlo2020}
{Malefahlo}, E., {Santos}, M.~G., {Jarvis}, M.~J., {White}, S.~V., \& {Zwart},
  J. T.~L. 2020, \mnras, 492, 5297

\bibitem[{{Mart{\'\i}nez-Sansigre} {et~al.}(2005){Mart{\'\i}nez-Sansigre},
  {Rawlings}, {Lacy}, {Fadda}, {Marleau}, {Simpson}, {Willott}, \&
  {Jarvis}}]{martinez2005}
{Mart{\'\i}nez-Sansigre}, A., {Rawlings}, S., {Lacy}, M., {et~al.} 2005, \nat,
  436, 666

\bibitem[{{Mauch} {et~al.}(2013){Mauch}, {Kl{\"o}ckner}, {Rawlings}, {Jarvis},
  {Hardcastle}, {Obreschkow}, {Saikia}, \& {Thompson}}]{mauch2013}
{Mauch}, T., {Kl{\"o}ckner}, H.-R., {Rawlings}, S., {et~al.} 2013, \mnras, 435,
  650

\bibitem[{{Mauduit} {et~al.}(2012){Mauduit}, {Lacy}, {Farrah}, {Surace},
  {Jarvis}, {Oliver}, {Maraston}, {Vaccari}, {Marchetti}, {Zeimann},
  {Gonz{\'a}les-Solares}, {Pforr}, {Petric}, {Henriques}, {Thomas}, {Afonso},
  {Rettura}, {Wilson}, {Falder}, {Geach}, {Huynh}, {Norris}, {Seymour},
  {Richards}, {Stanford}, {Alexand er}, {Becker}, {Best}, {Bizzocchi},
  {Bonfield}, {Castro}, {Cava}, {Chapman}, {Christopher}, {Clements}, {Covone},
  {Dubois}, {Dunlop}, {Dyke}, {Edge}, {Ferguson}, {Foucaud}, {Franceschini},
  {Gal}, {Grant}, {Grossi}, {Hatziminaoglou}, {Hickey}, {Hodge}, {Huang},
  {Ivison}, {Kim}, {LeFevre}, {Lehnert}, {Lonsdale}, {Lubin}, {McLure},
  {Messias}, {Mart{\'\i}nez-Sansigre}, {Mortier}, {Nielsen}, {Ouchi}, {Parish},
  {Perez-Fournon}, {Pierre}, {Rawlings}, {Readhead}, {Ridgway}, {Rigopoulou},
  {Romer}, {Rosebloom}, {Rottgering}, {Rowan-Robinson}, {Sajina}, {Simpson},
  {Smail}, {Squires}, {Stevens}, {Taylor}, {Trichas}, {Urrutia}, {van Kampen},
  {Verma}, \& {Xu}}]{mauduit2012}
{Mauduit}, J.~C., {Lacy}, M., {Farrah}, D., {et~al.} 2012, \pasp, 124, 714

\bibitem[{{McAlpine} {et~al.}(2012){McAlpine}, {Smith}, {Jarvis}, {Bonfield},
  \& {Fleuren}}]{mcalpine2012}
{McAlpine}, K., {Smith}, D.~J.~B., {Jarvis}, M.~J., {Bonfield}, D.~G., \&
  {Fleuren}, S. 2012, \mnras, 423, 132

\bibitem[{{Moln{\'a}r} {et~al.}(2018){Moln{\'a}r}, {Sargent}, {Delhaize},
  {Delvecchio}, {Smol{\v{c}}i{\'c}}, {Novak}, {Schinnerer}, {Zamorani},
  {Bondi}, {Herrera-Ruiz}, {Murphy}, {Vardoulaki}, {Karim}, {Leslie},
  {Magnelli}, {Carollo}, \& {Middelberg}}]{molnar2018}
{Moln{\'a}r}, D.~C., {Sargent}, M.~T., {Delhaize}, J., {et~al.} 2018, \mnras,
  475, 827

\bibitem[{{Murphy}(2009)}]{murphy2009}
{Murphy}, E.~J. 2009, \apj, 706, 482

\bibitem[{{Murphy} {et~al.}(2011){Murphy}, {Condon}, {Schinnerer}, {Kennicutt},
  {Calzetti}, {Armus}, {Helou}, {Turner}, {Aniano}, {Beir{\~a}o}, {Bolatto},
  {Brandl}, {Croxall}, {Dale}, {Donovan Meyer}, {Draine}, {Engelbracht},
  {Hunt}, {Hao}, {Koda}, {Roussel}, {Skibba}, \& {Smith}}]{murphy2011}
{Murphy}, E.~J., {Condon}, J.~J., {Schinnerer}, E., {et~al.} 2011, \apj, 737,
  67

\bibitem[{{Murphy} {et~al.}(2008){Murphy}, {Helou}, {Kenney}, {Armus}, \&
  {Braun}}]{murphy2008}
{Murphy}, E.~J., {Helou}, G., {Kenney}, J.~D.~P., {Armus}, L., \& {Braun}, R.
  2008, \apj, 678, 828

\bibitem[{{Muzzin} {et~al.}(2009){Muzzin}, {Wilson}, {Yee}, {Hoekstra},
  {Gilbank}, {Surace}, {Lacy}, {Blindert}, {Majumdar}, {Demarco}, {Gardner},
  {Gladders}, \& {Lonsdale}}]{muzzin2009}
{Muzzin}, A., {Wilson}, G., {Yee}, H.~K.~C., {et~al.} 2009, \apj, 698, 1934

\bibitem[{{Nisbet}(2018)}]{nisbetthesis}
{Nisbet}, D.~M. 2018, PhD thesis, University of Edinburgh

\bibitem[{{Noeske} {et~al.}(2007){Noeske}, {Weiner}, {Faber}, {Papovich},
  {Koo}, {Somerville}, {Bundy}, {Conselice}, {Newman}, {Schiminovich}, {Le
  Floc'h}, {Coil}, {Rieke}, {Lotz}, {Primack}, {Barmby}, {Cooper}, {Davis},
  {Ellis}, {Fazio}, {Guhathakurta}, {Huang}, {Kassin}, {Martin}, {Phillips},
  {Rich}, {Small}, {Willmer}, \& {Wilson}}]{noeske2007}
{Noeske}, K.~G., {Weiner}, B.~J., {Faber}, S.~M., {et~al.} 2007, \apjl, 660,
  L43

\bibitem[{{Novak} {et~al.}(2017){Novak}, {Smol{\v{c}}i{\'c}}, {Delhaize},
  {Delvecchio}, {Zamorani}, {Baran}, {Bondi}, {Capak}, {Carilli}, {Ciliegi},
  {Civano}, {Ilbert}, {Karim}, {Laigle}, {Le F{\`e}vre}, {Marchesi},
  {McCracken}, {Miettinen}, {Salvato}, {Sargent}, {Schinnerer}, \&
  {Tasca}}]{novak2017}
{Novak}, M., {Smol{\v{c}}i{\'c}}, V., {Delhaize}, J., {et~al.} 2017, \aap, 602,
  A5

\bibitem[{{Oliver} {et~al.}(2012){Oliver}, {Bock}, {Altieri}, {Amblard},
  {Arumugam}, {Aussel}, {Babbedge}, {Beelen}, {B{\'e}thermin}, {Blain},
  {Boselli}, {Bridge}, {Brisbin}, {Buat}, {Burgarella},
  {Castro-Rodr{\'\i}guez}, {Cava}, {Chanial}, {Cirasuolo}, {Clements},
  {Conley}, {Conversi}, {Cooray}, {Dowell}, {Dubois}, {Dwek}, {Dye}, {Eales},
  {Elbaz}, {Farrah}, {Feltre}, {Ferrero}, {Fiolet}, {Fox}, {Franceschini},
  {Gear}, {Giovannoli}, {Glenn}, {Gong}, {Gonz{\'a}lez Solares}, {Griffin},
  {Halpern}, {Harwit}, {Hatziminaoglou}, {Heinis}, {Hurley}, {Hwang}, {Hyde},
  {Ibar}, {Ilbert}, {Isaak}, {Ivison}, {Lagache}, {Le Floc'h}, {Levenson},
  {Faro}, {Lu}, {Madden}, {Maffei}, {Magdis}, {Mainetti}, {Marchetti},
  {Marsden}, {Marshall}, {Mortier}, {Nguyen}, {O'Halloran}, {Omont}, {Page},
  {Panuzzo}, {Papageorgiou}, {Patel}, {Pearson}, {P{\'e}rez-Fournon}, {Pohlen},
  {Rawlings}, {Raymond}, {Rigopoulou}, {Riguccini}, {Rizzo}, {Rodighiero},
  {Roseboom}, {Rowan-Robinson}, {S{\'a}nchez Portal}, {Schulz}, {Scott},
  {Seymour}, {Shupe}, {Smith}, {Stevens}, {Symeonidis}, {Trichas}, {Tugwell},
  {Vaccari}, {Valtchanov}, {Vieira}, {Viero}, {Vigroux}, {Wang}, {Ward},
  {Wardlow}, {Wright}, {Xu}, \& {Zemcov}}]{oliver2012}
{Oliver}, S.~J., {Bock}, J., {Altieri}, B., {et~al.} 2012, \mnras, 424, 1614

\bibitem[{{Pannella} {et~al.}(2009){Pannella}, {Carilli}, {Daddi}, {McCracken},
  {Owen}, {Renzini}, {Strazzullo}, {Civano}, {Koekemoer}, {Schinnerer},
  {Scoville}, {Smol{\v{c}}i{\'c}}, {Taniguchi}, {Aussel}, {Kneib}, {Ilbert},
  {Mellier}, {Salvato}, {Thompson}, \& {Willott}}]{pannella2009}
{Pannella}, M., {Carilli}, C.~L., {Daddi}, E., {et~al.} 2009, \apjl, 698, L116

\bibitem[{{Pannella} {et~al.}(2015){Pannella}, {Elbaz}, {Daddi}, {Dickinson},
  {Hwang}, {Schreiber}, {Strazzullo}, {Aussel}, {Bethermin}, {Buat},
  {Charmandaris}, {Cibinel}, {Juneau}, {Ivison}, {Le Borgne}, {Le Floc'h},
  {Leiton}, {Lin}, {Magdis}, {Morrison}, {Mullaney}, {Onodera}, {Renzini},
  {Salim}, {Sargent}, {Scott}, {Shu}, \& {Wang}}]{pannella2015}
{Pannella}, M., {Elbaz}, D., {Daddi}, E., {et~al.} 2015, \apj, 807, 141

\bibitem[{{Pearson} {et~al.}(2017){Pearson}, {Wang}, {van der Tak}, {Hurley},
  {Burgarella}, \& {Oliver}}]{pearson2017}
{Pearson}, W.~J., {Wang}, L., {van der Tak}, F.~F.~S., {et~al.} 2017, \aap,
  603, A102

\bibitem[{{Pozzetti} {et~al.}(2010){Pozzetti}, {Bolzonella}, {Zucca},
  {Zamorani}, {Lilly}, {Renzini}, {Moresco}, {Mignoli}, {Cassata}, {Tasca},
  {Lamareille}, {Maier}, {Meneux}, {Halliday}, {Oesch}, {Vergani}, {Caputi},
  {Kova{\v{c}}}, {Cimatti}, {Cucciati}, {Iovino}, {Peng}, {Carollo}, {Contini},
  {Kneib}, {Le F{\'e}vre}, {Mainieri}, {Scodeggio}, {Bardelli}, {Bongiorno},
  {Coppa}, {de la Torre}, {de Ravel}, {Franzetti}, {Garilli}, {Kampczyk},
  {Knobel}, {Le Borgne}, {Le Brun}, {Pell{\`o}}, {Perez Montero},
  {Ricciardelli}, {Silverman}, {Tanaka}, {Tresse}, {Abbas}, {Bottini}, {Cappi},
  {Guzzo}, {Koekemoer}, {Leauthaud}, {Maccagni}, {Marinoni}, {McCracken},
  {Memeo}, {Porciani}, {Scaramella}, {Scarlata}, \& {Scoville}}]{pozzetti2010}
{Pozzetti}, L., {Bolzonella}, M., {Zucca}, E., {et~al.} 2010, \aap, 523, A13

\bibitem[{{Prescott} {et~al.}(2016){Prescott}, {Mauch}, {Jarvis}, {McAlpine},
  {Smith}, {Fine}, {Johnston}, {Hardcastle}, {Baldry}, {Brough}, {Brown},
  {Bremer}, {Driver}, {Hopkins}, {Kelvin}, {Loveday}, {Norberg}, {Obreschkow},
  \& {Sadler}}]{prescott2016}
{Prescott}, M., {Mauch}, T., {Jarvis}, M.~J., {et~al.} 2016, \mnras, 457, 730

\bibitem[{{Read}(2019)}]{readthesis}
{Read}, S.~C. 2019, PhD thesis, University of Hertfordshire

\bibitem[{{Read} {et~al.}(2018){Read}, {Smith}, {G{\"u}rkan}, {Hardcastle},
  {Williams}, {Best}, {Brinks}, {Calistro-Rivera}, {Chy{\.Z}y}, {Duncan},
  {Dunne}, {Jarvis}, {Morabito}, {Prandoni}, {R{\"o}ttgering}, {Sabater}, \&
  {Viaene}}]{read2018}
{Read}, S.~C., {Smith}, D.~J.~B., {G{\"u}rkan}, G., {et~al.} 2018, \mnras, 480,
  5625

\bibitem[{{Read} {et~al.}(2020){Read}, {Smith}, {Jarvis}, \&
  {G{\"u}rkan}}]{read2020}
{Read}, S.~C., {Smith}, D.~J.~B., {Jarvis}, M.~J., \& {G{\"u}rkan}, G. 2020,
  \mnras, 492, 3940

\bibitem[{{Retana-Montenegro} {et~al.}(2018){Retana-Montenegro},
  {R{\"o}ttgering}, {Shimwell}, {van Weeren}, {Prandoni}, {Brunetti}, {Best},
  \& {Br{\"u}ggen}}]{retana2018}
{Retana-Montenegro}, E., {R{\"o}ttgering}, H.~J.~A., {Shimwell}, T.~W.,
  {et~al.} 2018, \aap, 620, A74

\bibitem[{{R{\"o}ttgering} {et~al.}(2011){R{\"o}ttgering}, {Afonso}, {Barthel},
  {Batejat}, {Best}, {Bonafede}, {Br{\"u}ggen}, {Brunetti}, {Chy{\.z}y},
  {Conway}, {de Gasperin}, {Ferrari}, {Haverkorn}, {Heald}, {Hoeft}, {Jackson},
  {Jarvis}, {Ker}, {Lehnert}, {Macario}, {McKean}, {Miley}, {Morganti},
  {Oosterloo}, {Orr{\`u}}, {Pizzo}, {Rafferty}, {Shulevski}, {Tasse}, {van
  Bemmel}, {van der Tol}, {van Weeren}, {Verheijen}, {White}, \&
  {Wise}}]{rottgering2011}
{R{\"o}ttgering}, H., {Afonso}, J., {Barthel}, P., {et~al.} 2011, Journal of
  Astrophysics and Astronomy, 32, 557

\bibitem[{{Sabater} {et~al.}(2020){Sabater}, {Best}, {Tasse}, {Authors},
  {Authors}, \& {Authors}}]{sabater_lotss}
{Sabater}, J., {Best}, P., {Tasse}, C., {et~al.} 2020, \aap, Submitted (LoTSS
  SI)

\bibitem[{{Sabater} {et~al.}(2019){Sabater}, {Best}, {Hardcastle}, {Shimwell},
  {Tasse}, {Williams}, {Br{\"u}ggen}, {Cochrane}, {Croston}, {de Gasperin},
  {Duncan}, {G{\"u}rkan}, {Mechev}, {Morabito}, {Prandoni}, {R{\"o}ttgering},
  {Smith}, {Harwood}, {Mingo}, {Mooney}, \& {Saxena}}]{sabater2019}
{Sabater}, J., {Best}, P.~N., {Hardcastle}, M.~J., {et~al.} 2019, \aap, 622,
  A17

\bibitem[{{Sadler} {et~al.}(2002){Sadler}, {Jackson}, {Cannon}, {McIntyre},
  {Murphy}, {Bland-Hawthorn}, {Bridges}, {Cole}, {Colless}, {Collins}, {Couch},
  {Dalton}, {De Propris}, {Driver}, {Efstathiou}, {Ellis}, {Frenk},
  {Glazebrook}, {Lahav}, {Lewis}, {Lumsden}, {Maddox}, {Madgwick}, {Norberg},
  {Peacock}, {Peterson}, {Sutherland}, \& {Taylor}}]{sadler2002}
{Sadler}, E.~M., {Jackson}, C.~A., {Cannon}, R.~D., {et~al.} 2002, \mnras, 329,
  227

\bibitem[{{Sargent} {et~al.}(2010){Sargent}, {Schinnerer}, {Murphy}, {Carilli},
  {Helou}, {Aussel}, {Le Floc'h}, {Frayer}, {Ilbert}, {Oesch}, {Salvato},
  {Smol{\v{c}}i{\'c}}, {Kartaltepe}, \& {Sanders}}]{sargent2010}
{Sargent}, M.~T., {Schinnerer}, E., {Murphy}, E., {et~al.} 2010, \apjl, 714,
  L190

\bibitem[{{Schreiber} {et~al.}(2015){Schreiber}, {Pannella}, {Elbaz},
  {B{\'e}thermin}, {Inami}, {Dickinson}, {Magnelli}, {Wang}, {Aussel}, {Daddi},
  {Juneau}, {Shu}, {Sargent}, {Buat}, {Faber}, {Ferguson}, {Giavalisco},
  {Koekemoer}, {Magdis}, {Morrison}, {Papovich}, {Santini}, \&
  {Scott}}]{schreiber2015}
{Schreiber}, C., {Pannella}, M., {Elbaz}, D., {et~al.} 2015, \aap, 575, A74

\bibitem[{{Shimwell} {et~al.}(2017){Shimwell}, {R{\"o}ttgering}, {Best},
  {Williams}, {Dijkema}, {de Gasperin}, {Hardcastle}, {Heald}, {Hoang},
  {Horneffer}, {Intema}, {Mahony}, {Mandal}, {Mechev}, {Morabito}, {Oonk},
  {Rafferty}, {Retana-Montenegro}, {Sabater}, {Tasse}, {van Weeren},
  {Br{\"u}ggen}, {Brunetti}, {Chy{\.z}y}, {Conway}, {Haverkorn}, {Jackson},
  {Jarvis}, {McKean}, {Miley}, {Morganti}, {White}, {Wise}, {van Bemmel},
  {Beck}, {Brienza}, {Bonafede}, {Calistro Rivera}, {Cassano}, {Clarke},
  {Cseh}, {Deller}, {Drabent}, {van Driel}, {Engels}, {Falcke}, {Ferrari},
  {Fr{\"o}hlich}, {Garrett}, {Harwood}, {Heesen}, {Hoeft}, {Horellou},
  {Israel}, {Kapi{\'n}ska}, {Kunert-Bajraszewska}, {McKay}, {Mohan},
  {Orr{\'u}}, {Pizzo}, {Prandoni}, {Schwarz}, {Shulevski}, {Sipior}, {Smith},
  {Sridhar}, {Steinmetz}, {Stroe}, {Varenius}, {van der Werf}, {Zensus}, \&
  {Zwart}}]{shimwell2017}
{Shimwell}, T.~W., {R{\"o}ttgering}, H.~J.~A., {Best}, P.~N., {et~al.} 2017,
  \aap, 598, A104

\bibitem[{{Shimwell} {et~al.}(2019){Shimwell}, {Tasse}, {Hardcastle}, {Mechev},
  {Williams}, {Best}, {R{\"o}ttgering}, {Callingham}, {Dijkema}, {de Gasperin},
  {Hoang}, {Hugo}, {Mirmont}, {Oonk}, {Prandoni}, {Rafferty}, {Sabater},
  {Smirnov}, {van Weeren}, {White}, {Atemkeng}, {Bester}, {Bonnassieux},
  {Br{\"u}ggen}, {Brunetti}, {Chy{\.z}y}, {Cochrane}, {Conway}, {Croston},
  {Danezi}, {Duncan}, {Haverkorn}, {Heald}, {Iacobelli}, {Intema}, {Jackson},
  {Jamrozy}, {Jarvis}, {Lakhoo}, {Mevius}, {Miley}, {Morabito}, {Morganti},
  {Nisbet}, {Orr{\'u}}, {Perkins}, {Pizzo}, {Schrijvers}, {Smith}, {Vermeulen},
  {Wise}, {Alegre}, {Bacon}, {van Bemmel}, {Beswick}, {Bonafede}, {Botteon},
  {Bourke}, {Brienza}, {Calistro Rivera}, {Cassano}, {Clarke}, {Conselice},
  {Dettmar}, {Drabent}, {Dumba}, {Emig}, {En{\ss}lin}, {Ferrari}, {Garrett},
  {G{\'e}nova-Santos}, {Goyal}, {G{\"u}rkan}, {Hale}, {Harwood}, {Heesen},
  {Hoeft}, {Horellou}, {Jackson}, {Kokotanekov}, {Kondapally},
  {Kunert-Bajraszewska}, {Mahatma}, {Mahony}, {Mandal}, {McKean}, {Merloni},
  {Mingo}, {Miskolczi}, {Mooney}, {Nikiel-Wroczy{\'n}ski}, {O'Sullivan},
  {Quinn}, {Reich}, {Roskowi{\'n}ski}, {Rowlinson}, {Savini}, {Saxena},
  {Schwarz}, {Shulevski}, {Sridhar}, {Stacey}, {Urquhart}, {van der Wiel},
  {Varenius}, {Webster}, \& {Wilber}}]{shimwell2019}
{Shimwell}, T.~W., {Tasse}, C., {Hardcastle}, M.~J., {et~al.} 2019, \aap, 622,
  A1

\bibitem[{{Shirley} {et~al.}(2019){Shirley}, {Roehlly}, {Hurley}, {Buat},
  {Campos Varillas}, {Duivenvoorden}, {Duncan}, {Efstathiou}, {Farrah},
  {Gonz{\'a}lez Solares}, {Malek}, {Marchetti}, {McCheyne}, {Papadopoulos},
  {Pons}, {Scipioni}, {Vaccari}, \& {Oliver}}]{shirley2019}
{Shirley}, R., {Roehlly}, Y., {Hurley}, P.~D., {et~al.} 2019, \mnras, 490, 634

\bibitem[{{Smith} {et~al.}(2016){Smith}, {Best}, {Duncan}, {Hatch}, {Jarvis},
  {R{\"o}ttgering}, {Simpson}, {Stott}, {Cochrane}, {Coppin}, {Dannerbauer},
  {Davis}, {Geach}, {Hale}, {Hardcastle}, {Hatfield}, {Houghton}, {Maddox},
  {McGee}, {Morabito}, {Nisbet}, {Pand ey-Pommier}, {Prandoni}, {Saxena},
  {Shimwell}, {Tarr}, {van Bemmel}, {Verma}, {White}, \&
  {Williams}}]{smith2016}
{Smith}, D.~J.~B., {Best}, P.~N., {Duncan}, K.~J., {et~al.} 2016, in SF2A-2016:
  Proceedings of the Annual meeting of the French Society of Astronomy and
  Astrophysics, ed. C.~{Reyl{\'e}}, J.~{Richard}, L.~{Cambr{\'e}sy},
  M.~{Deleuil}, E.~{P{\'e}contal}, L.~{Tresse}, \& I.~{Vauglin}, 271--280

\bibitem[{{Smith} {et~al.}(2012){Smith}, {Dunne}, {da Cunha}, {Rowlands},
  {Maddox}, {Gomez}, {Bonfield}, {Charlot}, {Driver}, {Popescu}, {Tuffs},
  {Dunlop}, {Jarvis}, {Seymour}, {Symeonidis}, {Baes}, {Bourne}, {Clements},
  {Cooray}, {De Zotti}, {Dye}, {Eales}, {Scott}, {Verma}, {van der Werf},
  {Andrae}, {Auld}, {Buttiglione}, {Cava}, {Dariush}, {Fritz}, {Hopwood},
  {Ibar}, {Ivison}, {Kelvin}, {Madore}, {Pohlen}, {Rigby}, {Robotham},
  {Seibert}, \& {Temi}}]{smith2012}
{Smith}, D.~J.~B., {Dunne}, L., {da Cunha}, E., {et~al.} 2012, \mnras, 427, 703

\bibitem[{{Smith} {et~al.}(2011){Smith}, {Dunne}, {Maddox}, {Eales},
  {Bonfield}, {Jarvis}, {Sutherland}, {Fleuren}, {Rigby}, {Thompson}, {Baldry},
  {Bamford}, {Buttiglione}, {Cava}, {Clements}, {Cooray}, {Croom}, {Dariush},
  {de Zotti}, {Driver}, {Dunlop}, {Fritz}, {Hill}, {Hopkins}, {Hopwood},
  {Ibar}, {Ivison}, {Jones}, {Kelvin}, {Leeuw}, {Liske}, {Loveday}, {Madore},
  {Norberg}, {Panuzzo}, {Pascale}, {Pohlen}, {Popescu}, {Prescott}, {Robotham},
  {Rodighiero}, {Scott}, {Seibert}, {Sharp}, {Temi}, {Tuffs}, {van der Werf},
  \& {van Kampen}}]{smith2011}
{Smith}, D.~J.~B., {Dunne}, L., {Maddox}, S.~J., {et~al.} 2011, \mnras, 416,
  857

\bibitem[{{Smith} \& {Hayward}(2018)}]{smith2018}
{Smith}, D. J.~B. \& {Hayward}, C.~C. 2018, \mnras, 476, 1705

\bibitem[{{Smith} {et~al.}(2014){Smith}, {Jarvis}, {Hardcastle}, {Vaccari},
  {Bourne}, {Dunne}, {Ibar}, {Maddox}, {Prescott}, {Vlahakis}, {Eales},
  {Maddox}, {Smith}, {Valiante}, \& {de Zotti}}]{smith2014}
{Smith}, D.~J.~B., {Jarvis}, M.~J., {Hardcastle}, M.~J., {et~al.} 2014, \mnras,
  445, 2232

\bibitem[{{Smol{\v{c}}i{\'c}} {et~al.}(2017){Smol{\v{c}}i{\'c}}, {Novak},
  {Bondi}, {Ciliegi}, {Mooley}, {Schinnerer}, {Zamorani}, {Navarrete},
  {Bourke}, {Karim}, {Vardoulaki}, {Leslie}, {Delhaize}, {Carilli}, {Myers},
  {Baran}, {Delvecchio}, {Miettinen}, {Banfield}, {Balokovi{\'c}}, {Bertoldi},
  {Capak}, {Frail}, {Hallinan}, {Hao}, {Herrera Ruiz}, {Horesh}, {Ilbert},
  {Intema}, {Jeli{\'c}}, {Kl{\"o}ckner}, {Krpan}, {Kulkarni}, {McCracken},
  {Laigle}, {Middleberg}, {Murphy}, {Sargent}, {Scoville}, \&
  {Sheth}}]{smolcic2017}
{Smol{\v{c}}i{\'c}}, V., {Novak}, M., {Bondi}, M., {et~al.} 2017, \aap, 602, A1

\bibitem[{{Sudoh} {et~al.}(2020){Sudoh}, {Linden}, \& {Beacom}}]{sudoh2020}
{Sudoh}, T., {Linden}, T., \& {Beacom}, J.~F. 2020, arXiv e-prints,
  arXiv:2005.08982

\bibitem[{{Sutherland} \& {Saunders}(1992)}]{sutherland1992}
{Sutherland}, W. \& {Saunders}, W. 1992, \mnras, 259, 413

\bibitem[{{Tabatabaei} {et~al.}(2017){Tabatabaei}, {Schinnerer}, {Krause},
  {Dumas}, {Meidt}, {Damas-Segovia}, {Beck}, {Murphy}, {Mulcahy}, {Groves},
  {Bolatto}, {Dale}, {Galametz}, {Sandstrom}, {Boquien}, {Calzetti},
  {Kennicutt}, {Hunt}, {De Looze}, \& {Pellegrini}}]{tabatabaei2017}
{Tabatabaei}, F.~S., {Schinnerer}, E., {Krause}, M., {et~al.} 2017, \apj, 836,
  185

\bibitem[{{Tacchella} {et~al.}(2016){Tacchella}, {Dekel}, {Carollo},
  {Ceverino}, {DeGraf}, {Lapiner}, {Mand elker}, \& {Primack
  Joel}}]{tacchella2016}
{Tacchella}, S., {Dekel}, A., {Carollo}, C.~M., {et~al.} 2016, \mnras, 457,
  2790

\bibitem[{{Tasse} {et~al.}(2020){Tasse}, {Shimwell}, {Hardcastle}, {Authors},
  {Authors}, \& {Authors}}]{tasse_lotss}
{Tasse}, C., {Shimwell}, T., {Hardcastle}, M., {et~al.} 2020, \aap, Submitted
  (LoTSS SI)

\bibitem[{{Upjohn} {et~al.}(2019){Upjohn}, {Brown}, {Hopkins}, \&
  {Bonne}}]{upjohn2019}
{Upjohn}, J.~E., {Brown}, M. J.~I., {Hopkins}, A.~M., \& {Bonne}, N.~J. 2019,
  \pasa, 36, e012

\bibitem[{{van der Kruit}(1971)}]{vanderkruit1971}
{van der Kruit}, P.~C. 1971, \aap, 15, 110

\bibitem[{{van Haarlem} {et~al.}(2013){van Haarlem}, {Wise}, {Gunst}, {Heald},
  {McKean}, {Hessels}, {de Bruyn}, {Nijboer}, {Swinbank}, {Fallows},
  {Brentjens}, {Nelles}, {Beck}, {Falcke}, {Fender}, {H{\"o}randel},
  {Koopmans}, {Mann}, {Miley}, {R{\"o}ttgering}, {Stappers}, {Wijers},
  {Zaroubi}, {van den Akker}, {Alexov}, {Anderson}, {Anderson}, {van Ardenne},
  {Arts}, {Asgekar}, {Avruch}, {Batejat}, {B{\"a}hren}, {Bell}, {Bell}, {van
  Bemmel}, {Bennema}, {Bentum}, {Bernardi}, {Best}, {B{\^\i}rzan}, {Bonafede},
  {Boonstra}, {Braun}, {Bregman}, {Breitling}, {van de Brink}, {Broderick},
  {Broekema}, {Brouw}, {Br{\"u}ggen}, {Butcher}, {van Cappellen}, {Ciardi},
  {Coenen}, {Conway}, {Coolen}, {Corstanje}, {Damstra}, {Davies}, {Deller},
  {Dettmar}, {van Diepen}, {Dijkstra}, {Donker}, {Doorduin}, {Dromer}, {Drost},
  {van Duin}, {Eisl{\"o}ffel}, {van Enst}, {Ferrari}, {Frieswijk}, {Gankema},
  {Garrett}, {de Gasperin}, {Gerbers}, {de Geus}, {Grie{\ss}meier}, {Grit},
  {Gruppen}, {Hamaker}, {Hassall}, {Hoeft}, {Holties}, {Horneffer}, {van der
  Horst}, {van Houwelingen}, {Huijgen}, {Iacobelli}, {Intema}, {Jackson},
  {Jelic}, {de Jong}, {Juette}, {Kant}, {Karastergiou}, {Koers}, {Kollen},
  {Kondratiev}, {Kooistra}, {Koopman}, {Koster}, {Kuniyoshi}, {Kramer},
  {Kuper}, {Lambropoulos}, {Law}, {van Leeuwen}, {Lemaitre}, {Loose}, {Maat},
  {Macario}, {Markoff}, {Masters}, {McFadden}, {McKay-Bukowski}, {Meijering},
  {Meulman}, {Mevius}, {Middelberg}, {Millenaar}, {Miller-Jones}, {Mohan},
  {Mol}, {Morawietz}, {Morganti}, {Mulcahy}, {Mulder}, {Munk}, {Nieuwenhuis},
  {van Nieuwpoort}, {Noordam}, {Norden}, {Noutsos}, {Offringa}, {Olofsson},
  {Omar}, {Orr{\'u}}, {Overeem}, {Paas}, {Pand ey-Pommier}, {Pandey}, {Pizzo},
  {Polatidis}, {Rafferty}, {Rawlings}, {Reich}, {de Reijer}, {Reitsma},
  {Renting}, {Riemers}, {Rol}, {Romein}, {Roosjen}, {Ruiter}, {Scaife}, {van
  der Schaaf}, {Scheers}, {Schellart}, {Schoenmakers}, {Schoonderbeek},
  {Serylak}, {Shulevski}, {Sluman}, {Smirnov}, {Sobey}, {Spreeuw}, {Steinmetz},
  {Sterks}, {Stiepel}, {Stuurwold}, {Tagger}, {Tang}, {Tasse}, {Thomas},
  {Thoudam}, {Toribio}, {van der Tol}, {Usov}, {van Veelen}, {van der Veen},
  {ter Veen}, {Verbiest}, {Vermeulen}, {Vermaas}, {Vocks}, {Vogt}, {de Vos},
  {van der Wal}, {van Weeren}, {Weggemans}, {Weltevrede}, {White}, {Wijnholds},
  {Wilhelmsson}, {Wucknitz}, {Yatawatta}, {Zarka}, {Zensus}, \& {van
  Zwieten}}]{vanhaarlem2013}
{van Haarlem}, M.~P., {Wise}, M.~W., {Gunst}, A.~W., {et~al.} 2013, \aap, 556,
  A2

\bibitem[{{Wang} {et~al.}(2019){Wang}, {Gao}, {Duncan}, {Williams},
  {Rowan-Robinson}, {Sabater}, {Shimwell}, {Bonato}, {Calistro-Rivera},
  {Chy{\.z}y}, {Farrah}, {G{\"u}rkan}, {Hardcastle}, {McCheyne}, {Prandoni},
  {Read}, {R{\"o}ttgering}, \& {Smith}}]{wang2019}
{Wang}, L., {Gao}, F., {Duncan}, K.~J., {et~al.} 2019, \aap, 631, A109

\bibitem[{{Wang} {et~al.}(2014){Wang}, {Rowan-Robinson}, {Norberg}, {Heinis},
  \& {Han}}]{wang2014}
{Wang}, L., {Rowan-Robinson}, M., {Norberg}, P., {Heinis}, S., \& {Han}, J.
  2014, \mnras, 442, 2739

\bibitem[{{Williams} {et~al.}(2019){Williams}, {Hardcastle}, {Best}, {Sabater},
  {Croston}, {Duncan}, {Shimwell}, {R{\"o}ttgering}, {Nisbet}, {G{\"u}rkan},
  {Alegre}, {Cochrane}, {Goyal}, {Hale}, {Jackson}, {Jamrozy}, {Kondapally},
  {Kunert-Bajraszewska}, {Mahatma}, {Mingo}, {Morabito}, {Prandoni},
  {Roskowinski}, {Shulevski}, {Smith}, {Tasse}, {Urquhart}, {Webster}, {White},
  {Beswick}, {Callingham}, {Chy{\.z}y}, {de Gasperin}, {Harwood}, {Hoeft},
  {Iacobelli}, {McKean}, {Mechev}, {Miley}, {Schwarz}, \& {van
  Weeren}}]{williams2019}
{Williams}, W.~L., {Hardcastle}, M.~J., {Best}, P.~N., {et~al.} 2019, \aap,
  622, A2

\bibitem[{{Wilson} {et~al.}(2009){Wilson}, {Muzzin}, {Yee}, {Lacy}, {Surace},
  {Gilbank}, {Blindert}, {Hoekstra}, {Majumdar}, {Demarco}, {Gardner},
  {Gladders}, \& {Lonsdale}}]{wilson2009}
{Wilson}, G., {Muzzin}, A., {Yee}, H.~K.~C., {et~al.} 2009, \apj, 698, 1943

\bibitem[{{York} {et~al.}(2000){York}, {Adelman}, {Anderson}, {Anderson},
  {Annis}, {Bahcall}, {Bakken}, {Barkhouser}, {Bastian}, {Berman}, {Boroski},
  {Bracker}, {Briegel}, {Briggs}, {Brinkmann}, {Brunner}, {Burles}, {Carey},
  {Carr}, {Castander}, {Chen}, {Colestock}, {Connolly}, {Crocker}, {Csabai},
  {Czarapata}, {Davis}, {Doi}, {Dombeck}, {Eisenstein}, {Ellman}, {Elms},
  {Evans}, {Fan}, {Federwitz}, {Fiscelli}, {Friedman}, {Frieman}, {Fukugita},
  {Gillespie}, {Gunn}, {Gurbani}, {de Haas}, {Haldeman}, {Harris}, {Hayes},
  {Heckman}, {Hennessy}, {Hindsley}, {Holm}, {Holmgren}, {Huang}, {Hull},
  {Husby}, {Ichikawa}, {Ichikawa}, {Ivezi{\'c}}, {Kent}, {Kim}, {Kinney},
  {Klaene}, {Kleinman}, {Kleinman}, {Knapp}, {Korienek}, {Kron}, {Kunszt},
  {Lamb}, {Lee}, {Leger}, {Limmongkol}, {Lindenmeyer}, {Long}, {Loomis},
  {Loveday}, {Lucinio}, {Lupton}, {MacKinnon}, {Mannery}, {Mantsch}, {Margon},
  {McGehee}, {McKay}, {Meiksin}, {Merelli}, {Monet}, {Munn}, {Narayanan},
  {Nash}, {Neilsen}, {Neswold}, {Newberg}, {Nichol}, {Nicinski}, {Nonino},
  {Okada}, {Okamura}, {Ostriker}, {Owen}, {Pauls}, {Peoples}, {Peterson},
  {Petravick}, {Pier}, {Pope}, {Pordes}, {Prosapio}, {Rechenmacher}, {Quinn},
  {Richards}, {Richmond}, {Rivetta}, {Rockosi}, {Ruthmansdorfer}, {Sand ford},
  {Schlegel}, {Schneider}, {Sekiguchi}, {Sergey}, {Shimasaku}, {Siegmund},
  {Smee}, {Smith}, {Snedden}, {Stone}, {Stoughton}, {Strauss}, {Stubbs},
  {SubbaRao}, {Szalay}, {Szapudi}, {Szokoly}, {Thakar}, {Tremonti}, {Tucker},
  {Uomoto}, {Vanden Berk}, {Vogeley}, {Waddell}, {Wang}, {Watanabe},
  {Weinberg}, {Yanny}, {Yasuda}, \& {SDSS Collaboration}}]{york2000}
{York}, D.~G., {Adelman}, J., {Anderson}, John~E., J., {et~al.} 2000, \aj, 120,
  1579

\bibitem[{{Yun} {et~al.}(2001){Yun}, {Reddy}, \& {Condon}}]{yun2001}
{Yun}, M.~S., {Reddy}, N.~A., \& {Condon}, J.~J. 2001, \apj, 554, 803

\bibitem[{{Zwart} {et~al.}(2014){Zwart}, {Jarvis}, {Deane}, {Bonfield},
  {Knowles}, {Madhanpall}, {Rahmani}, \& {Smith}}]{zwart2014}
{Zwart}, J. T.~L., {Jarvis}, M.~J., {Deane}, R.~P., {et~al.} 2014, \mnras, 439,
  1459

\bibitem[{{Zwart} {et~al.}(2015){Zwart}, {Santos}, \& {Jarvis}}]{zwart2015}
{Zwart}, J. T.~L., {Santos}, M., \& {Jarvis}, M.~J. 2015, \mnras, 453, 1740

\end{thebibliography}

\begin{appendix}

\section{Generating stacked PDFs by sampling}
\label{sec:sampling_pdfs}

To illustrate the way that we create stacked two-dimensional PDFs using random sampling, we have included the following simple example based on two model parameters of interest. We assume that both parameters have asymmetric uncertainties for different reasons. In the case of our first parameter (hereafter ``parameter 1"), we assume uncertainties which are normally distributed in linear space, but we wish to stack the PDF in log space (as in the case for \Lradio\ in our real data set). For our second parameter (``parameter 2") we proceed as for our \magphys\ estimates of stellar mass or SFR since  the \magphys\ PDFs do not in general have an analytic form. We therefore make the simplifying assumption that the underlying PDF is Gaussian distributed, and set the width of the distribution on the positive (negative) side using the difference between the 84th (16th) and 50th percentiles of the \magphys\ PDF. 

\begin{figure}
   \centering
	\subfigure[PDF sampling]{\includegraphics[height=0.65\columnwidth]{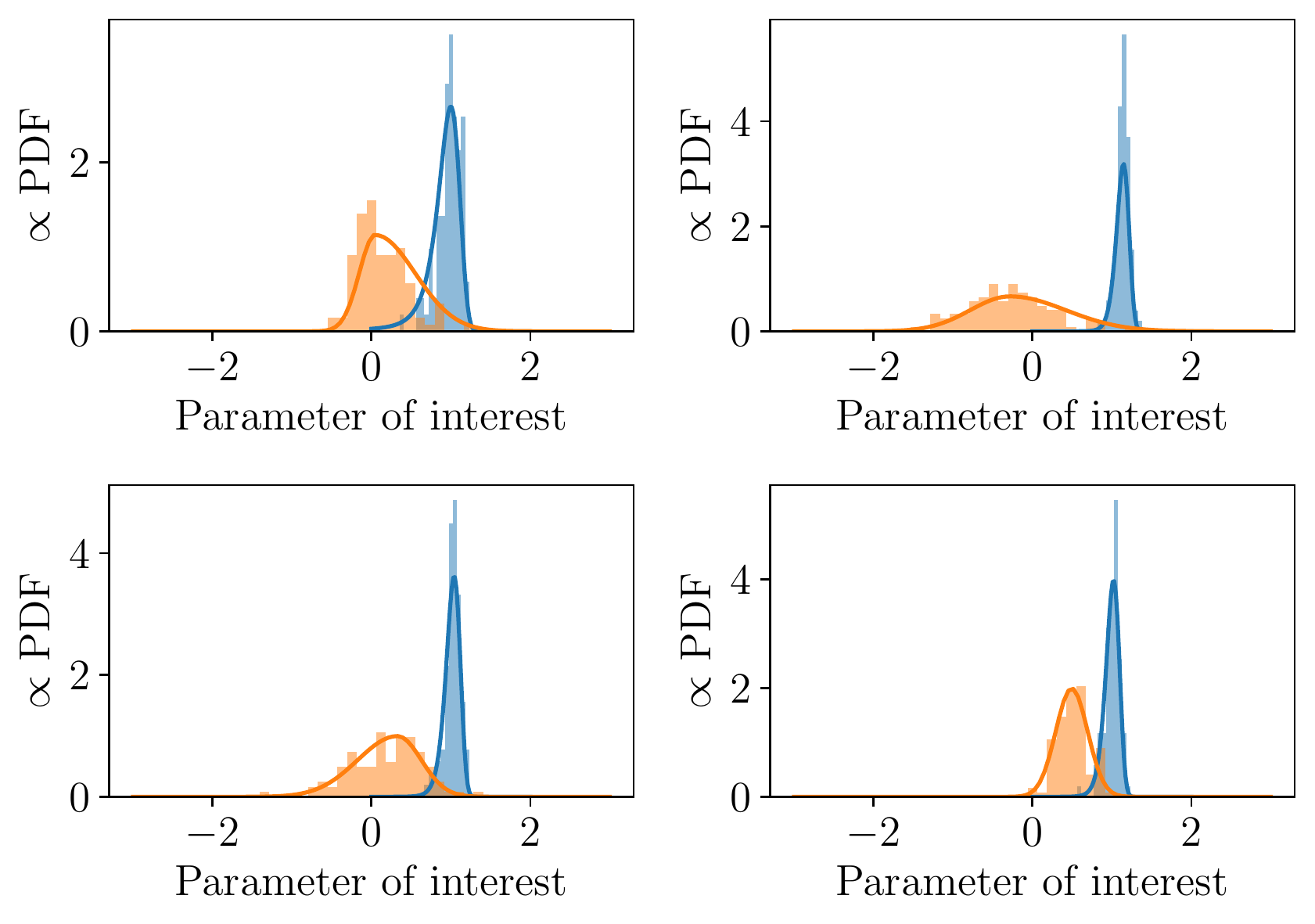}}
	\subfigure[Two-dimensional PDFs]{\includegraphics[height=0.65\columnwidth]{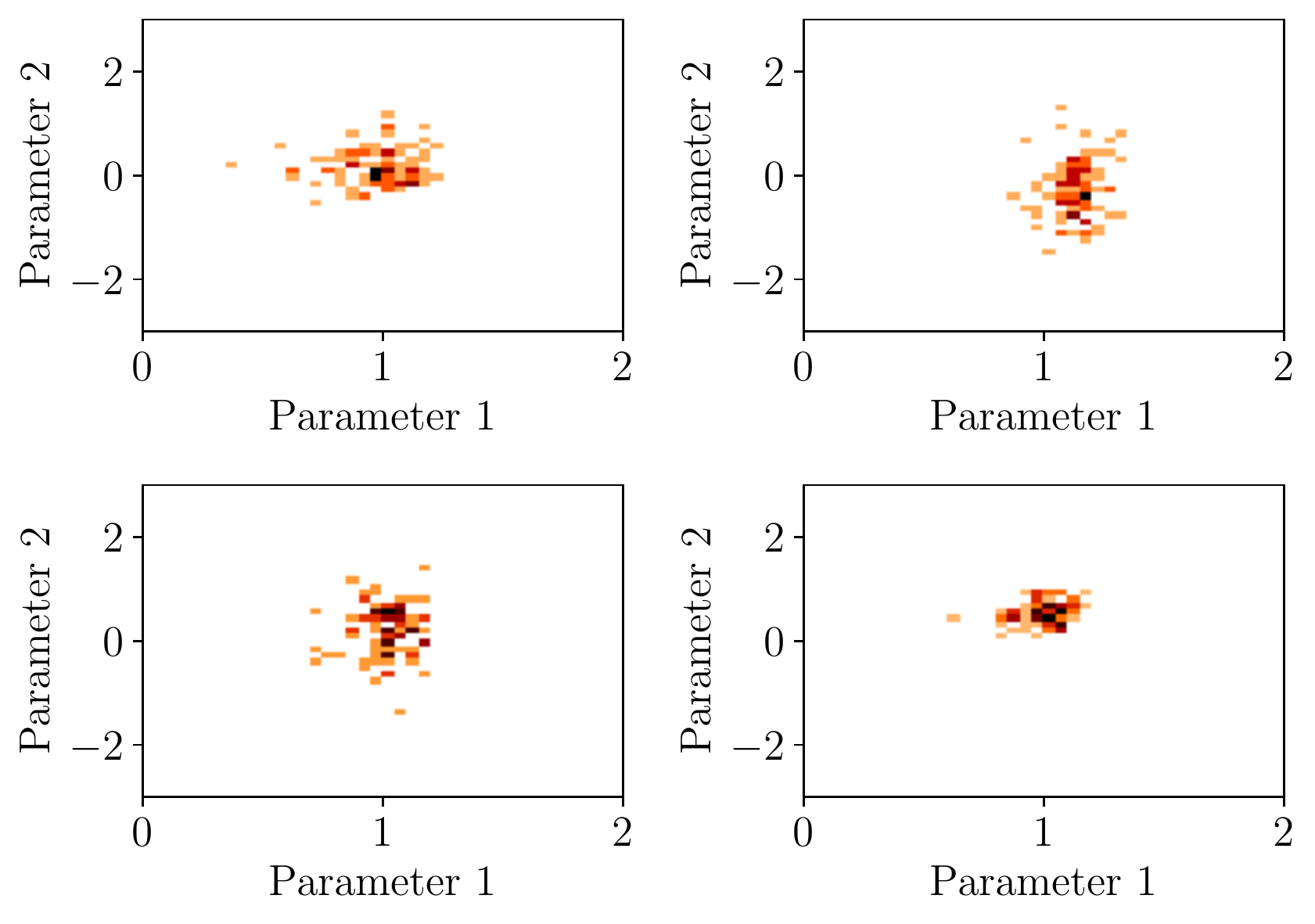}}
	\subfigure[Stacked PDF]{\includegraphics[height=0.65\columnwidth]{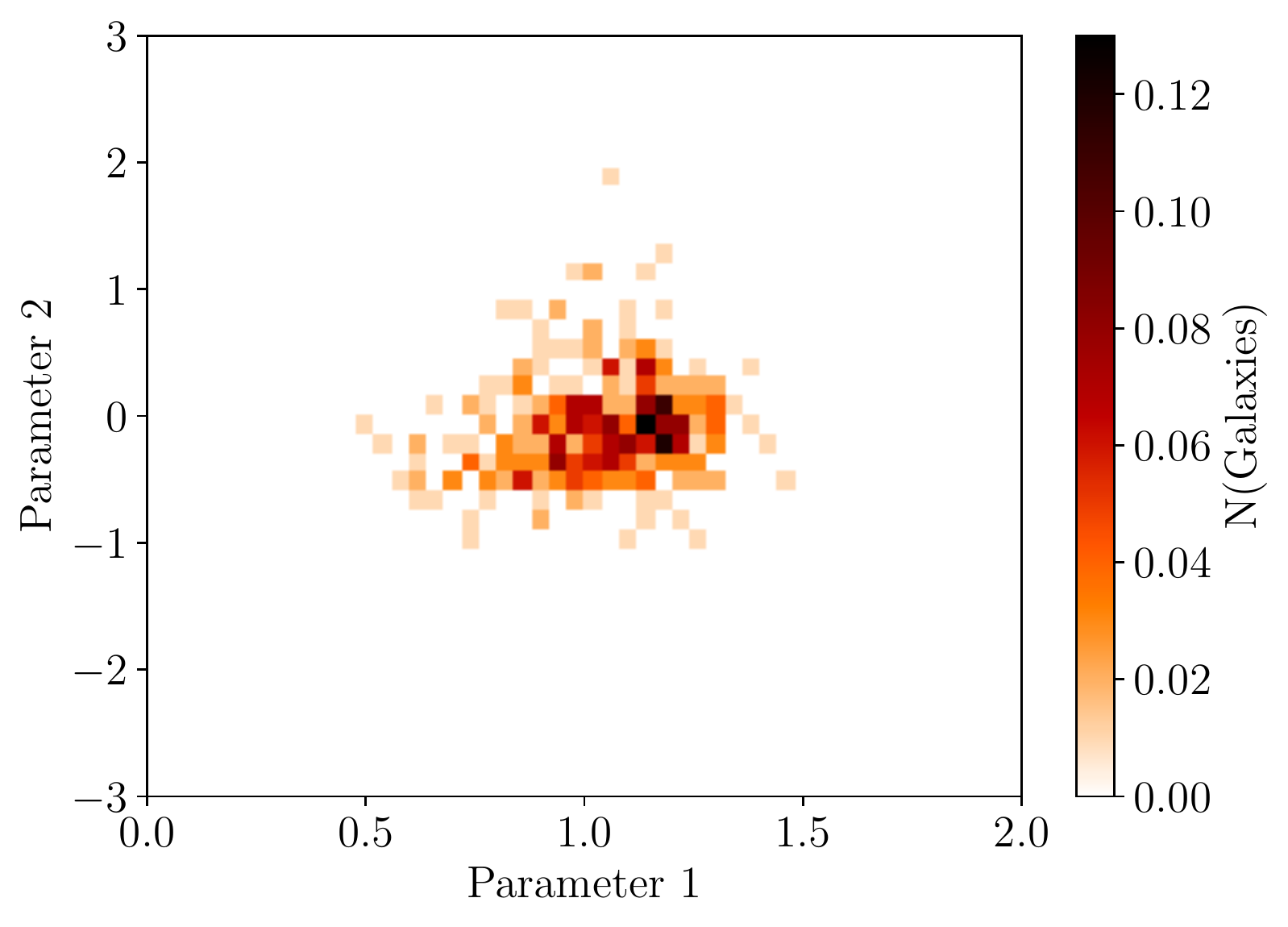}}
	\caption{Constructing PDFs for individual galaxies. In the top four panels we show the analytic PDFs assumed for parameter 1 (blue) and parameter 2 (orange). We then create 100 random samples drawn from each distribution and these results for each parameter are shown as the shaded histograms of the corresponding colours. In the middle panels, we show the two-dimensional PDFs for each hypothetical `galaxy', derived by assuming that the samples shown in the top panels for each parameter are independent. The bottom panel shows the stack of the individual two-dimensional PDFs for the four model galaxies and two indicative parameters shown in the above panels. The colour bar indicates the number of model galaxies that we expect in each bin - and the total obtained by summing all of the pixel values equals the total number of galaxies in the stack. Clearly, individual pixels can take non-integer values since we have sampled each galaxy 100 times. }
	\label{fig:sampling_pdfs}
\end{figure}

The four panels of figure \ref{fig:sampling_pdfs}\,(a) show the PDFs assumed for parameters 1 \&\ 2 in blue and orange, respectively, for some arbitrarily-chosen values, and for the hypothetical case of four galaxies that we wish to stack. We then generate 100 samples from the assumed PDF for each parameter, for each galaxy - histograms of these samples have been overlaid in figure \ref{fig:sampling_pdfs}\,(a). Since we assume that the uncertainties are independent, we can use these samples to create a two-dimensional PDF for each object, by creating a two-dimensional histogram using the same samples, and normalising. Examples of these 2D PDFs for the four hypothetical objects are displayed in figure \ref{fig:sampling_pdfs}\,(b). 
	
To study the relation between the two parameters for the full population (in this case, the hypothetical population of four galaxies, but in section \ref{sec:sfrl150} we use around 120,000 galaxies), we can then stack the PDFs by summing up the values of the individual two-dimensional PDFs in each bin. The results for our hypothetical data set are shown in figure \ref{fig:sampling_pdfs} (c) - with increasingly large numbers of galaxies, these PDFs become increasingly smooth, to the point that (as in figure \ref{fig:sfrl150}) the distribution appears continuous despite the individual galaxies being sampled only 100 times. Each pixel in the stack shows how many galaxies we would expect to find in that bin - and since we have sampled each galaxy multiple times and renormalised (to retain the correct total number of galaxies) these need not be integers, as shown in figure \ref{fig:sampling_pdfs}\,(c).

\section{Studying SFR-L150 with a 150\,MHz selected sample}
\label{sec:detections}

Many previous works have investigated SFR-radio luminosity relation using a sample identified at radio frequencies \citep[e.g.][]{bell2003,murphy2011,brown2017,calistro2017,wang2019}. To see the possible impact of this on our results for SFR-\Lradio, we have repeated our analysis, but instead including only those sources which are detected at 150\,MHz with $\ge 5\sigma$ significance. 

\begin{figure}
   \centering
	\includegraphics[width=1.01\columnwidth]{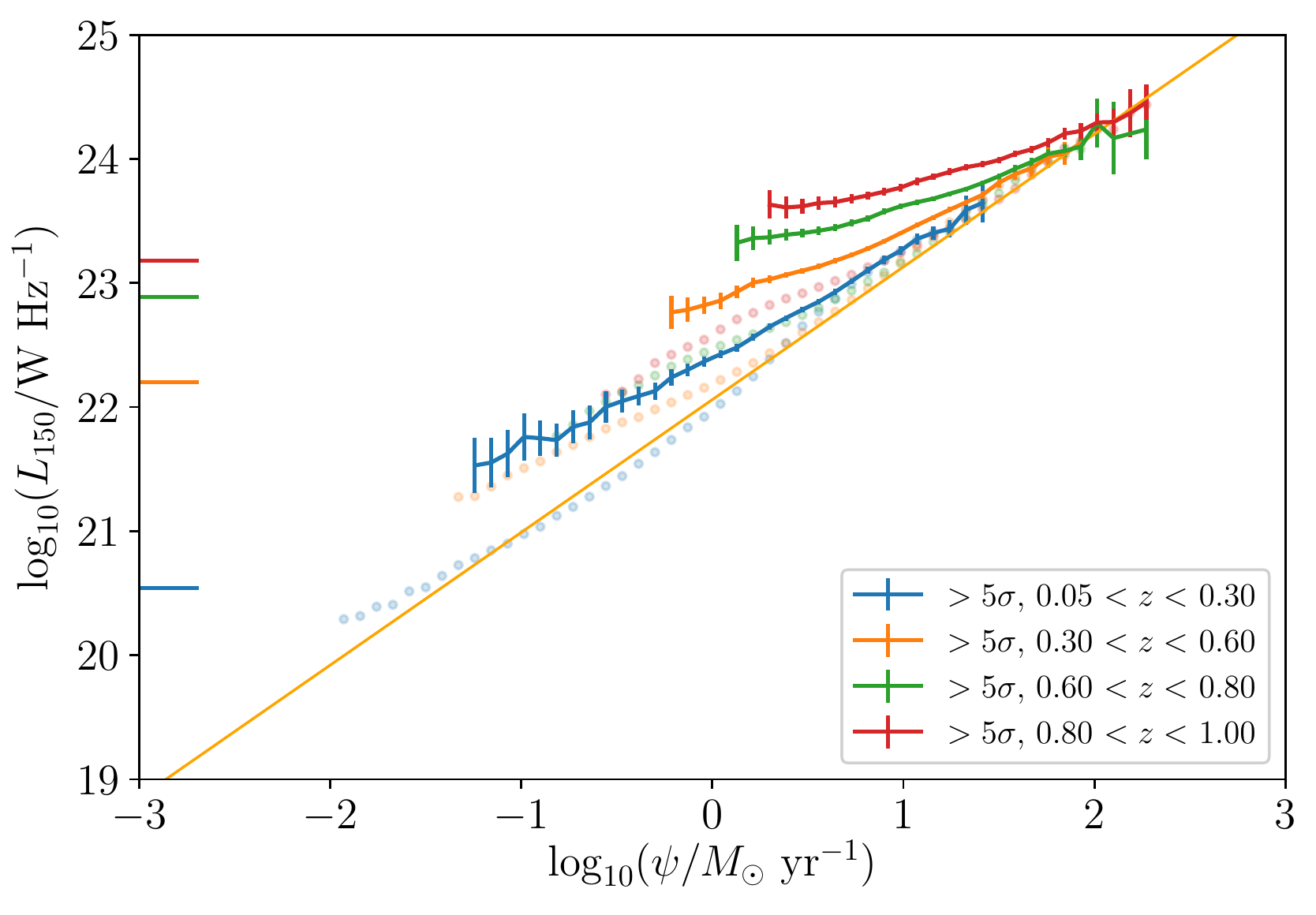}
      \caption{SFR-\Lradio\ plane, as in figure \ref{fig:sfrl150}. The coloured lines with error bars indicate the SFR-\Lradio\ relation recovered in each of four redshift bins detailed in the legend -- identical to those used in figure \ref{fig:slope_evolution}, and which are also reproduced here as the circles -- but including only those sources with $\ge 5\sigma$ 150\,MHz detections. Also overlaid is the best-fit relation from G18, and the horizontal bars adjacent to the left-hand vertical axis indicate the luminosity corresponding to 60\,$\mu$Jy at the lower bound of each redshift bin. }
         \label{fig:sfrl150_detections}
\end{figure}

Figure \ref{fig:sfrl150_detections} reveals that making this kind of selection gives results that are biased to higher L150 at a given SFR (relative to both our IRAC-selected sample shown in figure \ref{fig:slope_evolution}, and to the best-fit relation from G18 which is shown as the orange solid line). Dividing such a sample into four redshift bins also reveals apparent evolution in SFR-\Lradio\ -- in the sense of an apparent increase in \Lradio\ at a given SFR at higher redshift -- which is not recovered when the full IRAC-selected sample is considered. 

We conclude that great care is required when interpreting the results of this type of study based on radio-frequency selected samples, even when using the most sensitive radio data  in existence, such as the data used in this work from the LoTSS ELAIS-N1 deep field. 

\section{Supporting Simulations}
\label{sec:methodtests}

In order to test that our method of determining the median-likelihood SFR is able to recover the true SFR-\Lradio\ relation, we conduct a set of simulations, each based on sampling 120,000 model galaxies with a range of SFRs, stellar masses and redshifts sampled from among our real data. We assign each galaxy a true \Lradio\ value using an arbitrary mass-dependent SFR-\Lradio\ relation following equation \ref{eq:sfmass}, and simulate Gaussian scatter about that relation assuming a standard width (in dex). Using equation \ref{eq:sfmass} we are also able to simulate a stellar-mass independent SFR-\Lradio\ relation by setting $\gamma = 0$. 

We then generate a set of mock observed data by converting the noiseless values of \Lradio\ to true flux densities at 150\,MHz, before adding on Gaussian noise using a random number generator multiplied by the flux density uncertainty measured for the real sources in our LoTSS catalogue. We also simulate measuring the SFRs (and stellar masses if required) by resampling the ``true" values to give model observed values with uncertainties sampled in the same way from our real \magphys\ results. Our simulations account for AGN contamination, by randomly assigning galaxies an excess radio luminosity drawn from a log-normal distribution with a mean of 23.86 and a standard deviation of 0.91, based on the mean and standard deviation of the \Lradio\ of the flagged AGN in the Best et al. sample. The probability that a model galaxy is given such an excess luminosity is based on a mass-dependent probability using the luminosity-averaged results from \citet{sabater2019} shown in table \ref{tab:sabater_pagn}.

\begin{table}[]
\caption{The probability that a galaxy of a given stellar mass is assigned a radio-lumionsity excess in our simulations, based on the luminosity-averaged results of \citet{sabater2019}.}
\centering
\begin{tabular}{ll}
Mass range & P(AGN) \\
\hline
$\log_{10} (M/M_\odot) < 10.00$ & 0.00 \\
$10.00 < \log_{10} (M/M_\odot) < 10.50$ & 0.01 \\
$10.50 < \log_{10} (M/M_\odot) < 10.75$ & 0.02 \\
$10.75 < \log_{10} (M/M_\odot) < 11.00$ & 0.03 \\
$11.00 < \log_{10} (M/M_\odot) < 11.25$ & 0.04 \\
$11.25 < \log_{10} (M/M_\odot) < 11.50$ & 0.10 \\
$11.50 < \log_{10} (M/M_\odot) < 12.00$ & 0.30 
\end{tabular}
\label{tab:sabater_pagn}
\end{table}

\begin{figure}
   \centering
	\includegraphics[width=1.01\columnwidth]{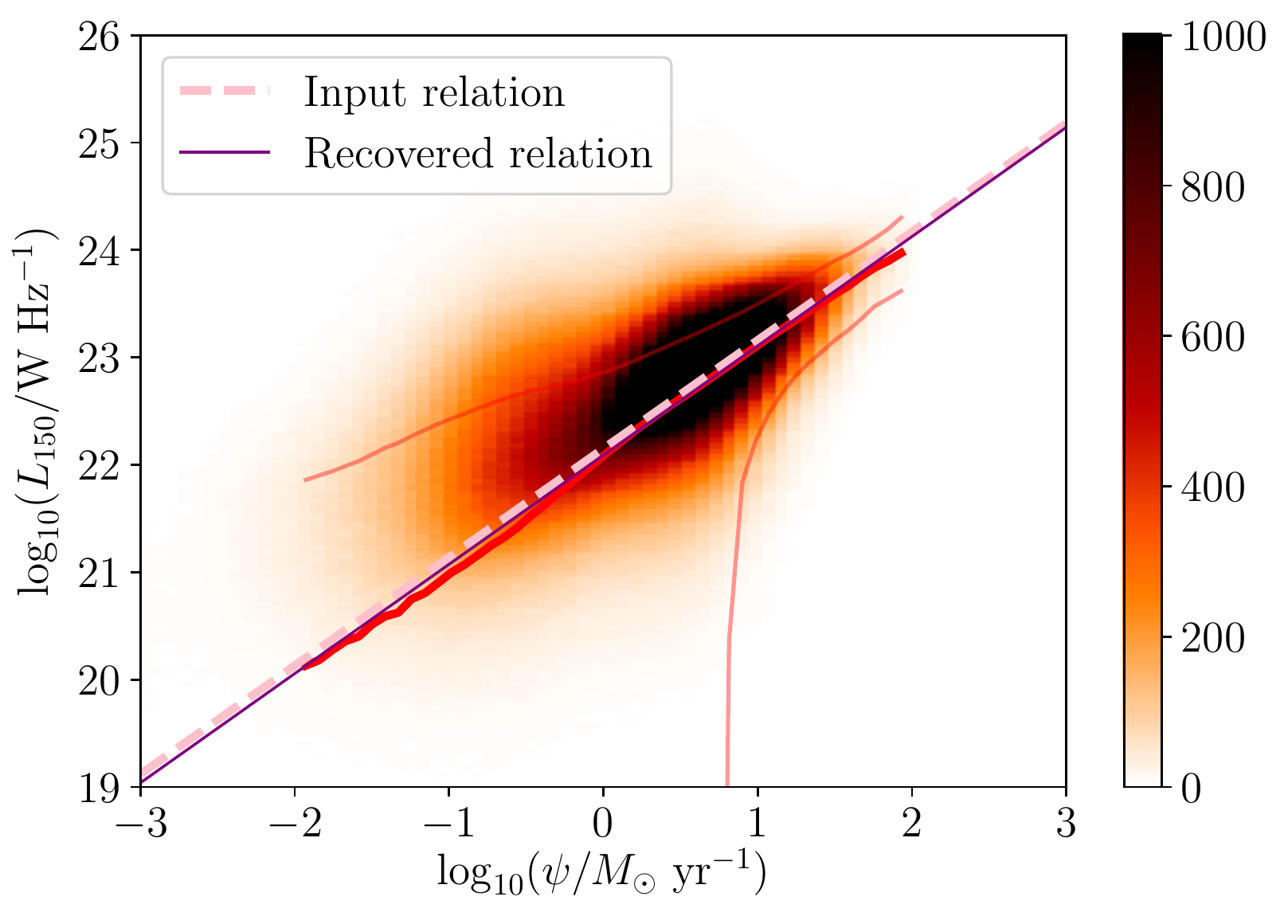}
      \caption{Heatmap showing the stacked two-dimensional PDF derived based on 120,000 galaxies with a redshift distribution sampled from our real data set, and accounting for the uncertainties in both SFR and \Lradio\ as in Figure \ref{fig:sfrl150}. The median likelihood estimate of the SFR-\Lradio\ relation is shown as the thick red line, while the recovered relation (in purple) is very close to the true SFR-\Lradio\ relation (dashed pink line). The colour bar to the right shows the effective number of galaxies in each bin. }
         \label{fig:simtest_heatmap}
\end{figure}

We then attempt to recover the known relation using the method discussed in section \ref{sec:sfrl150}. Figure \ref{fig:simtest_heatmap} shows an example visualisation of the two-dimensional PDF obtained from one of the simulations, assuming an SFR-\Lradio\ relation of the mass-independent form given in equation \ref{eq:sfrl150} with $\beta = 1.01$, $\log_{10} L_1 = 22.15$ (shown as the dashed pink line) and with scatter $\sigma_L = 0.25$\,dex. The thick red line shows the median likelihood estimate of the \Lradio\ in bins of SFR, while the purple line shows the best-fit to the data over the range $-1 < \psi < 1$, the same range used for the real data. It is clear that, just as in the real data, the median-likelihood estimate (thick red line) appears to be offset to lower SFRs than the apparent peak in the PDF, and suggests that our explanation for this effect in section \ref{sec:sfrl150} is plausible.

To better quantify the level of agreement between the input and output parameters in plausible realistic circumstances, we conducted 1000 Monte-Carlo simulations of 120,000 model sources, assuming a range of ``true" values, $0.5 \le \beta \le 1.0$, $21.5 \le \log_{10} L_C \le 22.5$ and $0.30 < \gamma < 0.60$ in ten equal steps each, and assuming a fixed scatter of $\sigma_L = 0.25$. 

\begin{figure*}
   \centering
	\includegraphics[width=0.667\textwidth]{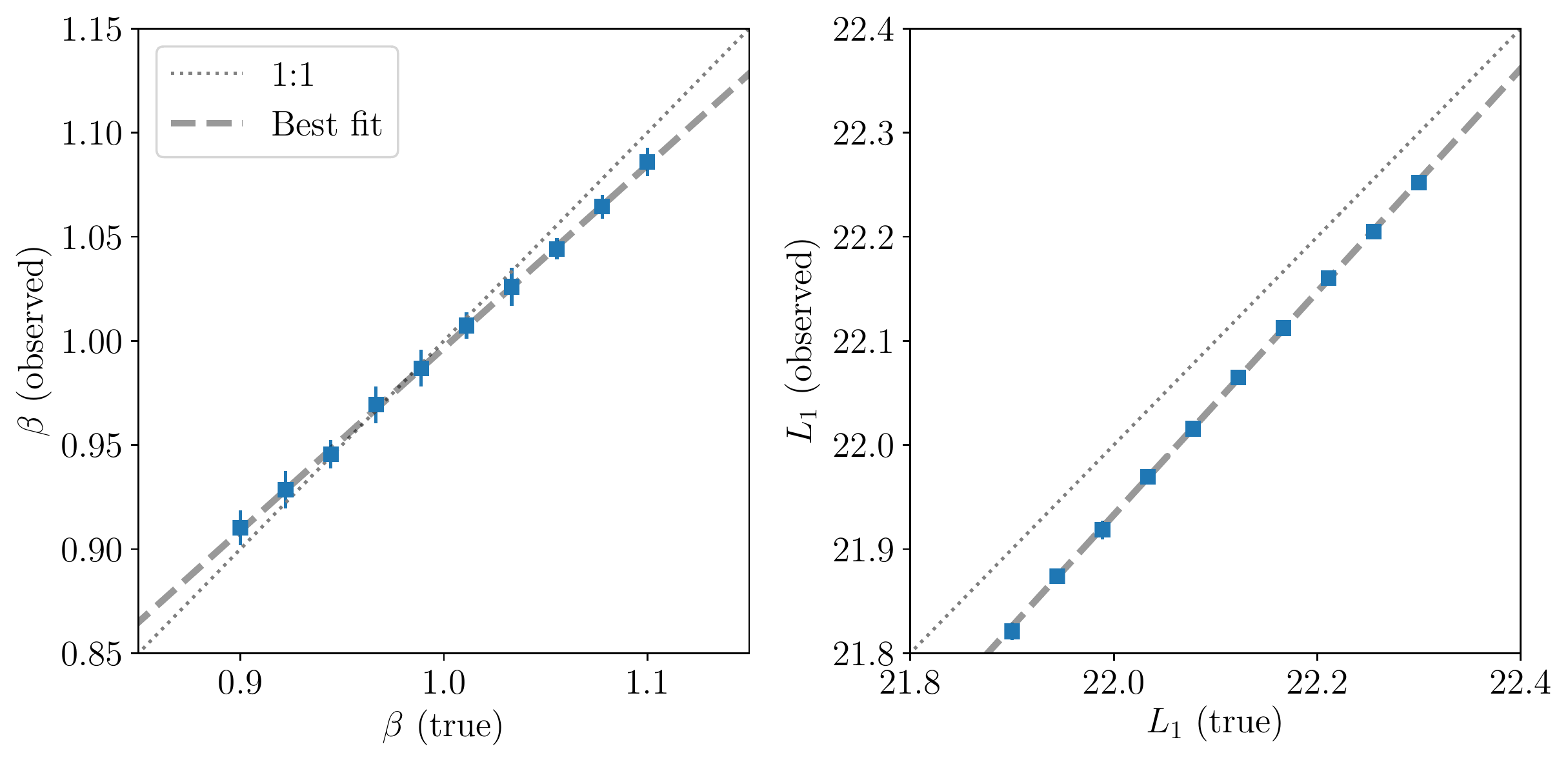}
      \caption{Simulations showing the ``observed" values of $\beta$ and $L_1$ obtained using the method described in section \ref{sec:sfrl150}, as a function of the known input values, assuming a fixed width of $\sigma_L = 0.25$ and mass-dependent AGN contamination following \citet{sabater2019}. The best-fit relations between the two sets of values -- shown as the dashed grey line -- are given in the text. The 1:1 line is shown as the dotted line. }
         \label{fig:simtest_inout}
\end{figure*}

\begin{figure*}
   \centering
	\includegraphics[width=\textwidth]{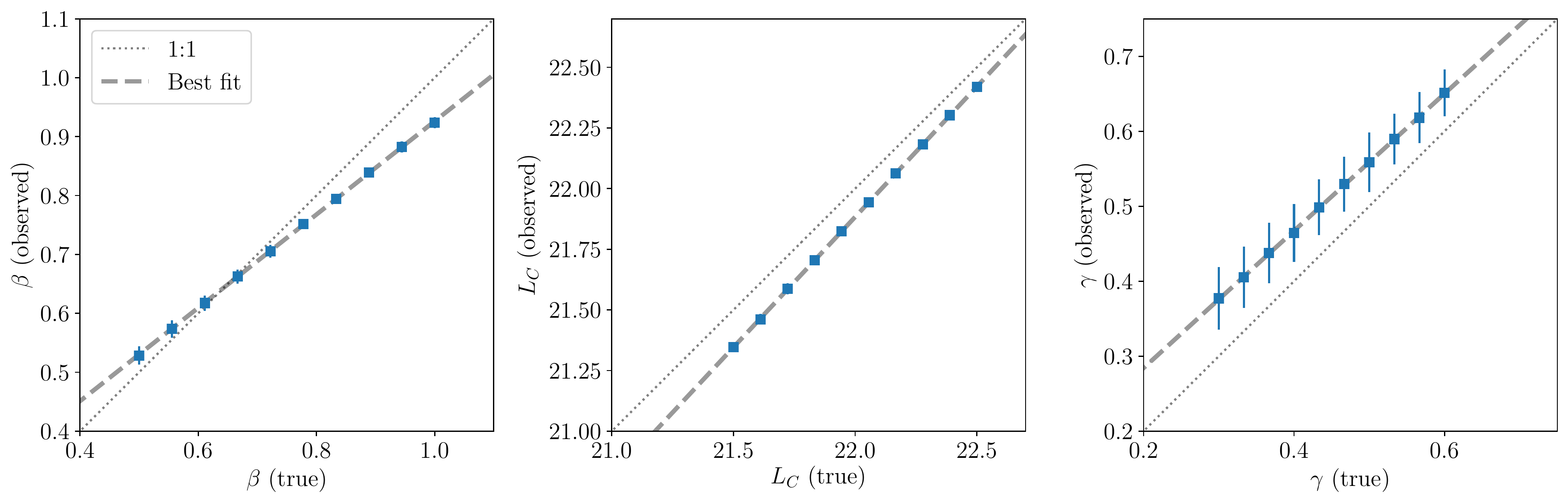}
      \caption{Simulations showing the ``observed" values of $\beta$ and $L_1$ obtained using the method described in section \ref{sec:sfrl150}, as a function of the known input values, assuming a fixed width of $\sigma_L = 0.25$ and AGN contamination following \citet{sabater2019}. The best-fit relations between the two sets of values are shown as the dashed grey lines, and parameterised as in the text. The dotted line in each panel shows the 1:1 relation.}
         \label{fig:simtest_inout_3d}
\end{figure*}

We conducted the simulations twice - once simulating the recovery of the true input parameters for the mass-independent SFR-\Lradio\ (results displayed in Figure \ref{fig:simtest_inout}) and once simulating the recovery of the mass-dependent version (results displayed in Figure \ref{fig:simtest_inout_3d}). In Figure \ref{fig:simtest_inout} the left panel shows the recovery of $\beta$, while the right panel shows how well $L_1$ is recovered. The 1:1 relation is indicated by the dotted line, while the best-fit relation between the input and ``observed'' values is shown as the dashed grey line. The best fit relationships between the input and ``observed'' values are: 

\begin{align}
\beta^\mathrm{true} &= 1.152\,\beta^\mathrm{obs} - 0.142, \\
\log_{10} \left( \frac{L_1^\mathrm{true}}{10^{22} W Hz^{-1}}\right) &= 0.936\,\log_{10} \left( \frac{L_1^\mathrm{obs}}{10^{22} W Hz^{-1}}\right) + 0.051,
\label{eq:offsets}
\end{align}

\noindent where the superscripts indicate the observed and ``true'' (i.e. model input) parameters. For the best-fit values quoted in section \ref{sec:sfrl150}, the difference in the parameters is small: $\Delta \beta \equiv \beta^\mathrm{true} - \beta^\mathrm{obs} = 0.016$ and $\Delta \log_{10} L_1 \equiv \log_{10} L_1^\mathrm{true} - \log_{10} L_1^\mathrm{obs} = 0.040$, with scatter about the best fit relation of $\sigma_\beta = 0.006$ and $\sigma_{\log_{10} L_1} = 0.006$, and we quote these systematic offsets and propagate the uncertainties on the values quoted in Section \ref{sec:sfrl150}.

Similarly for the 3D simulations, using 10 bins of $\beta$, $\log_{10} L_C$ and $\gamma$, the best fit relations are:

\begin{align}
\beta^\mathrm{true} &= 1.258\, \beta^\mathrm{obs} - 0.167, \\
\log_{10} \left(\frac{L_C^\mathrm{true}}{10^{22} W Hz^{-1}}\right) &= 0.931\, \log_{10} \left(\frac{L_C^\mathrm{obs}}{10^{22} W Hz^{-1}}\right) +0.112, \\
\gamma^\mathrm{true} &= 1.090\, \gamma^\mathrm{obs} - 0.108
\end{align}

For the best-fit values quoted in Section \ref{sec:massdep}, the difference in the quoted parameters is $\Delta \beta \equiv \beta^\mathrm{true} - \beta^\mathrm{obs} = 0.051$, $\Delta \log_{10} L_C \equiv \log_{10} L_C^\mathrm{true} - \log_{10} L_C^\mathrm{obs} = 0.104$ and $\Delta \gamma \equiv \gamma^\mathrm{true} - \gamma^\mathrm{obs} = -0.057$. The mean scatter about each best-fit relation dominates the uncertainties on these corrections, and the values are $\sigma_\beta = 0.011$, $\sigma_{\log_{10} L_C} = 0.016$ and $\sigma_\gamma = 0.037$.

\subsection{Flux density limit or mass dependence?}
\label{sec:upturn_test}

\begin{figure*}
  \centering
	\subfigure[Mass independent model]{\includegraphics[width=0.98\columnwidth]{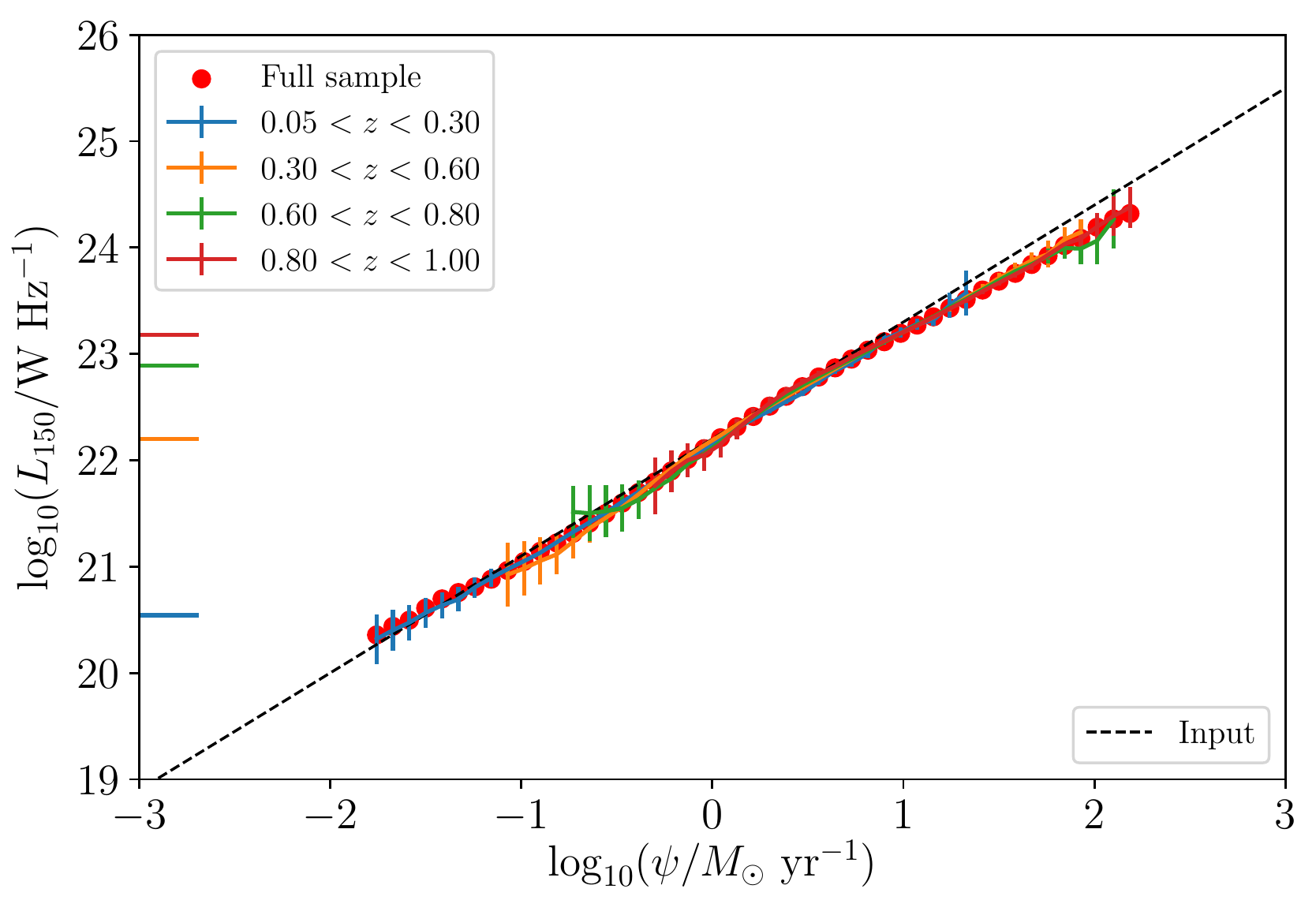}}
	\subfigure[Mass-dependent model]{\includegraphics[width=0.98\columnwidth]{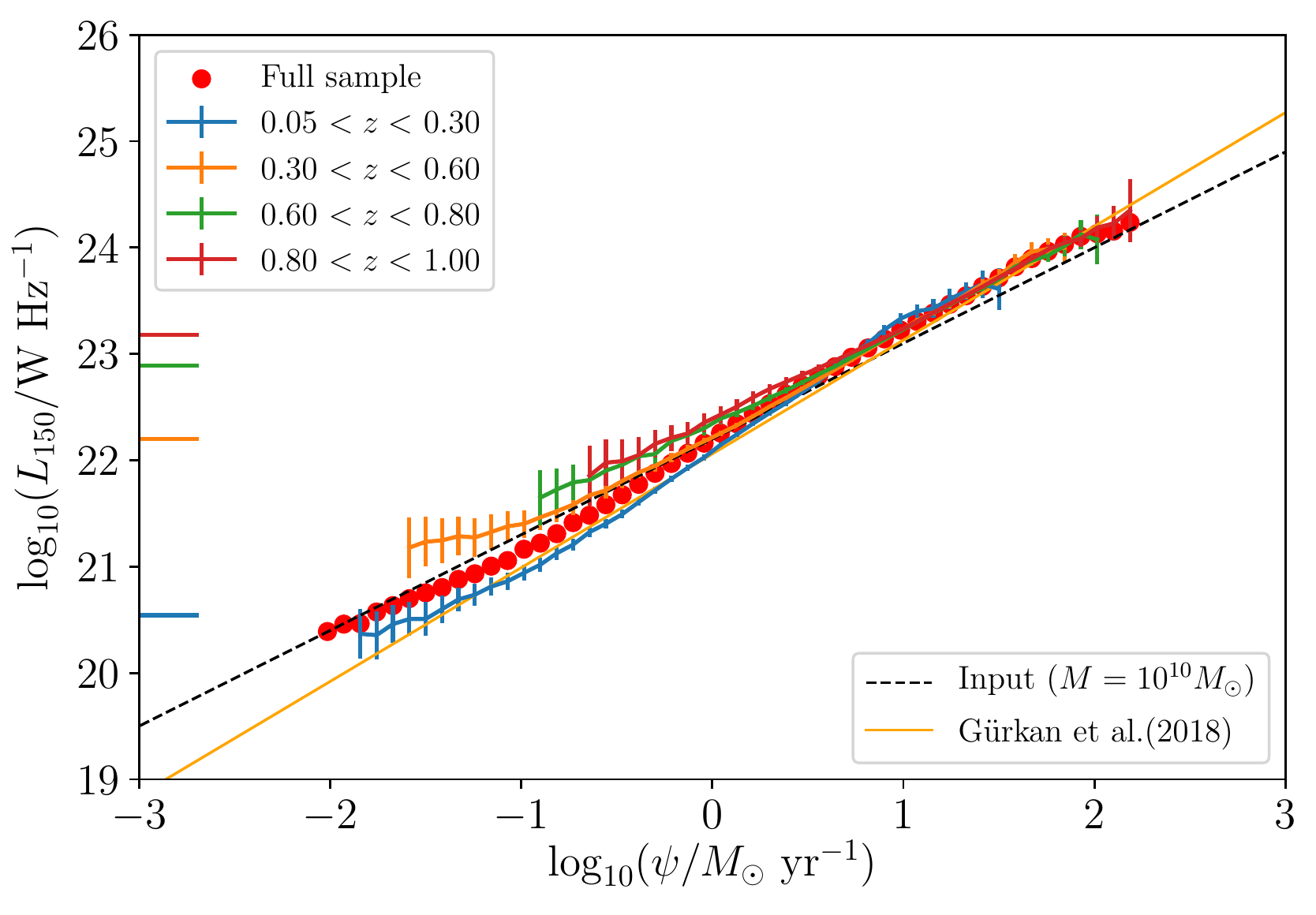}}
      \caption{Simulations of possible redshift evolution in the SFR-\Lradio\ relation using (a) a mass-independent input relation of the form given in Equation \ref{eq:sfrl150}, and (b) a mass-dependent input relation of the form given in Equation \ref{eq:sfmass}. The method used to obtain these figures is identical, and only the mass-dependent simulation reveals an upturn in \Lradio\ at lower SFRs, as seen for the real data set analysed in figure \ref{fig:slope_evolution} (see that caption for further details).}
         \label{fig:simtest_massdep}
\end{figure*}

Finally, to test whether the possible redshift-dependent upturn to higher \Lradio\ at lower SFRs shown in figure \ref{fig:slope_evolution} is real or an artefact of the finite signal to noise ratio available in the 150\,MHz data set, we perform two further tests. First, we created a simulation identical to the ones in the previous section with an input SFR-\Lradio\ relation that is independent of mass, and repeated the analysis of section \ref{sec:evolution} to see how well it was recovered. The left-hand panel of Figure \ref{fig:simtest_massdep} shows the results - while there is a slight bias towards a flatter SFR-\Lradio\ relation (as discussed in Appendix \ref{sec:methodtests}), there is no evidence of an upturn, including below the approximate flux density limits indicated for each redshift bin by the horizontal error bars adjacent to the left vertical axis.

In the right panel of Figure \ref{fig:simtest_massdep} we show the results of repeating an identical analysis using a second simulation, which includes a mass-dependent SFR-\Lradio\ relation of the form given in equation \ref{eq:sfmass}. The upturn to higher \Lradio\ towards lower SFRs is now clear, underlining our view that this effect -- if real -- is consistent with being a consequence of the mass-dependence in the SFR-\Lradio\ relation.

\end{appendix}
%\begin{thebibliography}{}

%\end{thebibliography}

\end{document}